\documentclass[12pt, a4paper]{article}
\usepackage[utf8]{inputenc}
\usepackage[T1]{fontenc}
\usepackage[absolute]{textpos}
\usepackage[top=2cm,bottom=2.5cm,left=2cm,right=2cm]{geometry}
\usepackage{amsmath,amssymb,mathtools,dsfont,emptypage,graphicx,dcolumn,bm,booktabs,colortbl,collcell,enumerate,float,verbatim,multirow,xparse}
\usepackage{cite}
\usepackage{tabularx}
\usepackage{multirow}
\usepackage{graphicx}
\usepackage{subcaption}
\usepackage{appendix}
\usepackage{slashed}
\usepackage{ragged2e}
\usepackage[labelfont=bf, labelsep=period]{caption}
\usepackage{contour}
\usepackage[colorlinks=true,urlcolor=myred,linkcolor=black,citecolor=mygreen]{hyperref}
\usepackage{cleveref}
\usepackage[normalem]{ulem}

\newcommand{\be}{\begin{equation}}
\newcommand{\ee}{\end{equation}}

\newcommand{\pfl}{p_{\rm fl}}
\newcommand{\efl}{e_{\rm fl}}
\newcommand{\calF}{{\cal F}}

\definecolor{amaranto}{RGB}{229,43,80}
\definecolor{myred}{cmyk}{0,1,1,0.55}
\definecolor{mygreen}{rgb}{0.27, 0.64, 0.48}
\definecolor{mygray}{gray}{.95}

\begin{document}

\rightline{DESY-26-040}
\rightline{CERN-TH-2026-058}

\begin{center}
{\bf\Large Slow-down 
of expanding bubbles in the early Universe} \\
[5mm]
\renewcommand*{\thefootnote}{\fnsymbol{footnote}}

Nabeen Bhusal$^{a}$
\footnote{\href{mailto:nabeen.bhusal@desy.de}
{nabeen.bhusal@desy.de}},
Simone Blasi$^{a}$
\footnote{\href{mailto:simone.blasi@desy.de}
{simone.blasi@desy.de}},
Thomas Konstandin$^a$
\footnote{\href{mailto:thomas.konstandin@desy.de}{thomas.konstandin@desy.de}},
\\
Enrico Perboni$^{a}$
\footnote{\href{mailto:enrico.perboni@desy.de}{enrico.perboni@desy.de}},
Jorinde van de Vis$^b$
\footnote{\href{mailto:jorinde.van.de.vis@cern.ch}{jorinde.van.de.vis@cern.ch}}
\\
\vspace{5mm}

$^{a}$\,{\it Deutsches Elektronen-Synchrotron DESY, Notkestr. 85, 22607 Hamburg, Germany}\\
$^{b}$\,{\it Theoretical Physics Department, CERN,
1 Esplanade des Particules, CH-1211 Geneva 23, Switzerland}
\end{center}

\centering
{\bf Abstract}\\
\justifying

We study slow-down effects for bubbles formed in a cosmological first-order phase transition (PT) focusing on deflagrations and hybrids, where the bubble wall is preceded by a shockwave of heated plasma. Slow-down has been observed in multi-bubble simulations together with a suppression of gravitational wave (GW) emission, mostly for slow walls. We study the impact of the shock waves on the wall velocity around percolation, by considering steady-state single-bubble solutions and incorporating the possible heating effects by two different mechanisms. First, we investigate the slow-down experienced by a bubble expanding into an impeding shockwave, where the temperature is higher than at nucleation, and the fluid is no longer at rest. Taking into account such heating and kinematic effects, we find that the most significant slow-down occurs for the fastest walls, and thus cannot explain the suppression of the GWs observed in the simulations. However, these effects are stronger for PTs with a sizeable change in degrees of freedom unlike what is usually implemented in simulations, suggesting that the degrees of freedom can be an important additional parameter for characterizing the GW spectrum. For the second slow-down mechanism, we study heated droplets of false vacuum that shrink towards the end of the PT. By implementing a suitable boundary condition motivated by energy conservation, we show how the droplet velocity, interpreted here as the late-time velocity of the bubble walls, can be predicted from the properties of the initial deflagration/hybrid, in remarkable agreement with numerical simulations. Droplets are found to shrink more slowly for stronger PTs and slower deflagrations, with mild dependence on the change of degrees of freedom. Such slow droplets naturally correlate with a suppression of GWs, while geometrical properties such as the shock width play an important role as well.

\vspace{2em}

\newpage
\tableofcontents
\vspace{2em}
\vspace{2em}

\renewcommand*{\thefootnote}{\arabic{footnote}}
\setcounter{footnote}{0}

\newpage
\section{Introduction}

Since the advent of gravitational wave detection experiments and their remarkable successes in recent years \cite{LIGOScientific:2018mvr, KAGRA:2021vkt, LIGOScientific:2025slb, NANOGrav:2023gor, EPTA:2023fyk, Reardon:2023gzh, Xu:2023wog}, the prospect of observing a cosmological background of gravitational waves (GWs) has become an exciting frontier in modern cosmology. Understanding the precise gravitational wave signatures arising from the dynamics of the early Universe is therefore of central importance. Among the most compelling sources of such a stochastic gravitational wave background are first-order phase transitions, which have long been recognized as promising candidates \cite{Witten:1984rs, Kosowsky:1991ua, Kamionkowski:1993fg}. In addition to generating gravitational waves, these transitions may also offer solutions to other outstanding cosmological puzzles, such as the origin of the baryon asymmetry of the Universe \cite{Kuzmin:1985mm} and the seeding of primordial magnetic fields \cite{Vachaspati:1991nm}.

Large scale numerical simulations performed in \cite{Hindmarsh:2013xza, Hindmarsh:2015qta, Hindmarsh:2017gnf} for weak and intermediate phase transitions demonstrated that the GW spectrum sourced by sound waves could be parameterized in terms of the bubble wall velocity and energy budget, obtained from a computation of a bubble in isolation.
This greatly simplified the prediction for the GW spectrum, which could thus be described by a fit function that only depends on a handful of microscopic and single-bubble parameters \cite{Caprini:2019egz}.
However, subsequent numerical simulations for stronger phase transitions \cite{Cutting:2019zws} have reported suppression of the GW signal compared to \cite{Caprini:2019egz}—sometimes even pushing it below the expected sensitivity range of upcoming experiments such as LISA \cite{amaroseoane2017laserinterferometerspaceantenna}.
The suppression of the GW amplitude observed in \cite{Cutting:2019zws} is accompanied by the generation of hot droplets of false vacuum, and vortical motion. 
The effect is most prominent for deflagration and hybrid solutions.

In this work, we explore several possible explanations of this suppression. 
Deflagrations and hybrid solutions feature a shock wave that precedes the bubble wall, and that can have a (significantly) higher temperature than the surrounding plasma.
We observe that heating of the plasma region above the nucleation temperature, $T_N$, contributes to the slowing down of bubbles nucleating or expanding in the vicinity of other bubbles. 
In extreme cases, this heating can even stop the growth entirely.
Besides, kinematic effects related to the bulk flow of the impeding shock can play an important role as well. 
Within this framework, 
we observe the strongest slow-down effects for initial velocities closest to the Jouguet velocity (the transition from hybrids to detonations), in contrast to numerical simulations where the strongest suppression in Ref.\,\cite{Cutting:2019zws} was instead observed for the smallest wall velocity considered, $\xi_w = 0.24$.
Nevertheless, an important outcome of this analysis is that the heating and kinematic effects described above turn out to depend significantly on the change of relativistic degrees of freedom between the two phases. In particular, these effects are stronger for equations of state more similar to the one of the Standard Model than the one used \emph{e.g.} in Ref.\,\cite{Cutting:2019zws}. In general, this suggests that the GW spectrum may not just be parameterized in terms of the usual four parameters (duration and strength of the phase transition, nucleation temperature, and wall velocity), but should also take into account the jump in the degrees of freedom.

To better understand the suppression of the GW spectrum observed in the simulations, we then provide a complementary approach based on the properties of droplet solutions, following up on the analysis of Ref.\,\cite{Cutting:2022zgd}.
Droplets are heated pockets of false vacuum forming around the time of percolation after the shock waves of the different bubbles have started to collide. These are characterized by a qualitatively different hydrodynamics compared to the standard expanding bubbles, as for instance the fluid and the wall interface are moving in opposite directions\,\cite{Rezzolla:1995kv,Rezzolla:1995br, Kurki-Suonio:1995yaf, Cutting:2022zgd,Barni:2024lkj}. Nevertheless, a droplet can represent the late-time evolution of an initial bubble expanding as a deflagration or hybrid, after the heating from shock waves is taken into account. The droplet velocity is then identified with the terminal velocity of the bubble wall. Compared to previous studies where this quantity was determined from 1D simulations\,\cite{Cutting:2022zgd}, in this paper we show how the velocity of the droplet can actually be predicted by implementing a suitable boundary condition motivated by energy conservation to solve the fluid equations. Such boundary condition fixes the droplet inner temperature, which is in general unknown as it reflects neither the nucleation temperature nor the temperature inside of the shock wave. By implementing this procedure, we obtain remarkable agreement with all the available simulation data for the droplet/late-time wall velocity\,\cite{Cutting:2019zws,Cutting:2022zgd, Correia:2025qif}.
We also recast the numerical results presented in Ref.\,\cite{Cutting:2019zws} in terms of our predicted droplet velocity and the initial size of the shock to further characterize the origin of the GW suppression. The picture is consistent with slower droplets generally implying a stronger suppression. The size of the initial shock wave appears to play an important role as well, with larger shocks yielding a stronger suppression of the GWs. We also notice that a droplet solution is not always guaranteed to exist as a consistent late-time evolution of a deflagration or hybrid, in particular for initial walls that are too fast. On the other hand, we find that droplet solutions can be obtained also in scenarios with negligible friction, namely in the so-called local-thermal-equilibrium (LTE) approximation\,\cite{Ai:2021kak}. 

Throughout this paper, we will always consider steady-state self-similar (spherical) solutions of the hydrodynamical equations. 
As equations of state, we will consider the well-known \emph{bag model} customarily used in numerical simulations, as well as other equations of state from explicit particle physics models in the context of the electroweak phase transition. 
In our analysis, we will be mostly including friction between the bubble wall and the plasma in the limit of local interactions, namely by including the operator $\mathcal{K}(\phi) \propto u_\mu\partial^\mu \phi$, with either a constant coefficient, or factoring in an overall $\phi^2/T$ dependence. 
We will also derive perturbative constraints on the consistent use of this operator based on the analysis in Ref.\,\cite{Ekstedt:2025awx}, and further compare the results obtained within the local-friction approximation with the outcome of the Boltzmann equations using {\tt WallGo} \cite{Ekstedt:2024fyq}. 

\section{Equations of motion for the scalar field and the fluid}\label{sec:EOMs}

We are interested in the dynamics of the coupled scalar field and the fluid. Here, we summarize the relevant equations, following the analysis of Refs.~\cite{Konstandin:2014zta, Ekstedt:2025awx}.
In this work, we assume for simplicity that only a single scalar field $\phi$ is involved in the phase transition. In local thermal equilibrium, we can describe the plasma as a perfect fluid
\begin{equation}
    T^{\mu\nu}_{\rm fl} = \omega u^\mu u^\nu - g^{\mu \nu} \pfl,\label{eq:perfectFluid}
\end{equation}
with $\pfl$ the thermodynamic pressure, containing the contribution proportional to $T^4$, as well as (minus) the temperature-dependent contribution to the scalar effective potential. 
 The enthalpy is given by $\omega = T d\pfl/dT$, $u^\mu = \gamma (1,\vec v)$ the fluid velocity, and $g^{\mu\nu}$ the spacetime metric (which we will assume to be the Minkowski metric).
 $\gamma$ denotes the Lorentz factor.
 This form of the energy-momentum tensor is valid sufficiently far away from the wall. In Appendix \ref{app:hydro}, we discuss the hydrodynamic solutions that describe the fluid at scales much larger than the bubble wall width and the mean free path of the particles.

Inside the bubble wall, the scalar field is governed by the following equation of motion
\begin{equation} \label{eq: KG eq_Higgs}
    \Box \phi + \frac{dV_0}{d \phi} + \sum_i N_i \frac{d m_i^2}{d\phi} \int \frac{d^3 p}{(2\pi)^3} \frac{1}{2E_i} f_i(p^\mu, x^\mu) =  0,
\end{equation}
where $V_0$ denotes the zero-temperature potential, and the $f_i$ denote the distribution functions of the plasma particles. $N_i$ denotes the number of degrees of freedom of each species $i$, and $m_i$ denotes the mass.
The distribution functions can be split into a local equilibrium part and an out-of-equilibrium part
\begin{equation}
    f_i(p^\mu, x^\mu) = f_{i}^{\rm eq}(p^\mu, x^\mu) + \delta f_i(p^\mu, x^\mu),
\end{equation}
where the contribution of the equilibrium part can be combined with $V_0$ to give the thermally corrected one-loop effective potential $\mathcal F$, and $\delta f_i$, which gets sourced by the passing bubble wall, provides the friction on the wall.
The $f_i$ satisfy a Boltzmann equation, that can be solved numerically (which is typically done after linearizing in $\delta f_i$).
These deviations from equilibrium affect also the energy-momentum tensor of the fluid, 
which can also be determined from
\begin{equation}
    T^{\mu\nu}_{\rm fl} = \sum_i N_i \int\frac{d^4p}{(2\pi)^3}p^\mu p^\nu f_i(p^\mu,x^\mu) \delta(p^2 - m_i^2).
\end{equation}

In principle, the scalar-field equation of motion, the Boltzmann equations and the condition of energy-momentum conservation can all be solved to determine the scalar field profile, the particle distributions, the fluid velocity and temperature profiles, as well as the wall velocity, see e.g.~\cite{Moore:1995si, Konstandin:2014zta, Laurent:2022jrs, Ekstedt:2024fyq, Dorsch:2024jjl}.
In practice, however, such a computation is not always feasible, and large-scale numerical simulations often parameterize the bubble-fluid interactions by the effective operator $\mathcal{K}$ in the following way: 
\begin{equation}
    \partial_\mu T^{\mu\nu}_{\rm fl} = \partial^\nu \phi (\partial_\phi \pfl 
    + \mathcal K ) , \qquad  
    \partial_\mu T^{\mu\nu}_\phi = - \partial^\nu \phi (\partial_\phi \pfl 
    + \mathcal K ) , \qquad  
\end{equation}
where $T^{\mu\nu}_{\rm fl}$ corresponds to the perfect fluid energy-momentum tensor of \cref{eq:perfectFluid},\footnote{
This assumption neglects deviation from the perfect fluid in the bubble wall. Using {\tt WallGo} \cite{Ekstedt:2024fyq} for some representative benchmark points we have checked that including these deviations from equilibrium affect the wall velocity at the percent level only.
} and the scalar field energy-momentum tensor is given by
\begin{equation}
T^{\mu \nu}_{\phi} = \partial^\mu \phi \partial^\nu \phi - g^{\mu \nu} \left ( \frac{1}{2} \partial_\rho \phi \partial^\rho \phi - V_0(\phi)  \right).
\end{equation}
Correspondingly, the scalar field equation of motion is given by
\begin{equation}
    \Box \phi + \frac{\partial \mathcal F}{\partial \phi} + \mathcal K(\phi) = 0 \, ,\label{eq:scalarEOM}
\end{equation}
where we defined the free energy $\calF = V_0 - \pfl = -p$.

\begin{figure}[t!]
    \centering
    \includegraphics[width=0.49\linewidth]{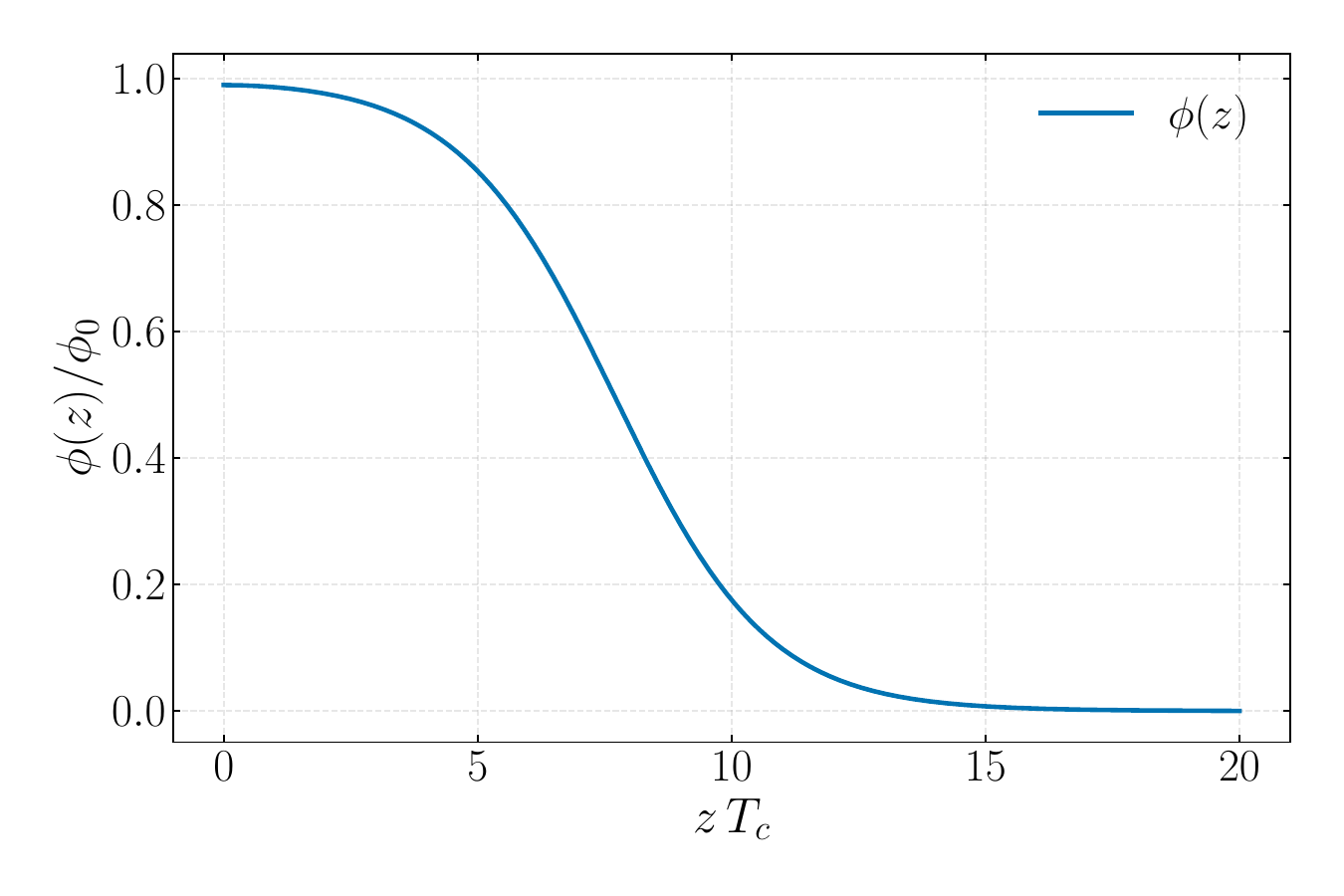}
    \caption{Field profile $\phi(z)$ for one of the solutions of the coupled plasma-wall system.}
    \label{fig: phi_profile}
\end{figure}

\begin{figure}[t!]
    \centering
    \includegraphics[width=0.49\linewidth]{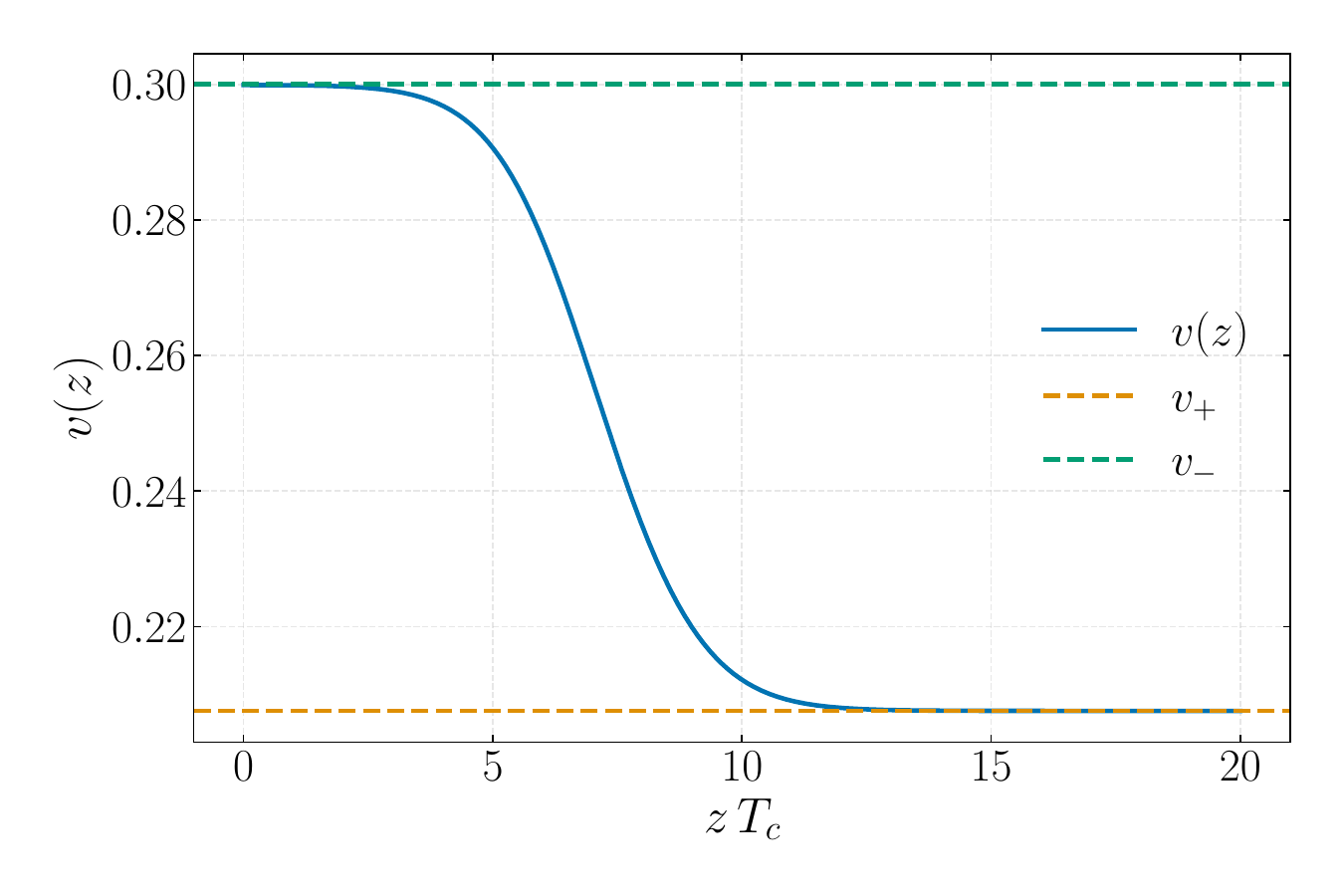}
    \hfill
    \centering
    \includegraphics[width=0.49\linewidth]{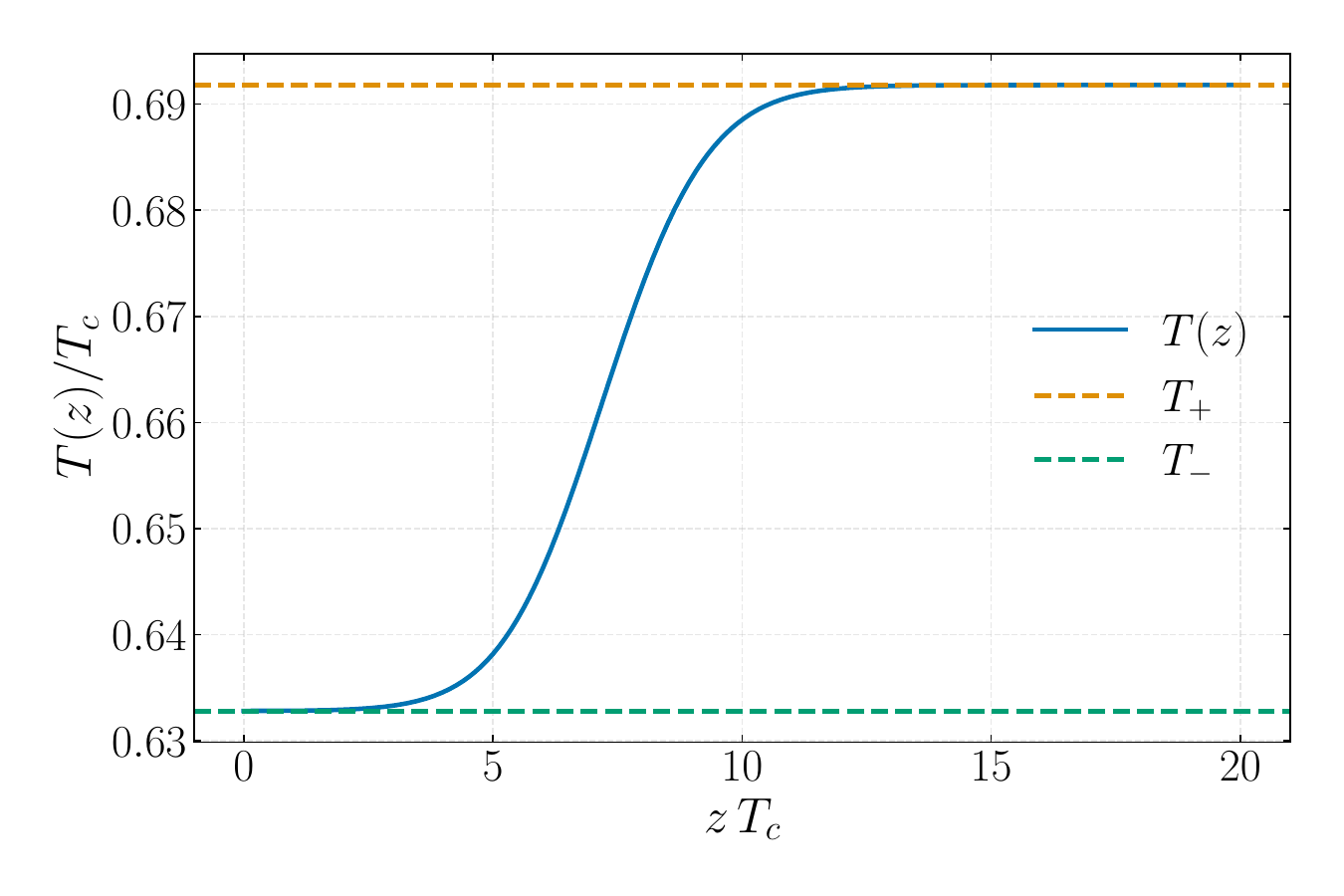}
    \caption{Temperature and velocity profile obtained from energy-momentum conservation. The asymptotic values of these quantities both in front and behind the wall obtained from the matching conditions are also plotted.}
    \label{fig. thermo_profiles}
 \end{figure}

We will assume that the bubbles are sufficiently large such that we can work in the planar limit, where, in the wall frame, everything becomes only a function of the distance to the bubble wall $z$.
Energy momentum conservation $\partial_\mu T^{\mu\nu} = 0$ then prescribes that
\be
    T^{z0} = \mathrm{k_2} , \hspace{2em} T^{zz} = \mathrm{k_2},
\ee
where $T^{\mu\nu} = T^{\mu\nu}_{\rm fl} + T^{\mu\nu}_\phi$ and
the two constants k$_1$ and k$_2$ can be determined knowing the hydrodynamic solutions far in front or far behind the wall (see Appendix \ref{app:hydro}). 
Inserting explicitly the form of the total energy-momentum tensor $T^{\mu \nu}$, the two conditions can be rewritten as 
\be \label{eq. non-linear-em}
    v^2 \, \gamma^2 \, \omega - \mathcal{F} + \frac{1}{2} \left ( \partial_z \phi \right )^2 = \mathrm{k}_1, \hspace{2em}
    v^2 \, \gamma^2 \, \omega = \mathrm{k}_2.
\ee
These two equations can be solved, once combined with the scalar field equation of motion, for finding $\phi(z)$ and the temperature and velocity profiles inside the bubble wall, i.e $T(z)$ and $v(z)$, as can be seen in Figs.~\ref{fig: phi_profile} and \ref{fig. thermo_profiles}.  Two different parameterizations for $\mathcal K$ appear commonly in the literature and will be compared in this work.

\paragraph*{Constant friction.}
A simple Lorentz-invariant choice which can be made, even if it does not lead to the correct behavior for highly relativistic bubble walls described by \cite{Bodeker:2009qy}, is to write
\be
    \mathcal{K}(\phi) = T_c \, \eta \, u^\mu \, \partial_\mu \phi,\label{eq:constfric}
\ee
where a factor of $T_c$ is inserted to keep $\eta $ dimensionless. This parameterization of the friction is the one used in most of the hydrodynamical simulations \cite{Hindmarsh:2013xza, Hindmarsh:2015qta, Cutting:2019zws, Correia:2025qif}.

\paragraph*{Field-dependent parameterization.} 
It has been shown in \cite{Huber:2013kj, Konstandin:2014zta, Ekstedt:2025awx}, that instead of the the typical constant friction parameter $\eta$, a better parameterization for the friction coefficient is 
\begin{equation}
    \mathcal{K}(\phi) = \frac{\tilde \eta \, \phi^2}{T} \, u^\mu \, \partial_\mu \phi.
    \label{eq: new_eta}
\end{equation}
This parameterization occurs naturally when dealing with models for which the masses are proportional to the vacuum expectation value (VEV) of the scalar field $\phi$, i.e. $m_i \propto \phi$.
The hydrodynamic simulations of \cite{Hindmarsh:2017gnf} used a field-dependent parameterization of the friction.

\section{Benchmark Models}
We introduce the benchmark models used for our numerical results.
To compare different models, we find it useful to introduce the phase transition strength defined at the nucleation temperature, $\alpha_N$ as in \cref{eq. alpha def.}.

\subsection{Standard Model with low cut-off}

In order to study a model that provides a realistic description of a Standard-Model-like electroweak phase transition,
we choose a simple extension of the SM with an effective $\phi^6$ operator, representing new physics coming into play at a scale $\Lambda$. This framework allows for a first-order phase transition with a Higgs mass compatible with present LHC data~\cite{Grojean:2004xa, Croon:2020cgk, Chala:2025aiz}.
Note however that the validity of the EFT has been questioned in \cite{Damgaard:2015con, deVries:2017ncy, Postma:2020toi}.
The new physics is supposed to contribute to the Higgs potential, providing a thermal barrier that induces a first order phase transition, but is assumed not to contribute significantly to the friction, thus allowing us to model the plasma as containing only SM particles. For this model, the high temperature expansion of the effective potential $V( \phi, T) $ can be written as
\begin{equation}
    V(\phi, T) \equiv \mathcal{F}(\phi, T) = V_0(\phi) - \frac{a}{3} T^4 + \frac{c}{2} \phi^2 T^2  \, , 
\end{equation}
and  
\begin{equation} \label{eq. parameters SM w/ low cutoff}
    a = g_* \dfrac{\pi^2}{ 30},
    \quad\text{and}\quad
    c = \frac{1}{16} \biggr(4 y_t^2 + 3 g^2 + 4 \frac{m_H^2}{v^2} - 12 \frac{v^2}{\Lambda^2} \biggr),
\end{equation}
with $g_*=106.75$ the SM effective number of relativistic degrees of freedom in the unbroken phase,
and $V_0(\phi)$ the tree-level zero temperature potential given by 
\begin{equation} \label{eq. zero-temp-pot}
    V_0(\phi) = - \frac{\mu^2}{2} \phi^2 + \frac{\Tilde{\lambda}}{4} \phi^4 + \frac{1}{8 \Lambda^2} \phi^6.
\end{equation}
The parameters of this potential are set in such a way that we recover the measured SM values for the Higgs mass $m_H = 125$ GeV and the Higgs VEV $v=246.22$ GeV, yielding
\begin{equation}
    \mu^2 = \frac{m_H^2}{2} - \frac{3}{4} \frac{v^4}{\Lambda^2}, 
    \hspace{2 em}
    \tilde{\lambda} = \frac{m_H^2}{2 v^2} - \frac{3}{2} \frac{v^2}{\Lambda^2}.
\end{equation}
We choose the mass of the $Z^0$ boson to be the same as the one of the $W^\pm$ bosons, i.e. approximating the coupling $g' \simeq 0$, an approximation that is also typically made when solving the Boltzmann equations for the top quark, the $W^\pm$ and the $Z^0$ bosons. 
It should be noted here that, in contrast to the pure SM scenario, the value of the quartic coupling $\Tilde{\lambda}$ is negative for regions of the parameter space where a first order phase transition occurs, and this is possible due to the presence of the $\phi^6$ term which stabilizes the potential. We only include thermal corrections up to order $\mathcal{O}(m^2 T^2)$ since for these classes of models the first-order nature of the phase transition does not rely on the presence of cubic thermal contributions of the bosons $\propto \phi^3 T$, as happens for example in SM like models with a lighter Higgs. 

\subsection{Bag Model}

A model that has been used extensively in the literature as an effective toy model, is the bag equation of state,
which parameterizes the symmetric and broken phase by pure radiation, with a temperature-independent vacuum energy difference, called the \emph{bag constant}.
Even though the model is very simple, it correctly captures the hydrodynamics of models with a speed of sound sufficiently close to $c_s^2 = 1/3$. 
In the electroweak phase transition, often $c_s^2 \sim 1/3$, due to the large amount of light degrees of freedom. 
In lattice simulations for gravitational waves \cite{Cutting:2019zws, Correia:2025qif}, the equation of state is supplemented with a scalar field, with the following effective potential
\[
V(\phi, T) = \mathcal{F}(\phi,T) = V_0(\phi) - a(\phi)  \, T^4 \,,
\]
with $V_0$ the zero temperature potential
defined as, 
\be
V_0(\phi) = \frac{1}{2} M^2 \phi^2 - \frac{1}{3} \mu \phi^3 + \frac{\lambda}{4} \phi^4 - V_c,
\ee
where $V_c$ is chosen such that $V_0 (\phi_b) = 0 $, and $\phi_b$ is the VEV of the scalar field $\phi$ in the broken phase at $T=0$. The potential energy difference between the broken and unbroken vacuum is denoted by $\Delta V_0 = V_0(0) - V_0 (\phi_b)$. 
The function $a(\phi)$ models the change in the degrees of freedom during the phase transition. It is expressed in terms of the potential difference $\Delta V_0$ and the 
critical temperature $T_c$.
\[
a(\phi) = a_0 - \frac{\Delta V_0}{T_c^4} \left[ 3 \, \left( \dfrac{\phi}{\phi_0} \right)^2 - 2 \, \left( \dfrac{\phi}{\phi_0} \right)^3 \right],
\]
where $a_0 = (\pi^2/90)g_*$.\footnote{There is a factor of 3 different between this definition of $a_0$ and the one of $a$ in \eqref{eq. parameters SM w/ low cutoff}; this is due to a different normalization used in the definition of the thermal contribution $\propto T^4$ in the free-energy $\mathcal{F}$.} Notice that this model is constructed in such a way that both $\phi = 0$ and $\phi = \phi_b$ are stationary points of the free-energy $\mathcal{F}$ for all $T$, i.e. that the VEV of the scalar field is temperature independent.

Thermodynamic quantities like enthalpy and energy density are then given by
\[
\omega = T \, \frac{\partial p}{\partial T} = 4 \, a(\phi)\, T^4, \hspace{0.5 cm} e = \omega - p = 3 \, a(\phi) \, T^4 + V_0(\phi).
\]
For the parameters of the zero-temperature potential $V_0(\phi)$, our choice reflects the one in \cite{Cutting:2019zws}, fixing $M^2 = 0.0427 \, T_c^2$, $\mu = 0.168 \, T_c$ and $\lambda = 0.0732$, in order to have $\phi_b = 2.0 \, T_c$. 
To match \cite{Cutting:2019zws}, we set the relative change in degrees of freedom to $\left[ a_0 - a(\phi_b) \right] / a_0 \equiv \delta a/a = 5.9 \, \times \, 10^{-3} $, but we will also consider the value $\delta a/a = 5.9 \times 10^{-2}$ for comparison, by fixing $M^2 = 0.1662 \, T_c^2$, $\mu = 0.312 \, T_c$ and $\lambda = 0.0732$.
At this point the only free parameters left in the model are $T_N$ and $T_c$, but since the only true dependence is on their ratio, we fix $T_c$ and vary $T_N$ to span over different values of the phase transition strength $\alpha_N$.\footnote{ 
Notice that $T_N$ is not the temperature where the phase transition actually would happen according the probability to nucleate bubbles. Changing the temperature
allows us to choose the strength of the phase transition $\alpha_N$ at will.}

\subsection{Standard Model coupled to a singlet}
We consider the Higgs coupled to a gauge singlet $s$, also called the xSM, with zero-temperature scalar potential:
\begin{equation}
\label{eq:Vhs}
    V_0(\phi, s) = -\frac{\mu^2}{2}\phi^2 + \frac{\lambda}{4} \phi^4 + \frac 1 2 \mu_s s^2 + \frac{\lambda_s}{4}s^4 + \frac{1}{4} \lambda_{hs}h^2 s^2.
\end{equation}
In this model, the phase transition can occur in two steps: first, the singlet obtains a vacuum expectation value, and in the second step, the singlet returns to zero, but the Higgs obtains its vacuum expectation value. 
The second step of the phase transition can be of first order.
Here, we consider a parameter point with $\lambda_{s} = 1.0$, $\lambda_{hs} = 0.9$ and a singlet mass $m_s = 120 \, {\rm GeV}$.
For further details of the effective potential implemented in our analysis, see \cite{Laurent:2022jrs, Ekstedt:2024fyq}, and also \cite{Blasi:2022woz,Agrawal:2023cgp,Blasi:2023rqi,Li:2023yzq,Bai:2025qch} for the consequences of the $\mathbb{Z}_2$-symmetric  potential in \eqref{eq:Vhs} for the phenomenology of the phase transition.
This model is considered in \cref{fig:vwvsvsWallGo}.

\subsection{Standard Model with a light Higgs mass}
Without new physics, the Standard Model does not feature a first order phase transition, but a smooth cross-over.
If the Higgs boson were lighter than 72 GeV however, the electroweak phase transition would be first order \cite{Kajantie:1996mn, Gurtler:1997hr}.
Despite being experimentally ruled out, the Standard Model with a light Higgs is an interesting toy model for a phase transition that is radiatively generated, with the appropriate particle content.  
This model is used in \cref{fig:vwvsvsWallGo}, where we use a Higgs mass of $34.0 \, {\rm GeV}$, and keep the masses of the gauge bosons and quarks at their measured values. For further details of the implementation, see \cite{Moore:1995si, Ekstedt:2024fyq}.

\section{Wall velocity with local friction} \label{sec. numerical setup}
In order to find the value of $\eta \, (\tilde \eta)$ (see \cref{sec:EOMs}), corresponding to a certain input wall velocity $\xi_w$ and strength $\alpha_N$,
we need to solve the coupled system of the three equations \eqref{eq:scalarEOM} and \eqref{eq. non-linear-em} and find the steady-states profiles of the scalar field $\phi(z)$, the temperature $T(z)$ and fluid velocity $v(z)$ across the phase transition front. To solve the coupled system we need to specify the boundary conditions: for the hydrodynamic quantities $T(z)$, $v(z)$ this is done by solving the matching conditions to find $T_-$ and $v_-$ (or equivalently $T_+,v_+$) for every value of the input $\xi_w$. For the scalar field $\phi(z)$ the conditions are 
\be
    \phi(-\infty) = \phi_0 (T_-) \hspace{2em} \text{and} \hspace{2em} \phi(\infty) = 0,
\ee
where $\phi_0(T)$ is the scalar VEV for the value of the temperature $T_-$ inside the bubble consistent with the matching conditions. 
The identification of the temperature and velocity of the plasma at $z = \pm \infty$ with the matching conditions of the hydrodynamic solution requires that the width of the wall is much smaller than the one of the shock/rarefaction wave, which in general is a good approximation.

Often in the literature, the field profile is described with a $tanh$-ansatz of the kind $\phi(z) = \frac{1}{2} \phi_0 \left[ 1 + \tanh(z/L_w )\right ]$.
In Refs.~\cite{Moore:1995si, Laurent:2022jrs} this was shown to be a reasonable approximation. 
Nevertheless, here we do not make assumptions about the shape of the scalar field, and solve for general $\phi(z)$.

\begin{figure}[t]
    \centering
    \includegraphics[width=0.48\linewidth]{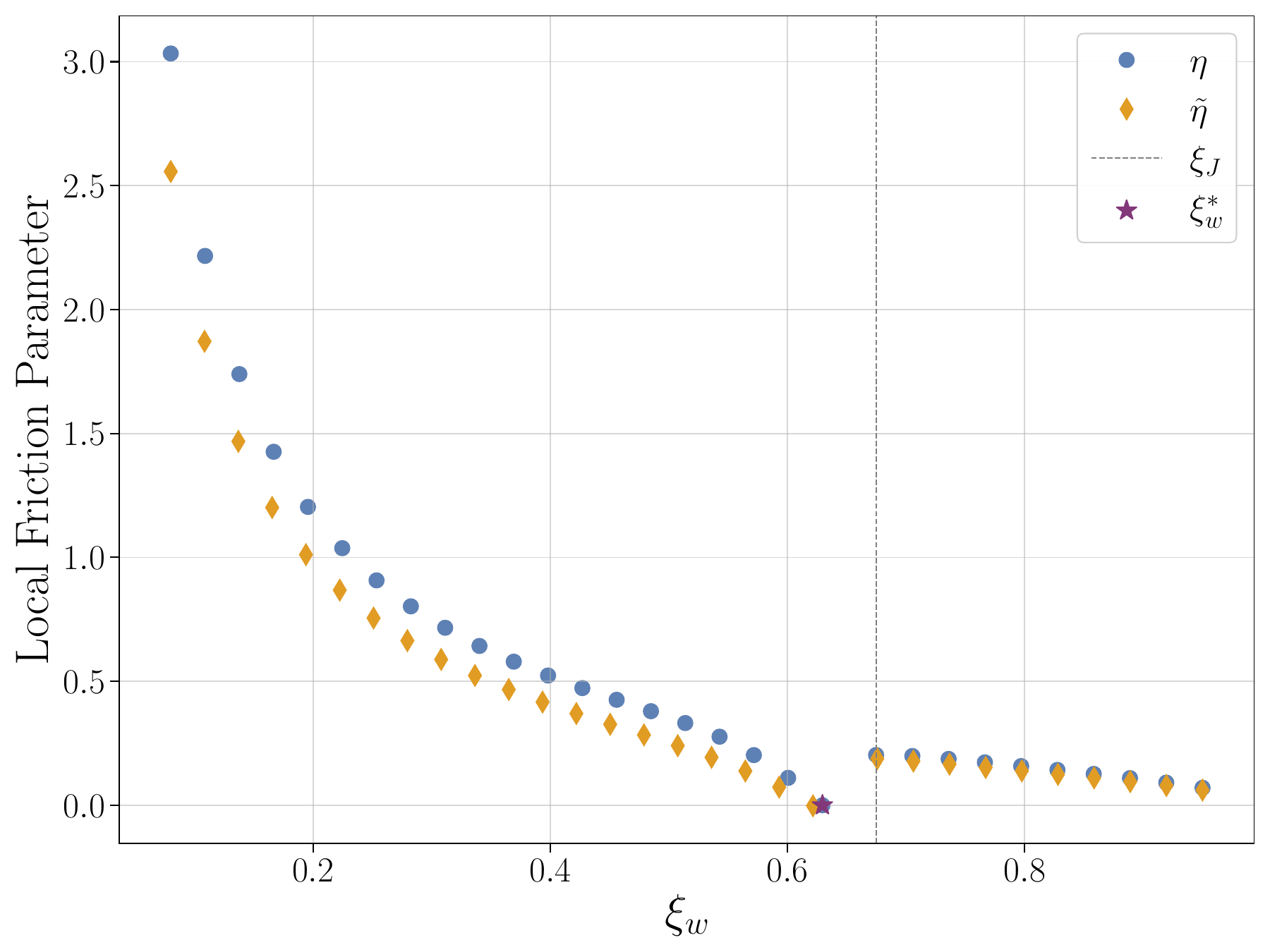}
    \hfill
    \includegraphics[width=0.48\linewidth]{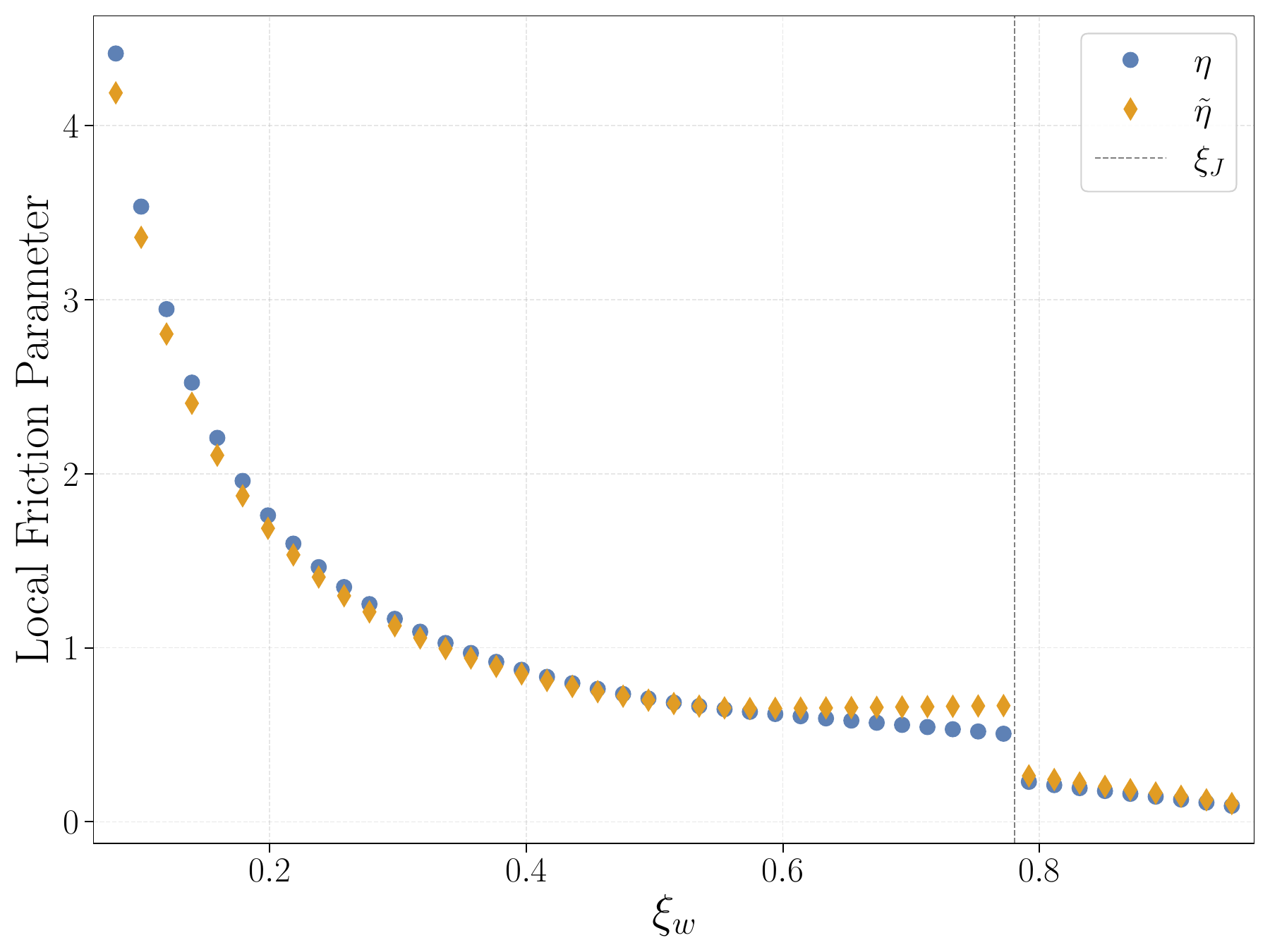}
    \caption{Plot of the friction coefficient $\eta$ in blue and $\tilde \eta$ in orange as a function of the wall velocity $\xi_w$. The left panel shows the values for the SM with a low-cutoff with $\alpha_N = 0.021$, corresponding to $\Lambda = 650 $ GeV. This model has a local thermal equilibrium solution represented in the figure by the purple star that is consistently found as the $\eta \to 0 $ limit in our procedure to solve the Klein-Gordon equation of motion. The right panel shows the friction coefficients for the Bag Model with $\delta a/a = 5.9 \times 10^{-3}$ and $\alpha_N = 0.11$.}
    \label{fig: eta_vs_vw_phi6_model}
\end{figure}

An example of a field profile $\phi(z)$ is shown in \cref{fig: phi_profile}, while examples for the fluid temperature and velocity are shown in \cref{fig. thermo_profiles}, with also their  asymptotic values obtained from the matching conditions \cref{eq:match1} and \cref{eq:match2}, plotted as dashed lines representing a consistency check. The resulting values of $\eta$ $(\tilde \eta)$ as a function of the wall velocity $\xi_w$ are shown for the Standard Model with low cut-off and the bag model in \cref{fig: eta_vs_vw_phi6_model}. The two different colors represent the two different parameterizations of the friction discussed in \cref{sec:EOMs}. It is important to point out a few aspects:
\begin{itemize}
    \item[I.] The vertical gray dashed line represents the Jouguet velocity $\xi_J$. Here, a discontinuity in the value of the friction parameters $\eta (\tilde \eta)$ appears. The reason is that there is a sudden change in the boundary conditions at $\xi_J$. Equivalently, the shock wave disappears at $\xi_w > \xi_J$, creating a physical jump from some $T_+ \to T_N$. No such discontinuity is present at $\xi_w = c_{s,b}$, the speed of sound of the broken phase, where the solution changes from deflagration to hybrid. 
    \item[II.] The purple star in the left panel of \cref{fig: eta_vs_vw_phi6_model} represents the local thermal equilibrium solution for the wall velocity $\xi_w^*$, which is computed independently without solving the KG equation of motion but instead using
    entropy conservation (see \cite{Ai:2021kak, Ai:2023see} and \cref{eq. 3rd matching lte} in the Appendix). Indeed, the value is consistent with the one found from the $\eta \, (\tilde \eta) \to 0$ limit in the coupled system of hydrodynamics and KG equations.
    Also notice that not all chosen parameter points possess a local-thermal-equilibrium (LTE) solution, with the right panel of \cref{fig: eta_vs_vw_phi6_model} being an example of this.
    
    \item[III.] The possibility of finding LTE solutions is limited to the region with $\xi_w < \xi_J$. Consequently, no stable detonation can be found without friction coming from the out-of-equilibrium fluctuations of the particles in the plasma.
    \item[IV.]  The two different parameterizations of the friction $\mathcal{K}(\phi)$ appear to differ quantitatively rather than qualitatively in the solutions for the friction coefficients. The impact of these changes in the friction parameter $\eta \, (\tilde \eta)$ on the correction to the wall velocity $\xi_w$ related to the heating of the plasma will be discussed in \cref{sec: temp and kin effects}.
\end{itemize}

\section{Slow-down in the background of an impeding shock}
\label{sec: temp and kin effects}

In this section, we present our first approach to describe the slow-down of the bubble walls arising for deflagration and hybrid solutions due to the presence of a shock wave. 
As detailed below, this approach takes into account the heating and kinematic effects acting on a bubble that tries to expand in the background of an impeding shock, where the temperature is larger than $T_N$ and the fluid already possesses a non-zero bulk motion.

\subsection{Heating and kinematic effects}

We consider here two effects to slow down the wall. One arises from the 
bulk motion in the shock and is model-independent. The other 
stems from the heating in the shock, which depends on the equation of state
in the symmetric phase.

\vskip .3 cm {\bf Heating effect.}
The latent heat is injected as thermal energy and bulk motion into the plasma, thus 
heating the plasma in front of the wall for deflagrations and hybrids.
The temperature-dependent potential $V(\phi,T)$ relevant to a bubble moving in the shocks of other bubbles thus gets modified as a consequence, and needs to be considered now at a different temperature $T \neq T_N$. In general, the temperature ranges between $T_+$ and $T_{\text{sh}}$, where $T_{\text{sh}}$ is the temperature of the plasma at the shock front.
The change in the potential shape is such that the free energy $\Delta \mathcal{F}$ available to the phase transition is reduced as the potential is modified towards symmetry restoration. In some cases, the plasma temperature in front of the wall, denoted by $T_+$ can even reach the critical temperature $T_c$. This means that the potential no longer exhibits the global minimum corresponding to the true vacuum and bubbles cannot grow into the symmetric phase.
Consequently, the driving pressure of a second bubble expanding into another bubble is reduced, leading to a slow-down and overall suppression of the GW signal. 
 
We determine the slow-down by recomputing the wall velocity, $\tilde \xi_w$, with a fictitious nucleation temperature, given by $T_+$ of the first bubble.
First, we repeat the numerical procedure described in \cref{sec. numerical setup}, to determine $\eta (\tilde \eta)$ for $\xi_w$ in a given model.
Then, we determine $\tilde \xi_w$, using the fictitious nucleation temperature. Now we keep $\eta (\tilde \eta)$ fixed, since it is a property of the model.
$\tilde \xi_w$ can be understood as an estimate of the wall velocity after slow-down due to heating. 
 
\vskip .3 cm {\bf Kinematic effect.}
On top of the heating effect, the plasma is being pushed away by the background shock. This means that a bubble nucleating or expanding into the shock region of another bubble will not suffer only from less driving pressure but also from the pushback of the plasma. In other words, the new wall velocity should be computed first considering a new, higher temperature and then Lorentz boosted to the reference frame moving with the shock wave.
In the extreme case where a bubble wall moves head-on through the shock of 
another bubble, the fluid moves with the velocity $\mu(\xi_w, v_+)$
and the new wall with the velocity $v_+$. 

\begin{figure}[t!]
    \centering
    \textbf{SM with low cut-off} \\
    \includegraphics[width=0.48\linewidth]{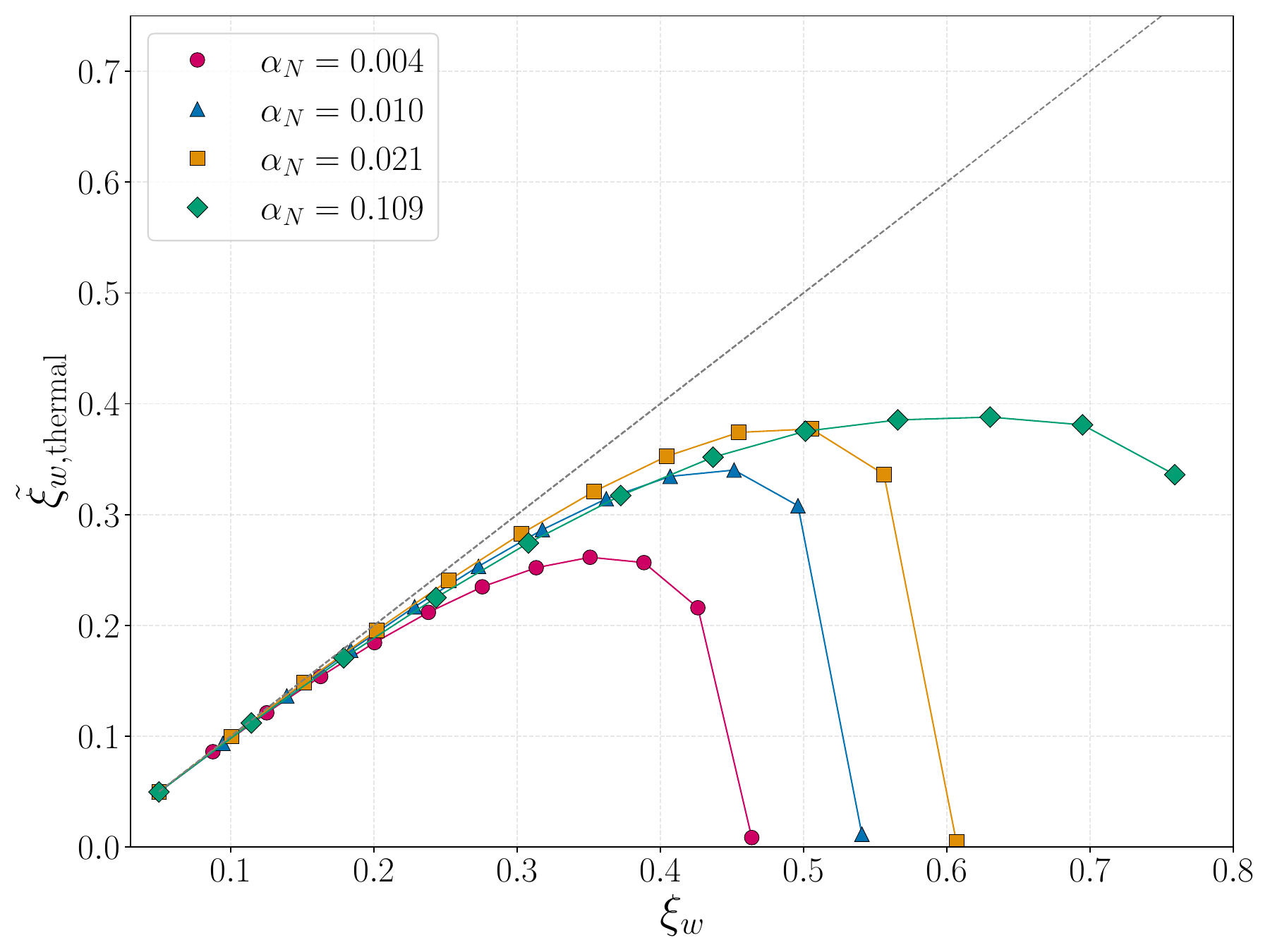}
    \hspace{1em}
    \includegraphics[width=0.48\linewidth]{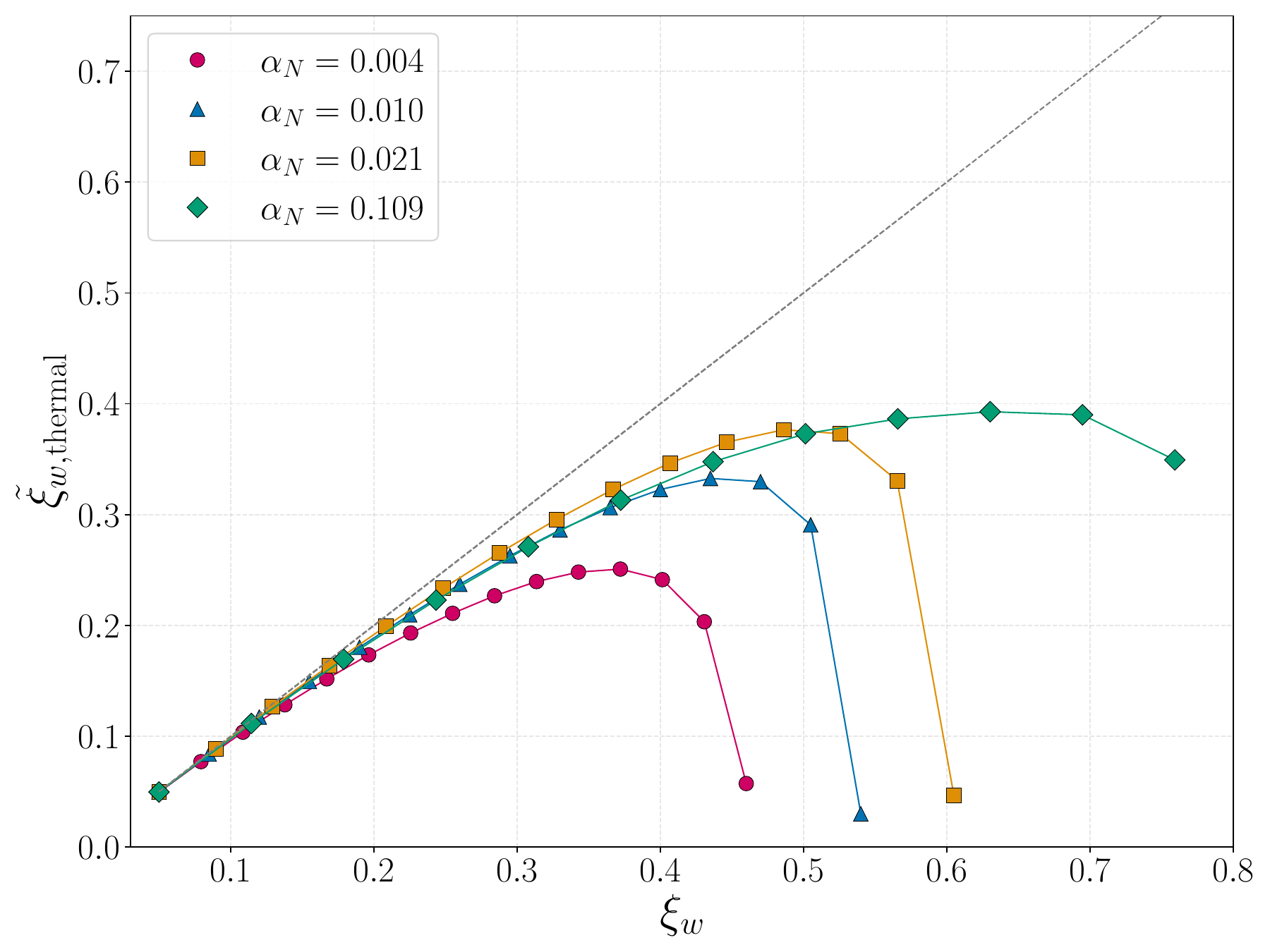}
    \\
    \includegraphics[width=0.48\linewidth]{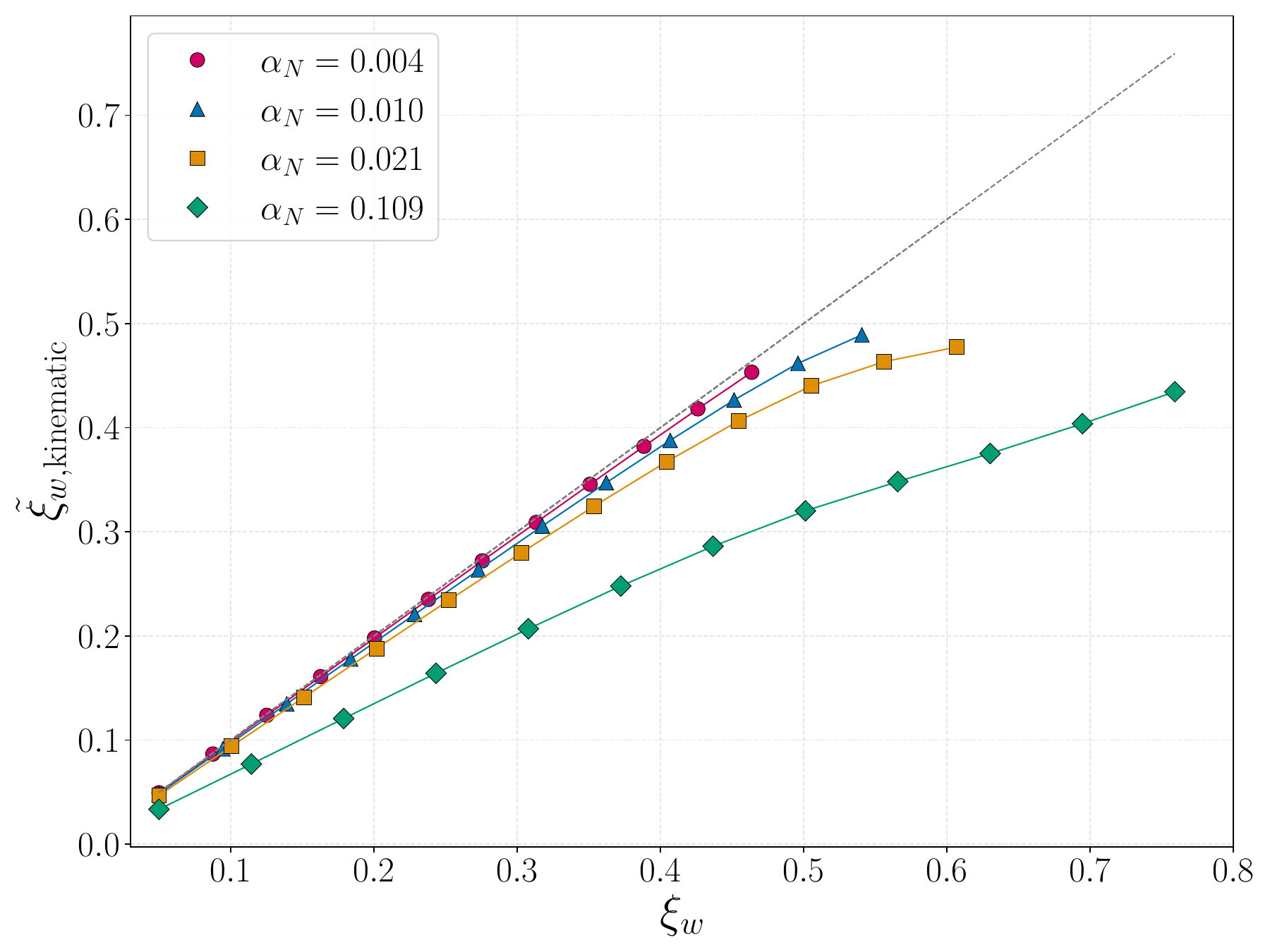}
    \hspace{1em}
    \includegraphics[width=0.48\linewidth]{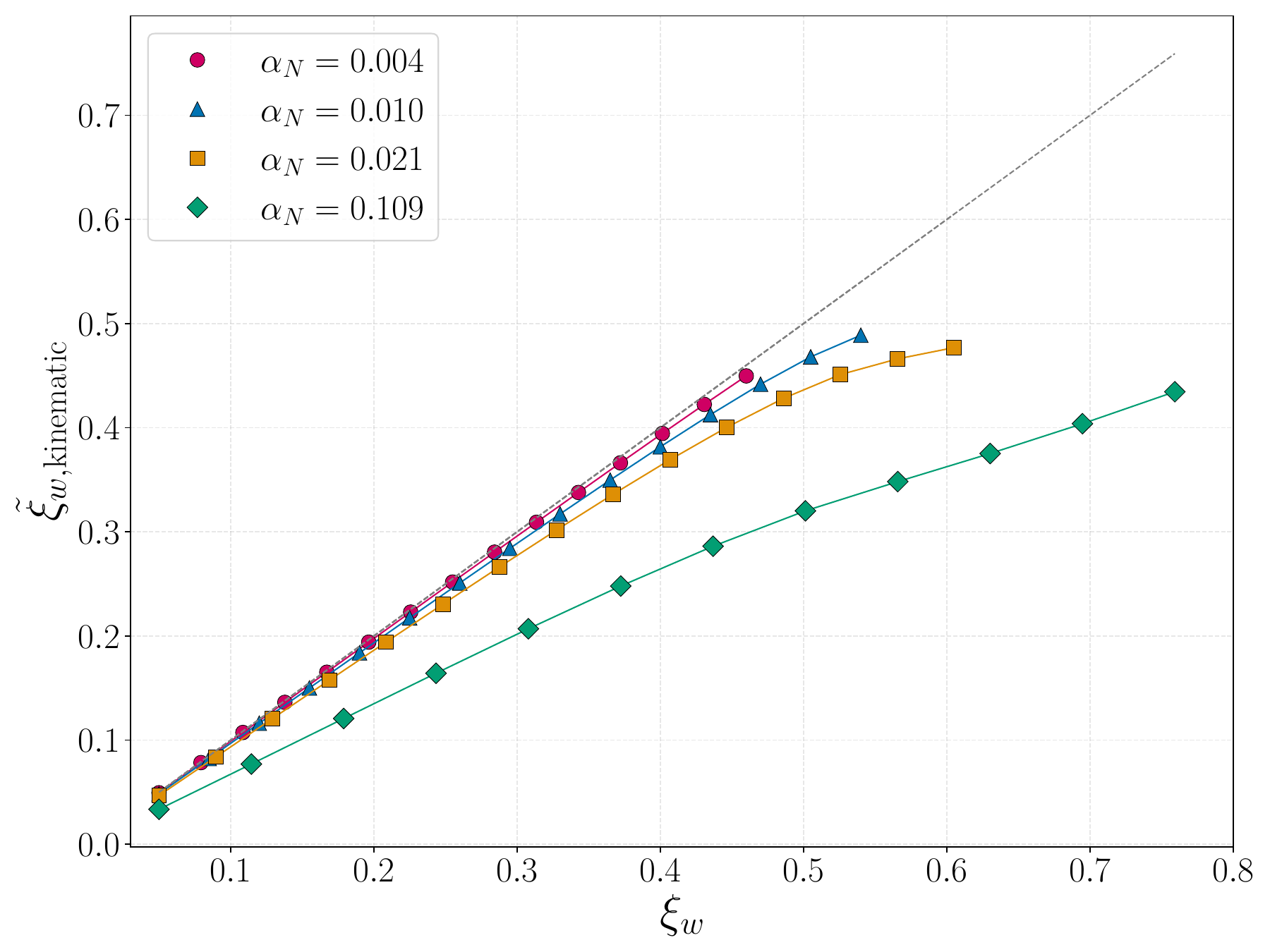}
    \caption{Heating and kinematic effects on the value of the bubble wall velocity $\tilde \xi_w$ as a function of the background wall velocity $\xi_w$ for different values of the phase transition strength $\alpha_N$ in the SM with low cut-off. The two different panels represents different friction parameterizations: on the left the results obtained considering the $\eta$ parameter constant, on the right the ones obtained with $\eta = \tilde \eta \, \phi^2 / T$.
    }
    \label{fig: SM_child_vs_parent_vw}
\end{figure}

\begin{figure}[t!]
    \centering
    \textbf{Bag model} \\
    \includegraphics[width=0.48\linewidth]{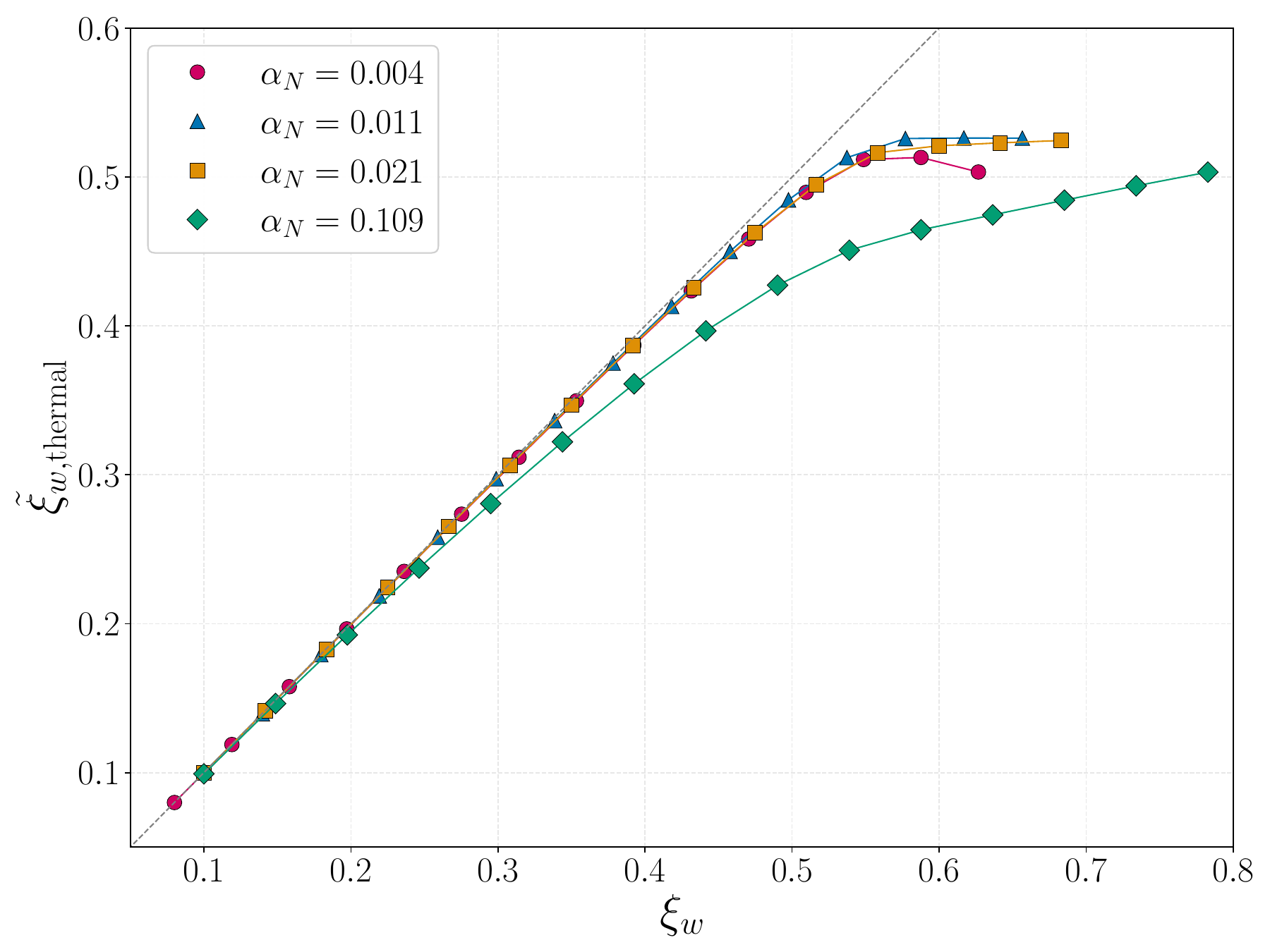}
    \hspace{1em}
    \includegraphics[width=0.48\linewidth]{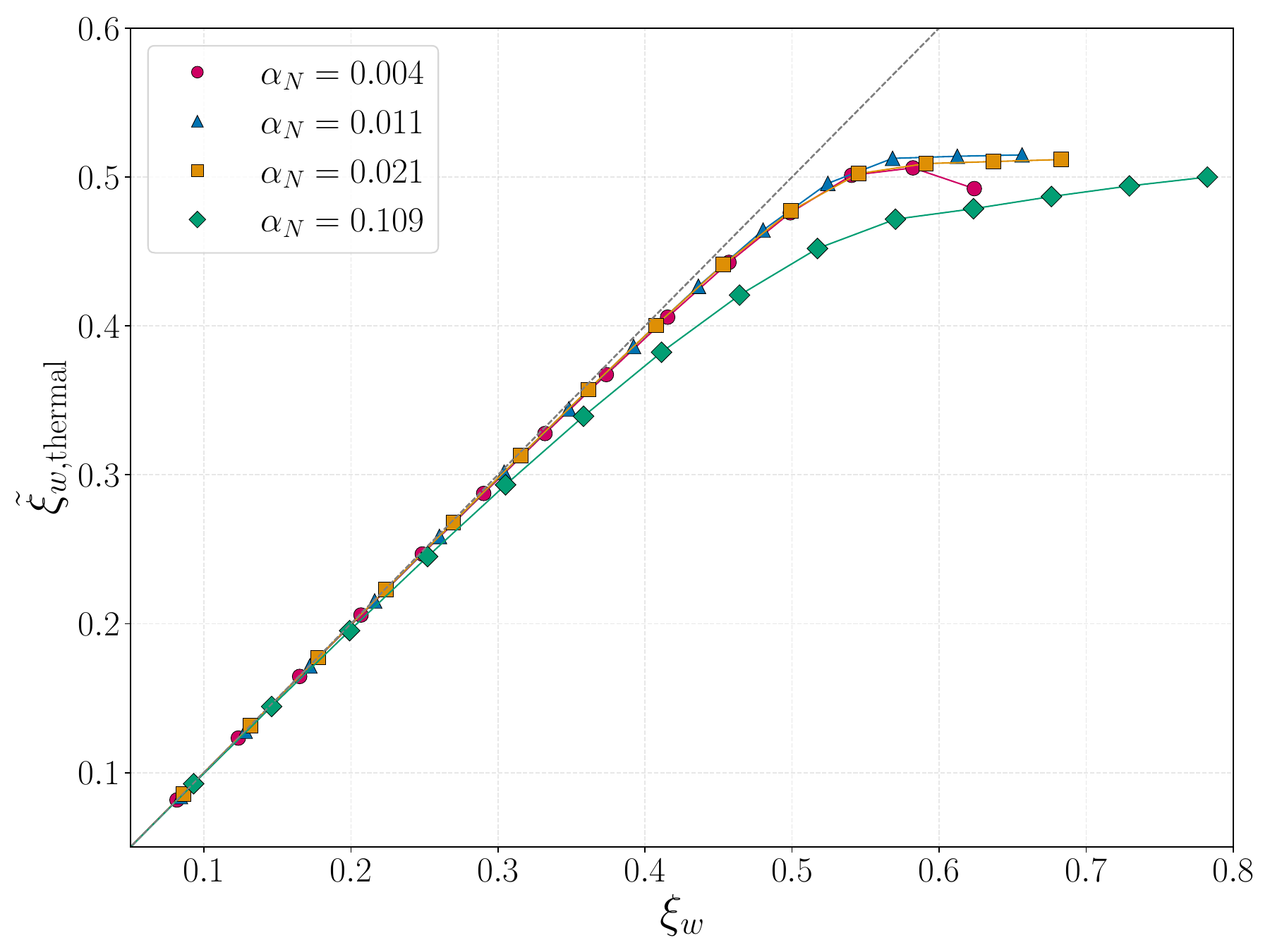}
    \\
    \includegraphics[width=0.48\linewidth]{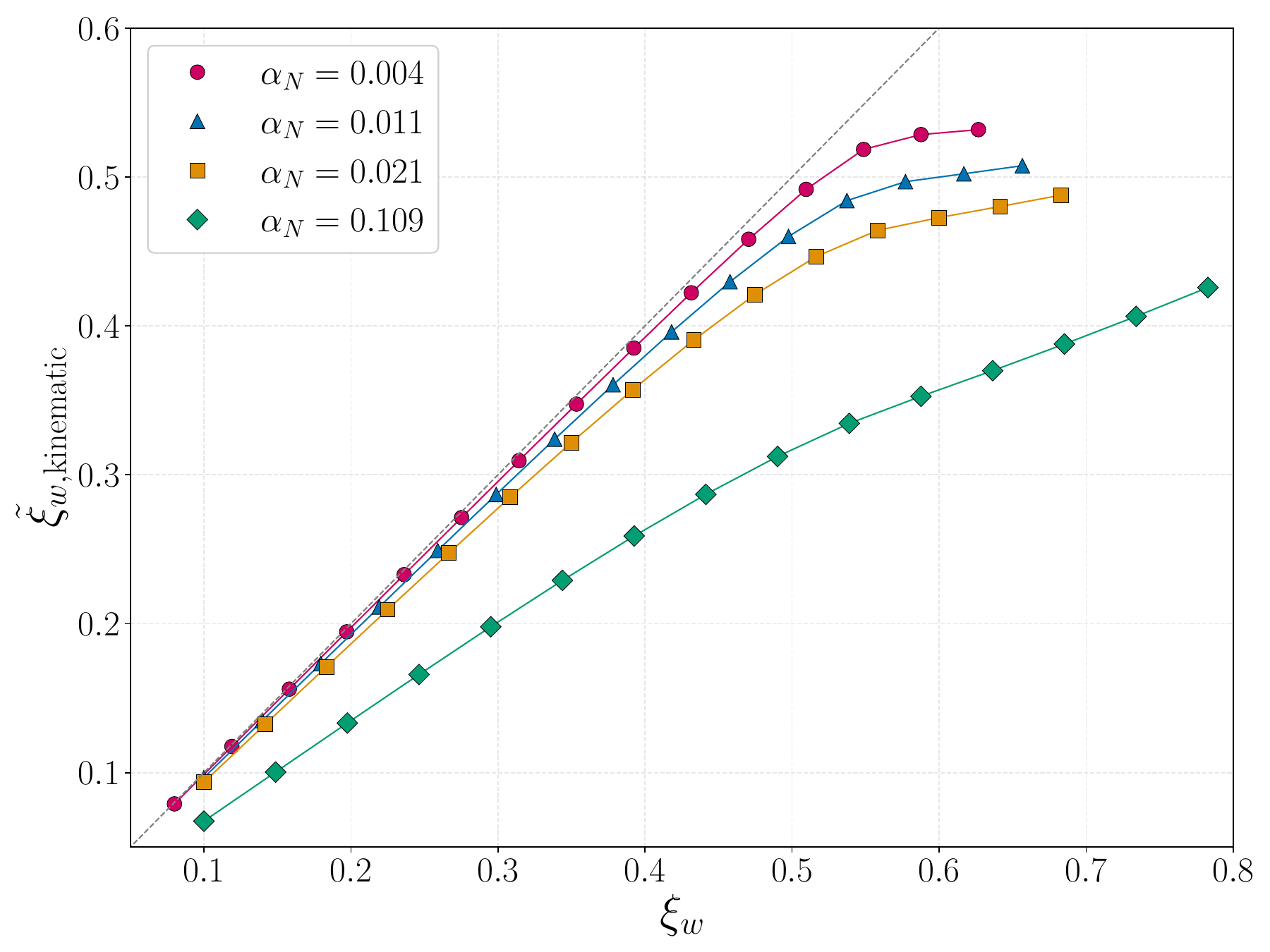}
    \hspace{1em}
    \includegraphics[width=0.48\linewidth]{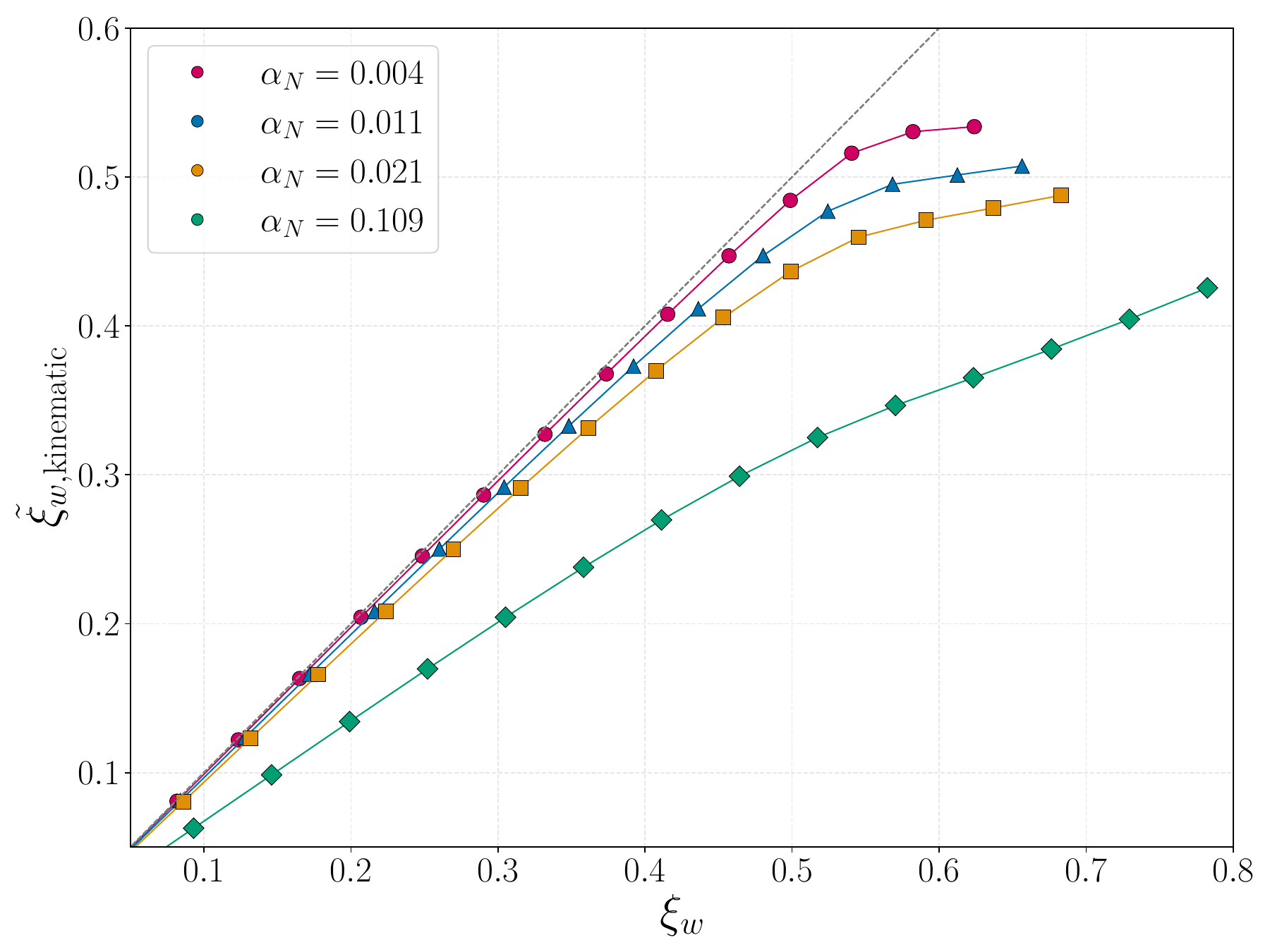}
    \caption{Same as \cref{fig: SM_child_vs_parent_vw} for the bag model, for $\delta a/a = 5.9 \times 10^{-3}$. }
    \label{fig: Bag_child_vs_parent_vw}
\end{figure}

\vspace{1em}
Both effects on the computation of $\tilde \xi_w$ are shown in \cref{fig: SM_child_vs_parent_vw} for the SM with low cut-off, and \cref{fig: Bag_child_vs_parent_vw} for the bag model. We show results for the two different parameterizations of the friction $\mathcal{K}(\phi)$, with the left panel for the constant $\eta$ parameterization in \cref{eq:constfric} and the right one for the field-dependent one $\tilde \eta$ in \cref{eq: new_eta}.
The upper panels show the corrections to the wall velocity coming from the heating in the plasma. 
The lower panels show instead the corrections to the wall velocity from the plasma being pushed by the background shock. 

A first aspect that should be noticed is that the heating effect strongly depends on the phase transition strength; for the SM with low cut-off and smaller $\alpha_N$ the critical point where $\tilde T_N = T_c$ can be reached at some values of $\xi_w$. This results in a complete suppression of nucleation or bubble growth.
For sufficiently large $\alpha_N$, the critical temperature is no longer reached. 
For the bag equation of state, the critical temperature is not reached either, and the suppression induced by thermal effects looks overall less significant, mainly due to the small change of degrees of freedom across the wall (see discussion below). We stress the significance of this model dependence of the slow-down effect. GW fitting formulas like \cite{Caprini:2019egz, Jinno:2022mie, Caprini:2024hue} parameterize the gravitational wave signal in terms of a handful of parameters: the phase transition strength $\alpha_N$, the wall velocity $v_w$, the inverse duration $\beta/H_*$ and the temperature $T_*$. In some cases the speeds of sound $c_s$ are also included \cite{Giese:2020rtr, Giese:2020znk}.
Here, we have found a strong indication that these parameters are not sufficient, and that additional information such as the change in degrees of freedom is required to accurately describe the heating effect and the resulting gravitational wave signal.

We also point out the very small difference between the results obtained for the two parameterizations of the local friction. The choice of $\mathcal K$ does not seem to influence strongly the correction of the wall velocity coming from the heating, $\tilde\xi_w$, as can be observed by comparing the left and right panels in \cref{fig: SM_child_vs_parent_vw} and \cref{fig: Bag_child_vs_parent_vw}. 
For small wall velocities the insensitivity of $\tilde \xi_w$ to the friction parameterization is easy to understand: the temperature $T_+$ at which $\tilde \xi_w$ is computed is very close to the original $T_N$. At $T_N$ the pressure difference from friction is identical in both parameterizations by construction, and therefore the results are also very close at $T_+$. 
For larger $\xi_w$, the difference between $T_+$ and $T_N$ is larger, and there is no reason to expect that the two friction terms are still identical. It should therefore be concluded, that the dominant effect is given by the hydrodynamic backreaction term, which is independent from the friction parameterization and is strongly temperature dependent. 
Since the heating effect is largely independent of the parameterization, in the remaining part of the paper we will consider only the constant parameterization of \cref{eq:constfric}, the one also used for the numerical simulations in \cite{Cutting:2019zws}. 

\subsection{Heating with friction computed with the Boltzmann equation}
\begin{figure}[t!]
    \centering
    \includegraphics[width=0.48\linewidth]{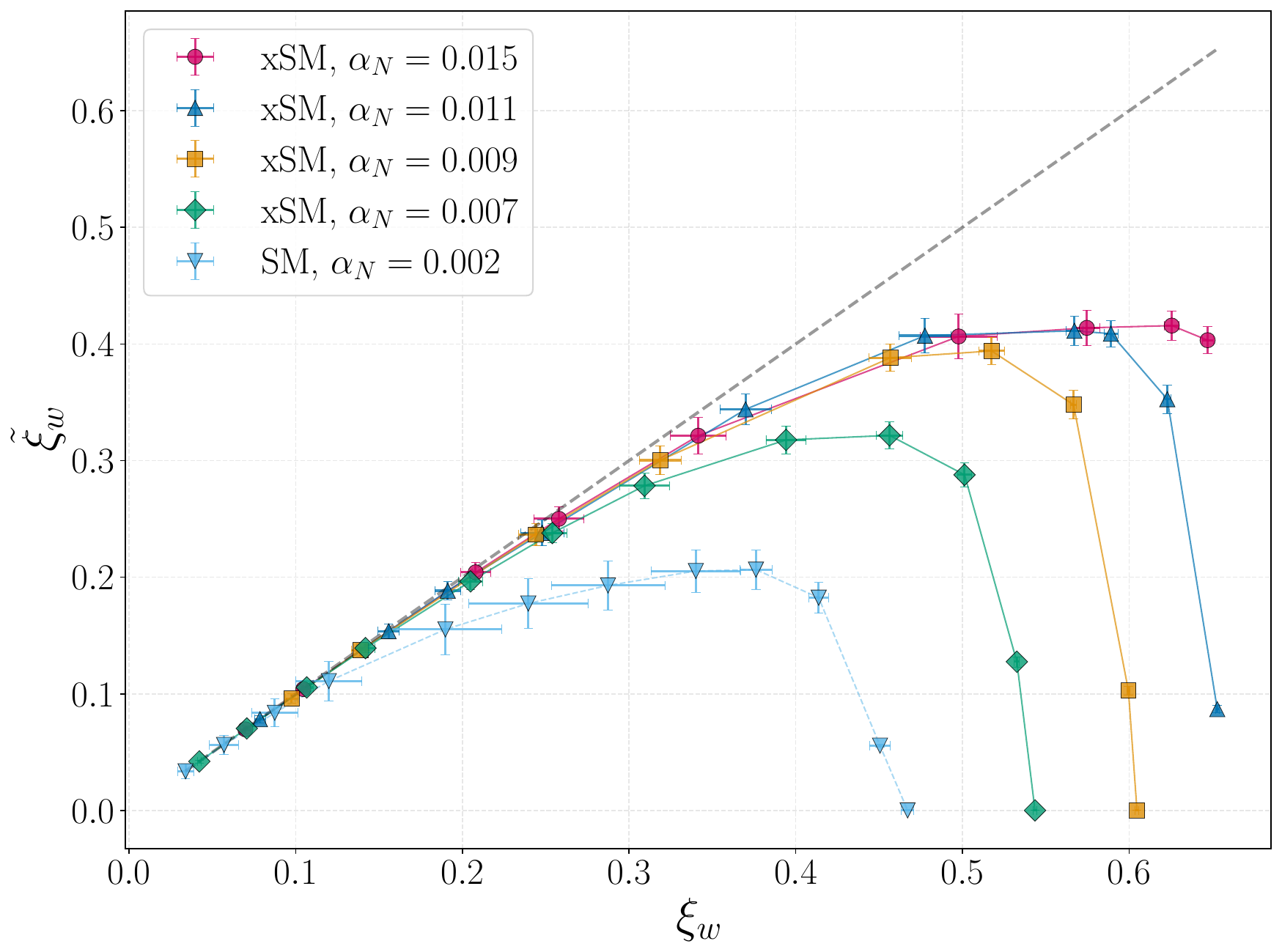}
    \includegraphics[width=0.48\linewidth]{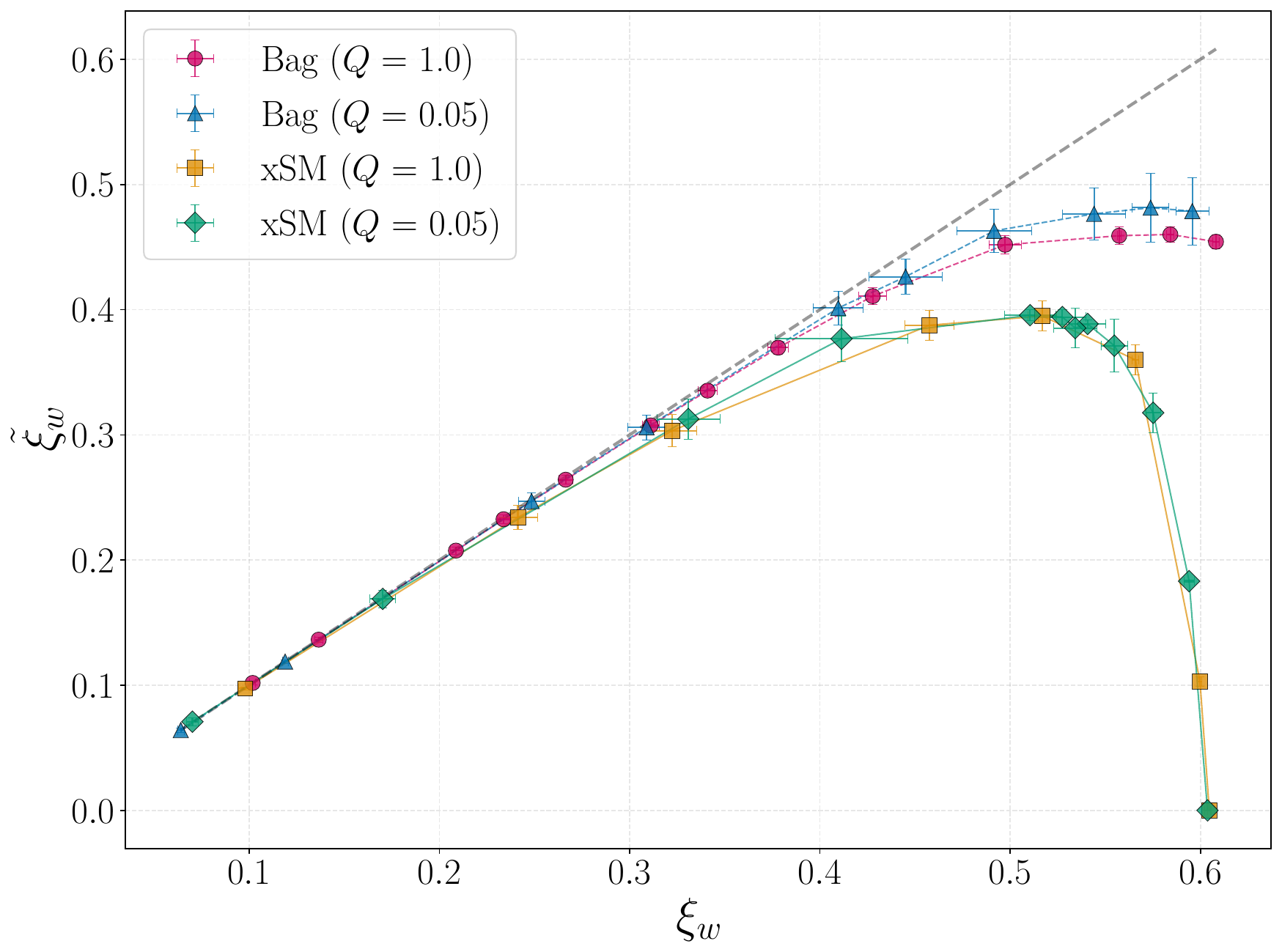}
    \caption{Heating effects on the value of the nucleated bubble wall velocity $\tilde \xi_w$ as a function of the wall velocity $\xi_w$, computed with {\tt WallGo}.  Left: different values of the phase transition strength $\alpha_N$ in xSM and the Standard Model with a $34.0 \, {\rm GeV}$ Higgs mass.
    Right: the bag model and the xSM, with an out-of-equilibrium top quark, for $\alpha = 0.004$ and $\alpha = 0.009$ respectively, and two different values of the quantity multiplying the collision terms $Q$. 
    The error bars reflect the truncation error in the solution to the Boltzmann equation.
    } 
    \label{fig:vwvsvsWallGo}
\end{figure}

For comparison, we also compute the value of $\tilde \xi_w$ including out-of-equilibrium effects in the Boltzmann equation, using {\tt WallGo} \cite{Ekstedt:2024fyq}.
The result is shown in the left panel of \cref{fig:vwvsvsWallGo}. The solid lines demonstrate results for a benchmark point in the xSM, the Standard Model coupled to a singlet. The phase transition strength is varied by treating the nucleation temperature as a free parameter. Only the out-of-equilibrium contribution of the top quark is included, and its collision terms only feature the leading logarithmic strong interactions.
The blue dashed line corresponds to the Standard Model with an unphysically light Higgs boson of $m_H = 34.0 \, {\rm GeV}$. Here, we include the out-of-equilibrium contributions of top quarks and weak gauge bosons, and include strong and weak leading logarithmic and power-enhanced \cite{vandeVis:2025plm} contributions in the collision terms.
In both scenarios, we use $N=11$ basis polynomials to expand $\delta f_i$ in the momentum direction.
Further details of the implementations can be found in \cite{Ekstedt:2024fyq}. In order to see how $\tilde \xi_w$ depends on $\xi_w$ and to compare with results of \cref{fig: SM_child_vs_parent_vw} and \cref{fig: Bag_child_vs_parent_vw}, we have implemented a new parameter {\tt frictionMultiplier}, 
which multiplies the friction term in the equation of motion of the scalar field and determines $\xi_w$. 
We see that the results for $\tilde \xi_w(\xi_w)$ are qualitatively similar to the results obtained with the friction terms $\eta$, $\tilde \eta$.

As demonstrated in \cite{Ekstedt:2025awx}, for small collisions terms, the shape of the friction deviates from the local approximation of \cref{eq: new_eta}, as the particle distributions that source the frictions have extended tails.
In the right panel of \cref{fig:vwvsvsWallGo} we test whether this deviation from the local shape of the friction affects the slowdown from heating.
We use the bag model, supplemented with an out-of-equilibrium top quark, and the xSM, considering the out-of-equilibrium effect from top quarks as well. 
To control the shape of the friction, we vary the {\tt collisionMultiplier} parameter in {\tt WallGo}, which we call $Q$ here. 
In order to obtain different values of the wall velocity $\xi_w$, we vary the {\tt frictionMultiplier}, just like above.
The friction of a top quark, with only strong interactions, is expected to marginally be described by \cref{eq: new_eta} for $Q=1$. We compare the heating effect for $Q = 1$ and $Q = 0.05$. 
From \cite{Ekstedt:2025awx} we can conclude that the friction of the top quark with $Q = 0.05$ is deviating significantly from the local shape. 
For the bag model, we use a momentum basis size of $N = 21$, and for the xSM a basis size of $N=11$ is sufficient.
In the bag model, and for $\xi_w \gtrsim 0.5$, we indeed see a small deviation between the results for $Q = 1.0$ and $Q = 0.05$, showing that the shape of the friction indeed (mildly) affects the amount of slow-down due to heating.
For the xSM however, the heating effect shows no dependence on the value of $Q$.
The reason is that the hydrodynamic backreaction is relatively more important for the xSM, as will be discussed in the following subsection. The backreaction term does not depend strongly on the shape of the friction, and therefore the relation of $\tilde \xi_w(\xi_w)$ does not depend on $Q$ either.

\subsection{Dependence of the strength of the heating effect on the ratio of degrees of freedom in the bag equation of state}
\begin{figure}[t!]
    \centering
    \textbf{Corrections to $\xi_w$ in the Bag Model for two different values of $\delta a/a$} \\
    \hspace{1em}
    \includegraphics[width=0.65\linewidth]{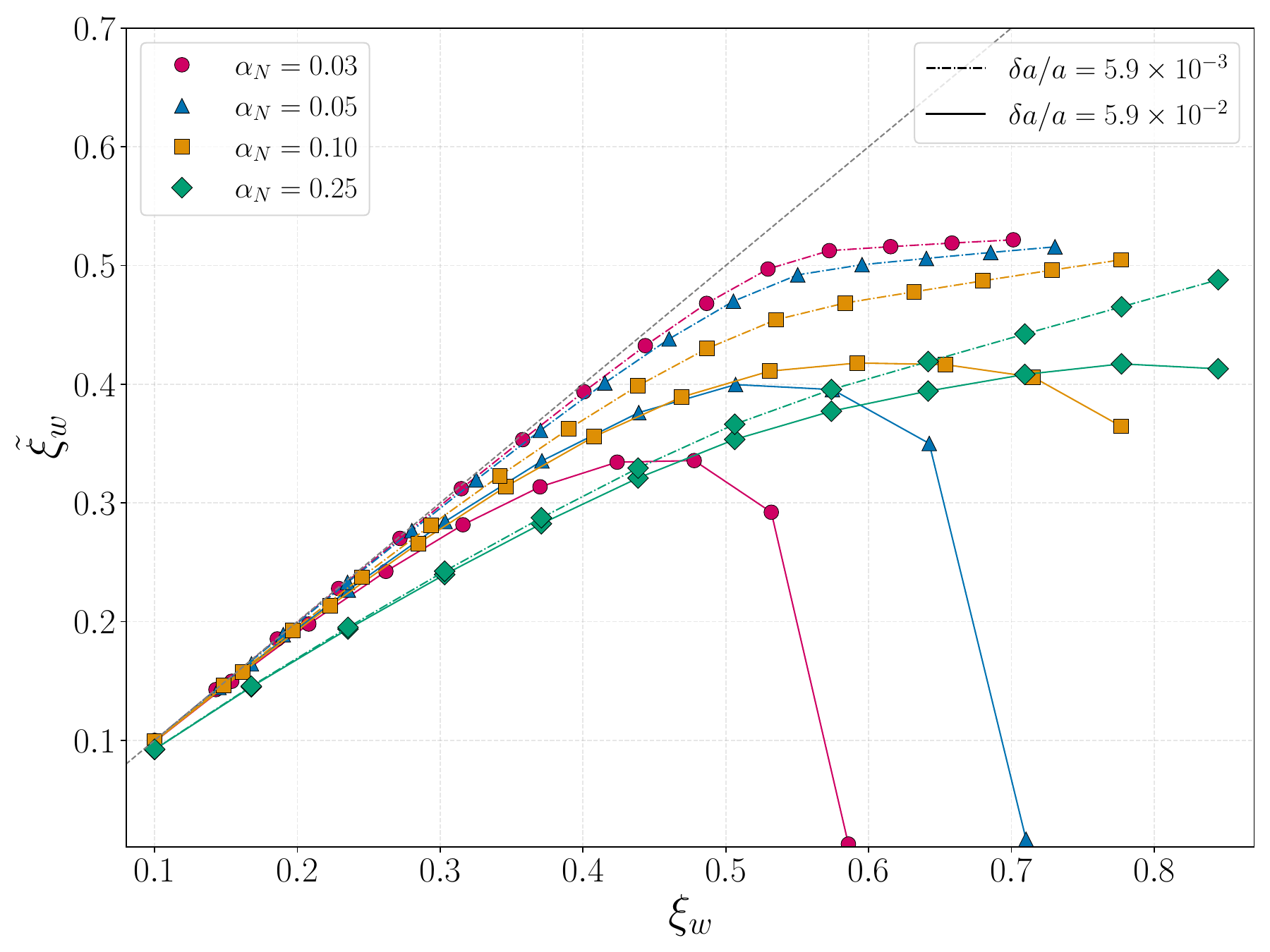} 
    \caption{Heating effects on the value of the nucleated bubble wall velocity $\tilde \xi_w$ as a function of the wall velocity $\xi_w$ in the Bag Model. Different colors represents different values of the phase transition strengths $\alpha_N$, while different lines represent a different value for the change in light degrees of freedom during the phase transition, i.e.~$\delta a / a$. } 
    \label{fig:heating_correction_comparing_delta_a}
\end{figure}

From a comparison of \cref{fig: SM_child_vs_parent_vw} and \cref{fig: Bag_child_vs_parent_vw}, we concluded that the heating effect is stronger in the Standard Model with low cut-off than in the bag model, even for comparable values of $\alpha_N$.
An important difference between these models is the difference in degrees of freedom between the symmetric and the broken phase. 
By comparing pairs of models with identical $\alpha_N$'s and comparable $\xi_w$'s, but different $\delta a/a$, in \cref{fig:heating_correction_comparing_delta_a}, we indeed see that the strength of the heating effect is sensitive to the value of $\delta a/a$, the difference of degrees of freedom between the two phases.
Here, we aim to understand why the heating effect is stronger for larger $\delta a/a$, and we argue that (in models well described by the bag equation of state), the ratio of degrees of freedom largely determines the strength of the heating.

The total pressure acting on the wall is given by 
\begin{equation}
P = \int_{-\infty}^{+\infty} dz\,\phi'
\left[ \Box \phi + \frac{\partial \mathcal{F}}{\partial \phi} + \mathcal{K}(\phi)
\right] ,\label{eq:pressureBalance}
\end{equation}
and for the physical wall velocity, the total pressure vanishes.
For the bag equation of state, the pressure balance equation becomes
\begin{equation}
	\Delta V_0 - \frac 1 3 \int da T^4 - \int dz \phi' \mathcal K(\phi) = 0 \, .\label{eq:pressureBalance_int}
\end{equation}

Let us take a fixed value of $\xi_w$, a fixed value of $\alpha_N$, but two different values of $\delta a / a$. 
Obtaining the same value of $\xi_w$ for the two different models, requires two different values of $\eta$.
The model with the smaller value of $\delta a/a$ requires a larger value of $\eta$ than the model with larger $\delta a/a$. 
A way to see this is provided by \cite{Ai:2024btx}, where entropy production is interpreted as an effective enhancement of $\delta a/a$. 
As an extreme case,
consider the limit when no friction is required and the entropy current 
is conserved across the wall (the so-called local thermal equilibrium solution)
\be
\left . s \, \gamma \, v \right|^+_- = 0  \, ,
\ee
where $s = a T^3$ is the entropy density. This can be written as the following ratio
\be
1=\frac{s_- \, \gamma_- \, v_-}{s_+ \, \gamma_+ \, v_+}
 = \left( \frac{\omega_-}{\omega_+}\right)^{3/4}
 \left(\frac{\gamma_- \, v_-}{\gamma_+ \, v_+}\right)
 \left(\frac{a_-}{a_+}\right)^{1/4} \, .
\ee
The first two factors on the right side are uniquely determined 
by hydrodynamics once the wall velocity is fixed and $\delta a/a$ can be determined 
for the LTE solution. Larger wall velocities (for fixed $\delta a/a$) or
larger $\delta a/a$ (for fixed $\xi_w$) would require negative friction 
and cannot be realized.

\vskip .3cm

Also away from the LTE solutions, the ratio $\delta a/a$ will
have an impact on the solution.
Two models described by the same $\alpha_N$ and $\xi_w$ can have a different $\delta a/a$ and friction coefficient $\eta$. Consequently, when the temperature changes, they move on different trajectories through the space of hydrodynamic solutions. More concretely, close to percolation when a large fraction 
of space can be filled by shock fronts, the dynamics will depend on 
how the heating and kinematic effect influence the wall dynamics. 
As a proxy, we will study in the following how the wall velocity 
changes under a change of ambient temperature, as reflected in the
heating effect.

Let us now focus on models where the equation of state is close to the bag equation of state, and the local friction form is a good approximation.
We want to demonstrate that, for a given $\alpha_N$ and $\xi_w$, the strength of the heating effect is set by $\delta a/ a$.
Consider the integrated relation for the pressure balance \cref{eq:pressureBalance_int}.
The first term $\Delta V_0$ is the vacuum energy that is temperature independent. 
Meanwhile, the second term in \cref{eq:pressureBalance} integrates to
\be
\frac13 \int da \, T^4 = \delta a \, T_*^4 \, ,
\ee
where by monotony of $a(z)$ the temperature $T_*$ denotes some temperature 
attained in the wall. For a weak phase transition, the temperature $T_* \simeq T_N$
while in general $T_- < T_* < T_+$ and $T_*$ depends weakly on the 
other parameters such as the strength of the PT, the wall velocity and 
the change of number of degrees of freedom. Hence, one expects
\be
\kappa_{T_*} \equiv \frac{d \log T_*}{d\log T_N} \simeq 1 \, .
\ee
Likewise one can define
\begin{equation}
     \int dz \phi' \mathcal K(\phi) \equiv \gamma_w \xi_w \, \bar{\mathcal K} ,
\end{equation}
and
\be
\kappa_{\xi} \equiv \frac{d \log (\gamma_w \xi_w)}{d\log T_N}  \, , \quad
\kappa_{\cal K} \equiv \frac{d \log \bar{\mathcal K}  }{d\log T_N} \, .
\ee
Depending on the modeling of the friction (see \cref{sec:EOMs}) one 
obtains $\kappa_{\cal K}\simeq 0 $ or $\kappa_{\cal K}\simeq -1$.
Also here, some subleading effects will enter, for example from
the temperature dependence of the Higgs wall thickness or
of the Higgs VEV in the broken phase.

Still, in essence, the pressure balance equation reads
\be
\Delta V_0 - \delta a \,  T_*^4 = \gamma_w \xi_w  \bar{\mathcal K} \, .
\ee
and the temperature derivative of this relation reads
\be
\kappa_\xi = -\frac{4 \chi \kappa_{T_*}}{1-\chi} - \kappa_{\cal K} \, ,
\ee
where we introduced the parameter
\be
\chi = \frac{\delta a T_*^4}{\Delta V_0} =  \frac{T_*^4}{T_c^4}\,,
\ee
that quantifies the impact of the thermal pressure compared to
friction. 

For $\chi\ll1$, the pressure budget is dominated by 
the friction term. In this case, the wall velocity dependence on the 
temperature comes solely from the temperature dependence of the 
friction term itself. This seems somewhat unphysical, especially 
since an increase in temperature tends to increase the 
wall velocity when $\kappa_{\cal K}<0$. Still, this is essentially 
the situation when $\delta a \ll a$ and the thermal pressure is small. 

In the regime of sizable $\chi$, an increase in temperature will
lead to a sizable increase in thermal pressure which acts against the 
vacuum pressure. Accordingly, the wall velocity will be reduced. 
In this regime, the change of number of degrees of freedom  $\delta a/a$
will have a strong impact on the heating effect.

In summary, for the bag equation of state, the thermal pressure 
stems from the change of degrees of freedom across the wall.
Lattice simulations often operate in the limit 
of small thermal pressure where friction is compensating the 
vacuum pressure rather than thermal effects. In this case, the 
heating effects does not depend on how the number of degrees of 
freedom change across the wall.

\subsection{Friction and perturbativity}
\label{sec:upperbound}

In \cite{Ekstedt:2025awx}, it was shown that an upper limit on the value of the phenomenological friction parameter $\eta (\tilde \eta)$ could be derived by comparing the expression of the local friction to the one coming from the runaway limit, also known as B\"odeker--Moore pressure \cite{Bodeker:2009qy}. We will reformulate this bound as an lower bound on the 
couplings. 

To compare the two frictions one needs to account for the different spatial dependence of the sources of friction.
In order to compare the different mechanisms we will integrate the Higgs EOM in the wall frame \cref{eq: KG eq_Higgs} after multiplying it by $\phi'$, exactly like in \cref{eq:pressureBalance}.
This yields the pressure balance condition on the two sides of the wall
\begin{equation}
\Delta \mathcal{F} + P_{\rm{LTE}} + P_{\rm{out}} = 0 ,
\end{equation}
where
\begin{equation}
\Delta \mathcal{F} = \int dz\, \partial_z \mathcal{F}(T,\phi), \quad
\end{equation}
describes the free energy release
and $P_{\rm{LTE}}$ denotes the pressure difference in local thermal 
equilibrium from the temperature change across the wall
\begin{equation}
    P_{\rm{LTE}} = -\int dz\, T' \partial_T \mathcal{F} \, .
\end{equation}
The integrated scalar damping term
\begin{equation}
    P_{\rm{out, \,  loc}} = \int dz \, \phi' \, \mathcal{K}(\phi),
\end{equation}
contains the contributions to the pressure from out-of-equilibrium physics.

In \cite{Ekstedt:2025awx},
it was observed that the pressure from  the plasma actually never exceeded the pressure from the ultra-relativistic limit. This pressure is given by the Bödeker--Moore result
\begin{equation}
P_{\rm out, BM}
= \sum_X P_{\rm out,X},
\qquad
P_{\rm out,X} =
\begin{cases}
\displaystyle \frac{1}{24} m_X^2(\phi_b) T^2 & \text{(bosons)},\\[6pt]
\displaystyle \frac{1}{48} m_X^2(\phi_b) T^2 & \text{(fermions)} ,
\end{cases}
\end{equation}
per degree of freedom.
The pressure in the plasma has two contributions: 
one coming from the equilibrium distributions -- the so-called $J_{b,f}$ functions -- and the other corresponding to the local friction term. In our notation, this bound then reads~\cite{Ekstedt:2025awx}
\begin{equation}
P_{\rm out, \, loc} \le P_{\rm out,\,BM} - \Delta J_{b,f},
\end{equation}
where $\Delta J_{b,f} \equiv J_{b,f}(m,T) - J_{b,f}(0,T)$. 

The right-hand side of this equality is a function of the masses, spins and number of degrees of 
freedom from the particles in the plasma, while the left-hand side can be inferred 
using the information presented in \cref{fig: eta_vs_vw_phi6_model}. Hence, these limits can be interpreted for a given model as functions of the Yukawa couplings and the number of degrees of freedom that become massive, i.e. $N_X$, during the phase transitions. 

\begin{figure}[t] 
    \centering
    \includegraphics[width=0.65\linewidth]{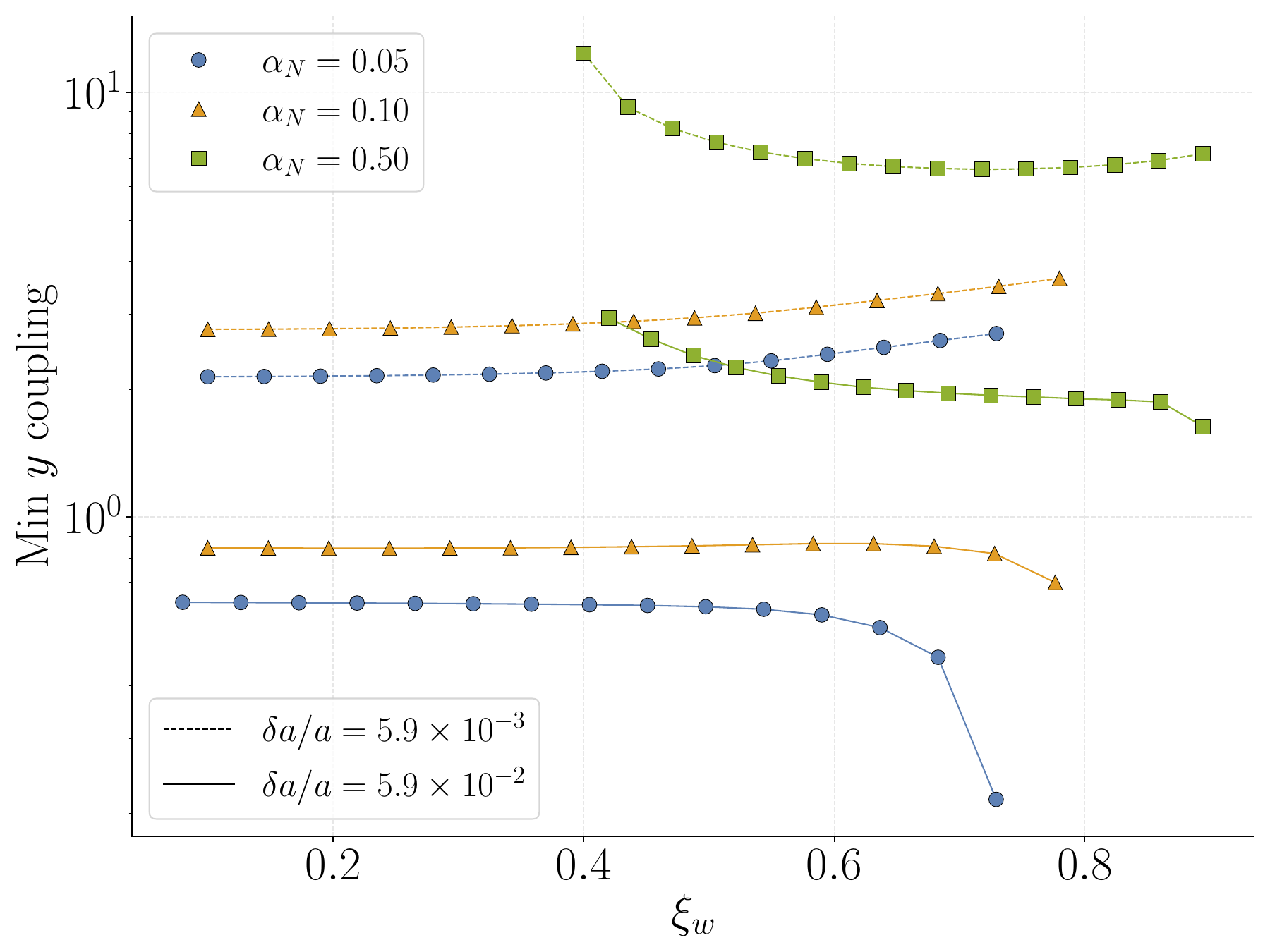}
    \caption{Lower bounds on the \textit{universal} Yukawa couplings $y$ for which the maximum value of the local friction parameter $\eta$ could be obtained, corresponding to the Bödeker--Moore friction. The plot shows the results for different values of the phase transition strength $\alpha_N$ and for the bag model in two different situations: the case of  $\delta a /a = 5.9 \times 10^{-2}$ is shown with continuous lines and aims to resemble the particle content of the SM, with a number $N_X \sim 10$ of bosonic degrees of freedom gaining a mass; the one of  $\delta a /a = 5.9 \times 10^{-3}$ represents the benchmark chosen in \cite{Cutting:2019zws}, with $N_X \sim 1$ the number of degrees of freedom gaining mass.
    }
\label{fig:upper_limits_da=0.06}
\end{figure}

The corresponding lower limits on the Yukawa couplings are shown in \cref{fig:upper_limits_da=0.06}, for different values of the phase transition strength $\alpha_N = \{ 0.05, 0.1, 0.5 \}$ and two different choices for the relative change in the light degrees of freedom given by $\delta a / a = \{5.9 \times 10^{-2}, 5.9\times 10^{-3}\}$. 
The larger choice of $\delta a/a$, represented by the continuous lines in the plot, aims to mimic an EW-like first order phase transition, with top quarks and $W^\pm, Z^0$ bosons acquiring non-negligible masses. The second value, denoted by dashed lines, represents the one considered in the numerical simulations of \cite{Cutting:2019zws}. We can observe how the bound becomes more stringent as we increase the strength of the phase transitions: one would need higher values of the Yukawa coupling to obtain a sufficiently big friction that can stop the wall at the corresponding value of $\xi_w$. 
In fact, for $\alpha_N$ and $\delta a/a = 5.9\times 10^{-3}$, the minimum value of the Yukawa coupling is so large, that it casts doubt on the validity of perturbation theory in the corresponding particle physics model.
We should notice that the value of $\alpha_N$ is not the only driver of the bound; the bound on the Yukawa tends to push the model toward the perturbative limit when going to higher values of $\delta a/a$. 
In essence, deflagrations in strong phase transitions either require non-perturbative couplings (to enhance friction) or a sizable change of number of degrees of freedom. The latter would enhance heating effects.

\section{Slow-down due to droplet formation}
\label{sec:droplets}

In \cite{Cutting:2019zws,Cutting:2022zgd, Correia:2025qif}, simulations demonstrated the presence of heated, droplet-like\footnote{
The term `droplet' carries some potential for confusion, as the same term was used in \cite{Witten:1984rs} to describe metastable regions of quark phase in a transition with phase coexistence. 
The current case is different, as the droplets disappear on a time scale much shorter than a Hubble time.
Nevertheless, from now on, we continue to follow the terminology of \cite{Cutting:2019zws, Cutting:2022zgd, Correia:2025qif} and thus refer to our regions of false vacuum as droplets.
}, patches of the false vacuum in a sea of newly percolated true vacuum in the final stage of the phase transition. 
The simulation carried out in Ref.\,\cite{Correia:2025qif} indicates that these droplets can occupy a significant $\mathcal{O}{(30)}\%$ fraction of the whole volume. Eventually, droplets shrink away as the phase transition completes and the true vacuum is established everywhere. 

The late--time formation of droplets leads to a different wall velocity compared to the initial expansion of the bubbles following nucleation, and can thus explain the slow-down of the bubble walls seen in numerical simulations\,\cite{Cutting:2022zgd,Correia:2025qif}.
In this section, we shall pursue this second mechanism that causes slow-down in the final stages of the phase transition. 
This picture can also help elucidating the suppression of the GW emission compared to the theoretical expectations based on the fluid kinetic energy observed in\,\cite{Cutting:2019zws}. 
In fact, it is reasonable to expect that droplets will suppress GW production: First, the wall velocity 
for the shrinking droplets is smaller than the one of the expanding bubbles, and thus the fluid kinetic energy is reduced. Second, the droplet will eventually convert kinetic energy into heat which will reduce the kinetic energy that resides in sound waves after percolation. 

In the following, we will outline the droplet hydrodynamics and explain how the bubble wall velocity at late times can be predicted from simple theoretical arguments from the thermodynamics of the phase transition. In this regard, we will make a comparison with all the available results from numerical simulations\,\cite{Cutting:2019zws,Cutting:2022zgd,Correia:2025qif}, and find very good agreement, as shown in \cref{fig:multivsdroplet}.
Finally, we will also discuss how the droplet analysis relates to the heating and kinematic effects in the background of a shock discussed in the previous sections.

\subsection{Hydrodynamics of droplets} \label{subsec. hydro droplets}

Droplets are self--similar solutions to the hydrodynamics equations \cref{eq. spherical velocity profile} with negative wall velocities in the plasma frame, $\xi_d <0$, describing the evolution of a collapsing object. The fluid velocity, $v(\xi)$ with $\xi = -r/t<0$, is instead positive in the same frame, indicating that the fluid is actually expelled by the shrinking droplet. In the following, we will neglect the effects related to the droplet surface tension and finite width, which are expected to play a role only in the very final stage of evaporation\,\cite{Kurki-Suonio:1995yaf}. 

The possible solutions of this kind have been discussed in\,\cite{Rezzolla:1995kv,Rezzolla:1995br, Kurki-Suonio:1995yaf, Cutting:2022zgd} as well as in\,\cite{Barni:2024lkj}.
Here, we will focus on droplets where the fluid is at rest inside the droplet but has a nontrivial profile in the broken phase up to $\xi = -c_s$.
In terms of the usual matching conditions for the fluid velocity in the wall frame, one then has $v_+ = |\xi_d|$.
The fluid gets extracted by the droplet forming a rarefaction wave in the broken phase, where it will move at a velocity $v_-$ in the wall frame, jumping from a temperature $T_+$ to $T_-$ given by the usual matching conditions in \cref{eq:match1} and \cref{eq:match2}. 

\begin{figure}[t!]
    \centering
    \includegraphics[width=0.65\linewidth]{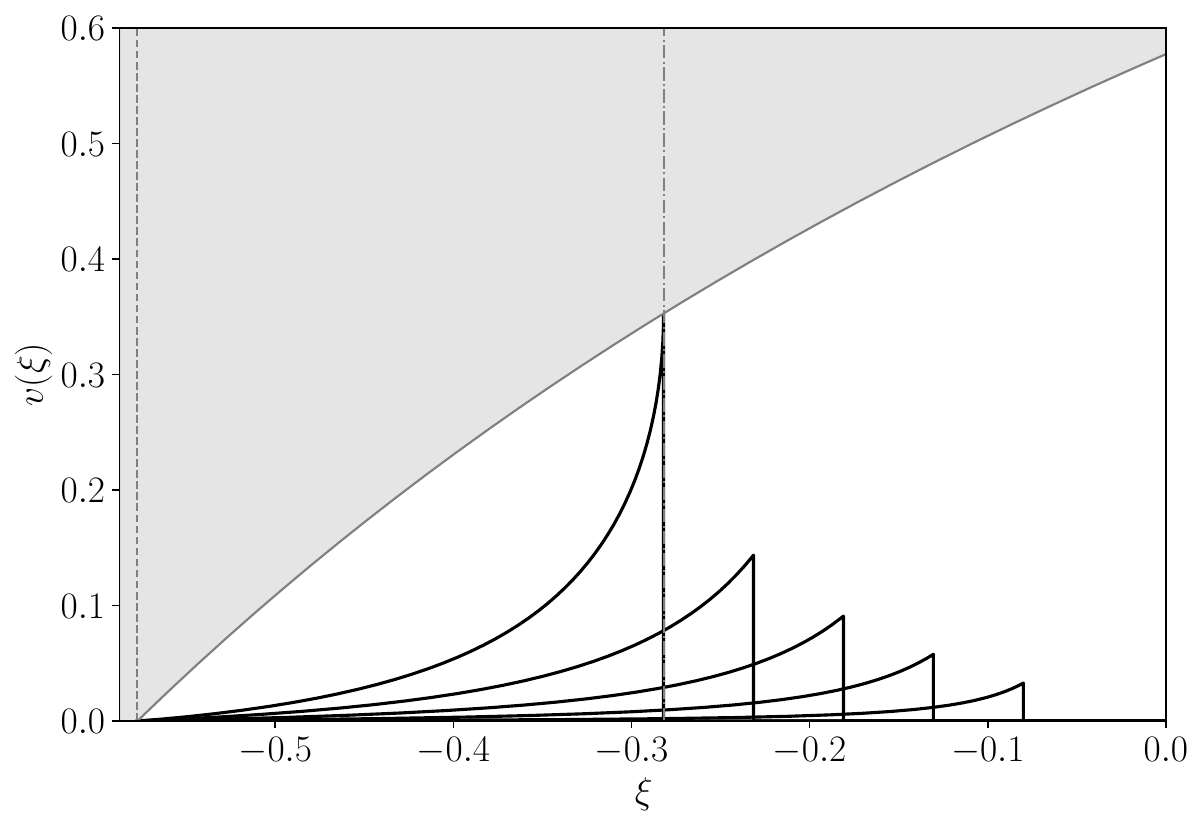}
    \caption{Self--similar solutions for the fluid velocity outside the droplet occurring in the late stage of a phase transition with strength $\alpha_N = 0.15$. The different curves correspond to the different values of the local--friction parameter $\eta$, or equivalently different wall velocities for the original deflagration. The gray line is given by $v(\xi) =\mu(\xi,-c_s)$, and represents the maximum velocity that the fluid outside the droplet can have while being consistent with the hydrodynamical boundary condition of interest, $v(-c_s)=0$. 
    The vertical dashed-dotted line represents the maximum allowed (absolute) droplet velocity for the chosen value of $\alpha_N$.
    Notice that, similarly to what happens for detonations, the velocity field smoothly vanishes at the fixed point $\xi_w = - c_s$, represented by the vertical dashed line.}
\label{fig:droplet_self_similar}
\end{figure}

The parameter space for the droplet velocity and the fluid velocity expelled by the droplet is bounded by $\xi_d > - c_s$ and by the requirement $v_- < c_s$. This is plotted as a black line in \cref{fig:droplet_self_similar}, where the white region corresponds to consistent droplet solutions. 
As we shall see, when searching for droplets that can consistently appear in the final stage of the phase transition, one obtains a reduced set of solutions. This is indicated by the vertical dashed line in \cref{fig:droplet_self_similar}, which represents the largest possible droplet velocity, $|\xi_d|$, for a phase transition with strength $\alpha_N = 0.15$.

\begin{figure}[t!]
    \centering
    \includegraphics[width=0.49\linewidth]{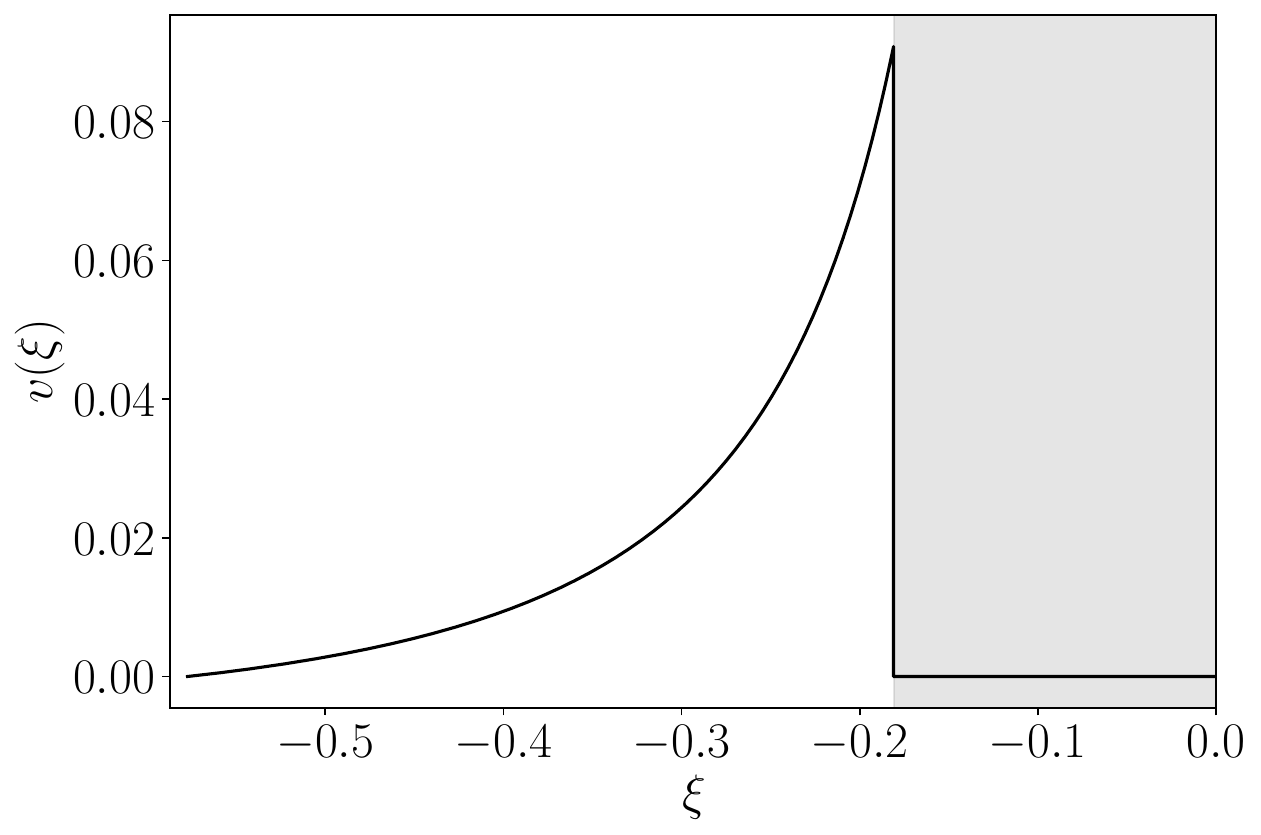}
    \hfill
    \includegraphics[width=0.49\linewidth]{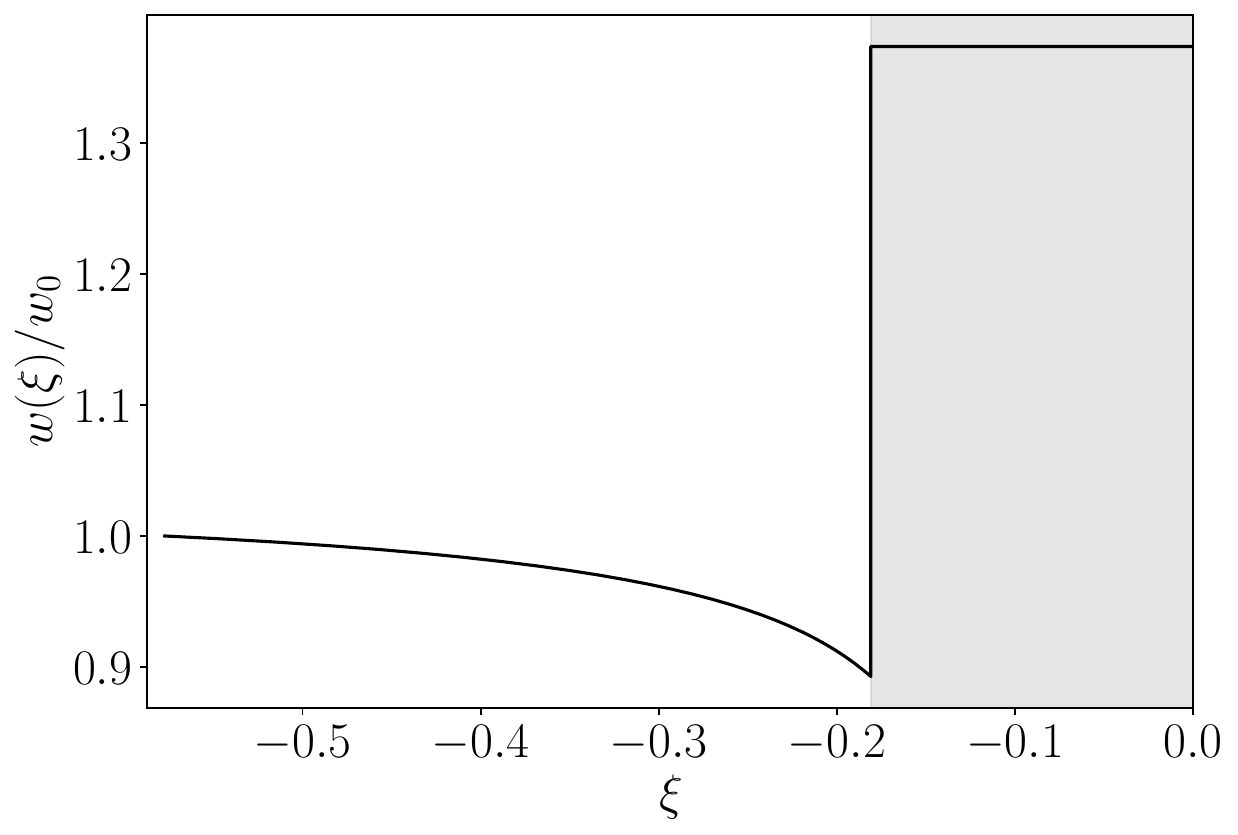} 
    
    \caption{Characteristic self--similar profiles for the fluid velocity (left) and enthalpy (right) inside (gray region) and outside (white region) the droplet, obtained for the specific case of a phase transition with strength $\alpha_N = 0.15$ and $\eta \simeq 1.3$.} 
    \label{fig:droplet_temperature}
\end{figure}

It is instructive to investigate how the droplet solution satisfies energy--momentum conservation along the typical profiles shown in \cref{fig:droplet_self_similar} (see also \cref{fig:droplet_temperature}). By taking into account the different phases inside and outside of the droplet, as well as the kinetic energy of the fluid in the region $-\xi^\prime \in (-c_s, -|\xi_d|)$, energy conservation reads:
\begin{align}
\label{eq:enconsdroplet}
    \frac{1}{3}\left( V_0 + \frac{3}{4} \omega_+ \right)|\xi_d|^3 + \int_{|\xi_d|}^{c_s}\xi^{\prime\,2} d\xi^\prime \left[ \gamma^2(-\xi^\prime) - \frac{1}{4} \right]\omega(-\xi^\prime)& = \frac{1}{4} \omega_0 c_s^3, 
\end{align}
where $\gamma^2(-\xi^\prime) = 1/(1-v^2(-\xi^\prime))$, $\omega_0 = \omega(-c_s)$, and we have used the bag equation of state with constant speed of sound, $c_s^2=1/3$. 
Differently than the usual expanding bubbles, droplets are shrinking. This means that droplets and their fluid shells are becoming smaller and smaller with time, until the droplets completely evaporate. The final state of the evolution is then essentially given by a vanishing fluid velocity and a constant value of the enthalpy everywhere given by $\omega_0$ in \cref{eq:enconsdroplet}, which in the true vacuum is related to the energy density as $e_0 =  (3/4)\omega_0$. In essence, the droplet converts the kinetic energy surrounding it and the latent heat 
into thermal energy. 

\begin{figure}[ht]
\centering
\includegraphics[width=0.65\linewidth]{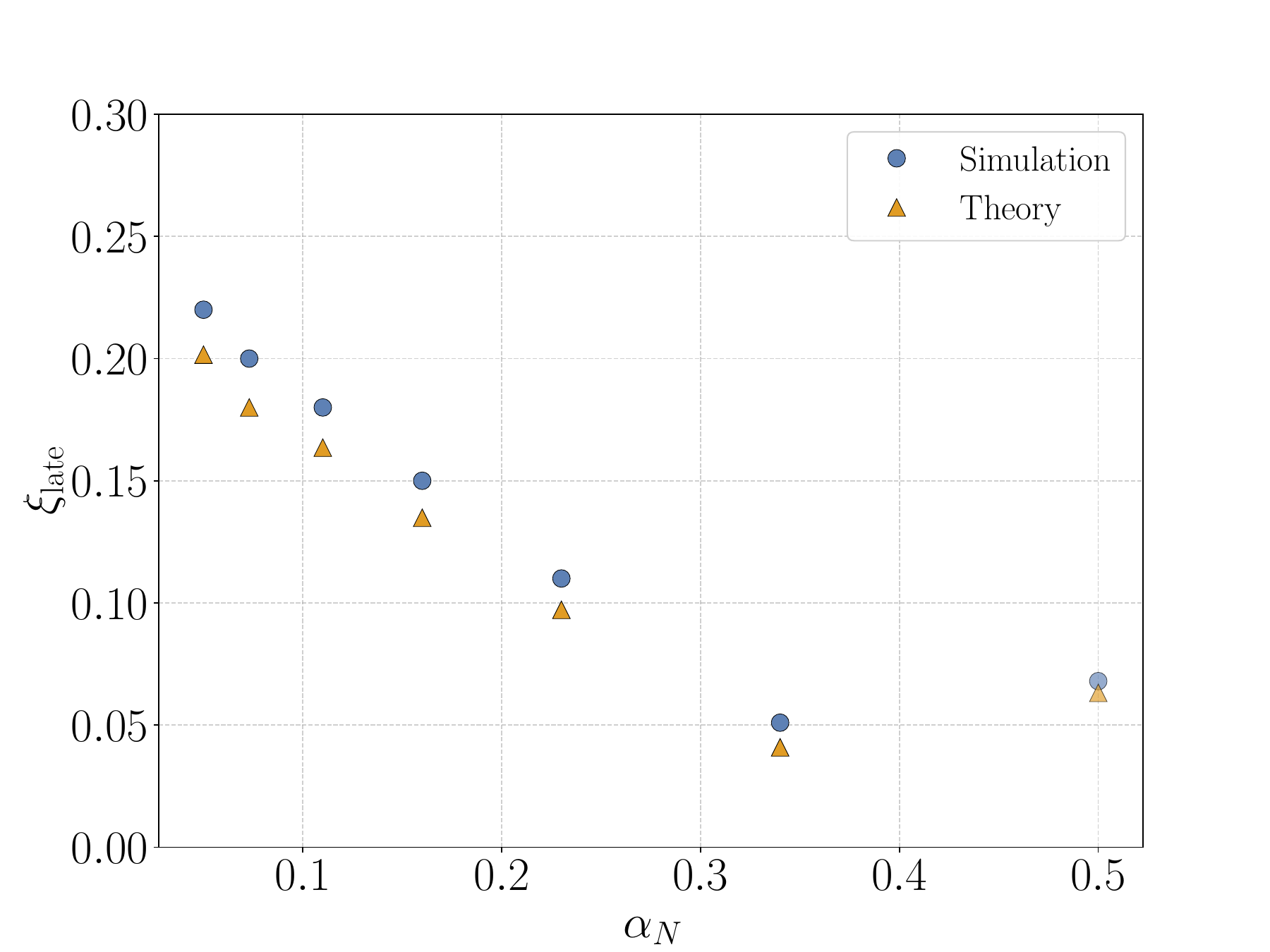}
\caption{Late-time wall velocity as observed in multi-bubble simulations from Refs.\,\cite{Cutting:2019zws,Cutting:2022zgd, Correia:2025qif} (blue dots) with the wall velocity of the initial deflagration fixed to $\xi_w = 0.24$ (except the point with $\alpha_N=0.5$, where $\xi_w = 0.44$), and our theoretical prediction based on the formation of droplets with boundary conditions specified in \cref{eq:encons} (orange triangles).
}
\label{fig:multivsdroplet}
\end{figure}

This very fact suggests a way to identify the appropriate boundary conditions for an admissible droplet solution arising in the late stage of a first order phase transition.
In fact, the temperature inside the droplet is not known a priori, since the plasma has been heated up by the bubble collisions at the time of droplet formation. Also the temperature right outside of the droplet is in principle unknown, as it no longer reflects the temperature $T_-$ inside the original expanding bubble.
What can still be used, however, is (total) energy conservation. Given that the final state after droplet evaporation is given by the energy density $e_0 = (3/4) \omega_0$ as discussed above, one needs to require:
\begin{equation}
\label{eq:encons}
\frac{3}{4} \omega_0 + K = e_0 + K \overset{!}{=} e_N = \frac{3}{4} \omega_N(1+ \alpha_N),
\end{equation}
where $e_N$ is the false-vacuum energy density at the nucleation temperature. In this relation, we have introduced $K$ as the kinetic energy in the fluid after the transition has completed.
This energy can be in the form of sound waves, and the factor $K$ allows us to take into account the regions where droplets have not formed and the bubble collisions have occurred in the usual way. In the following, however, we shall set $K=0$, as we have checked that values of $K$ with $K/\omega_0 \lesssim10 \%$ (typical for weak and intermediate phase transitions) introduce a negligible correction to our predictions.

By assuming that the broken phase already reaches a uniform temperature at the moment that the droplets are formed, one can determine the temperature far away from the droplet as 
\be
\label{eq:T0}
    T_0 = \left( \frac{e_N - V_0(\phi_b)}{3 \, a(\phi_b) }\right)^{\frac{1}{4}},
\ee
where we have used the bag equation of state.
This relation fixes the boundary conditions for the hydrodynamics of droplet collapse, and allows us to predict the droplet velocity, interpreted here as the late-time velocity of the bubble walls, for a given value of $e_N$ at nucleation.

The results of such computations (see the next subsection for more details) are shown in \cref{fig:multivsdroplet} against simulation data from\,\cite{Cutting:2019zws,Cutting:2022zgd, Correia:2025qif}. As we can see, the droplet solution with the boundary condition in \cref{eq:encons} is able to reproduce the late-time evolution of the bubble walls that are slowed down by heating effects in very good agreement with all the available numerical simulations. This confirms that the impact of the kinetic fraction $K$ is numerically small.

\subsection{Wall velocities from droplets}
\label{sec:etanonzerodroplets}

The droplet velocity that will be realized in the final stage of a first order phase transition can be determined by following a similar procedure to the one outlined in \cref{sec. numerical setup}. 
For a given model, $\alpha_N$ and $\xi_w$ related to the original expanding bubble fix the value of the local--friction parameter $\eta$. As $\eta$ does not depend on the nature of the hydrodynamical solution, we can solve the Higgs equation \cref{eq:scalarEOM} determining the droplet velocity $\xi_d$ by using the same value of $\eta$, with the boundary condition in \cref{eq:T0}. 
We have also checked that this procedure can be extended to the case of local thermal equilibrium with $\eta=0$, showing that LTE deflagrations can, in principle, consistently end up in LTE droplets, see Appendix \ref{sec:dropletLTE} for more details.

\begin{figure}[t!]
    \centering
    \includegraphics[width=0.65\linewidth]{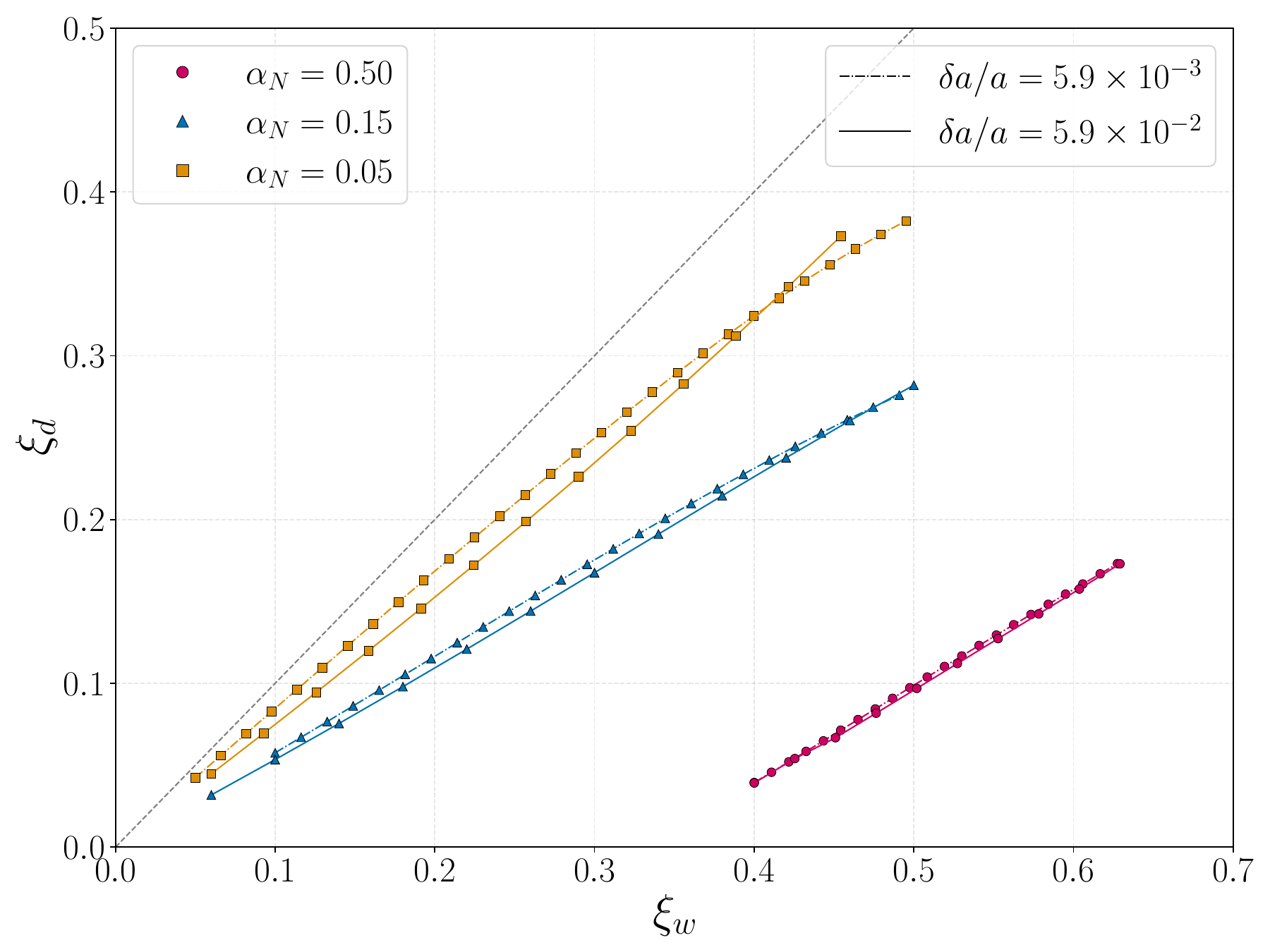}
    \caption{Droplet velocities, $\xi_d$, for different values of the strength parameter $\alpha_N$ in the bag model, compared to the wall velocity, $\xi_w$, of the original deflagration/hybrid (diagonal dashed line). In this case the Higgs equation of motion is solved considering the $\eta$ parameter constant, i.e. the friction modeled as $\mathcal{K}(\phi) = T_c \, \eta \, u^\mu \partial_\mu \phi$. Solid and dashed lines correspond to different values for the relativistic degrees of freedom, $\delta a/a$, leading however to similar results for the values considered here.}
    \label{fig: suppression_droplet}
\end{figure}

Our general results for the bag model are shown in \cref{fig: suppression_droplet}. As we can see, droplets are generically slower than the original deflagration/hybrid solutions, with droplet velocities becoming more and more suppressed when $\alpha_N$ is increased. For $\alpha_N =0.5$, the slowest deflagration actually corresponds to the one simulated in Ref.\,\cite{Correia:2025qif}, with the droplet velocity matching the simulation results as shown in \cref{fig:multivsdroplet}.
For a fixed $\alpha_N$, the largest value of $|\xi_d|$ corresponds to the maximum velocity that a droplet can have consistently with hydrodynamics, as discussed in \cref{subsec. hydro droplets}. 
Let us also notice that, in contrast to the slow-down effects observed in \cref{sec: temp and kin effects}, the droplet velocity $\xi_d$ is mostly fixed by $\alpha_N$ and $\xi_w$, and depends only weakly on $\delta a/a$, as can be seen by comparing the lines with $\delta a/a= 5.9 \times 10^{-3}$ and $\delta a/a = 5.9 \times 10^{-2}$.
Notice that, for each value of $\alpha_N$, there exists a maximum initial wall velocity $\xi_w$ above which a consistent droplet solution ceases to exist, \emph{e.g.} $\xi_w \simeq 0.5$ for $\alpha_N = 0.05$ and $\alpha_N = 0.15$. This is due to the the existence of a an upper bound on $|\xi_d|$ for fixed $\alpha_N$, as shown in \cref{fig:droplet_self_similar}.
The lower end for $\xi_w$ is instead simply determined by the requirement of a consistent deflagration for fixed $\alpha_N$. 

\begin{figure}[t!]
    \centering
    \includegraphics[width=0.65\linewidth]{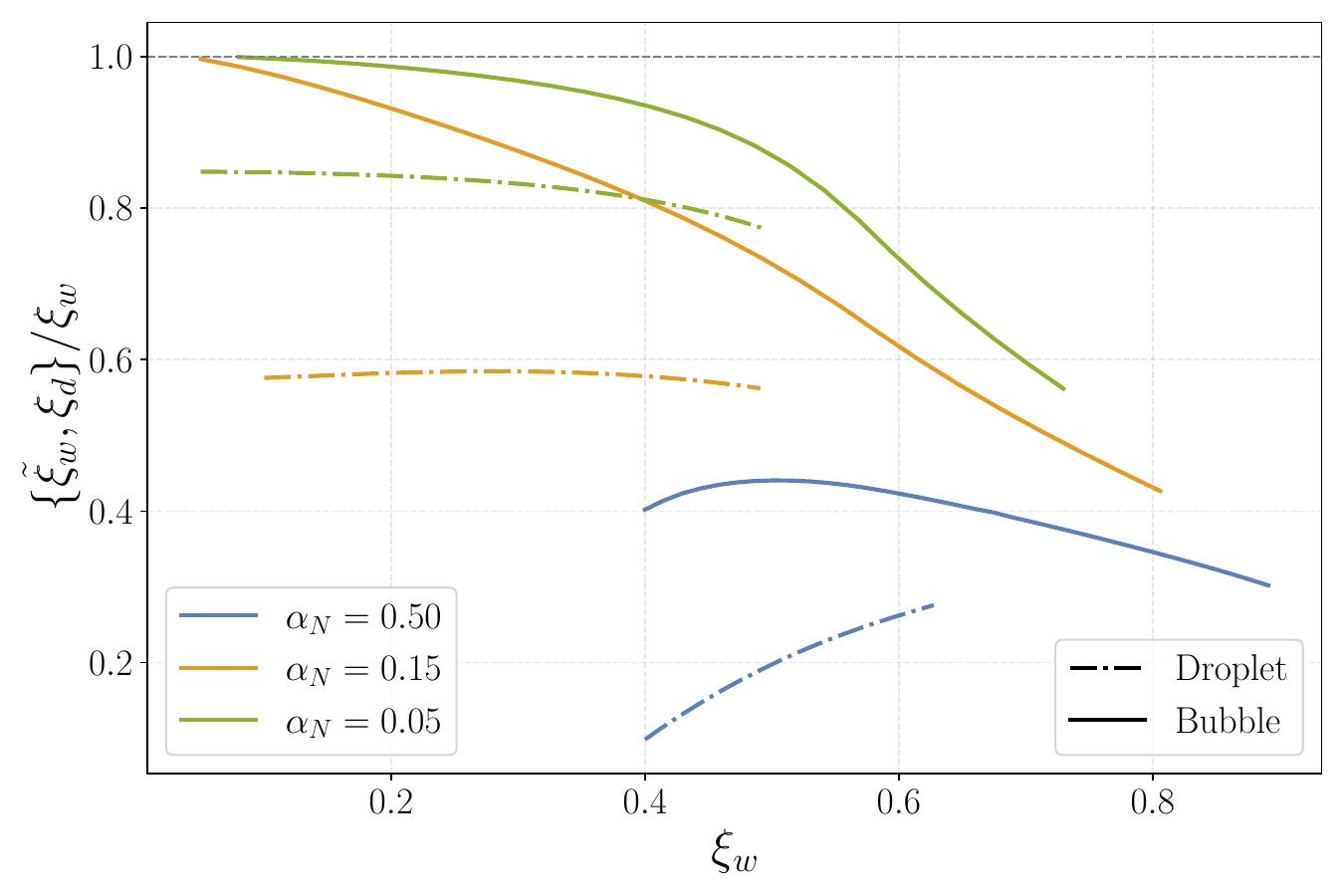}
    \caption{
    Ratio between the late--time wall, $\tilde \xi_w$, (solid lines) and droplet velocity, $\xi_d$, (dashed lines) computed by taking into account heating effects in the plasma according to the analysis of \cref{sec: temp and kin effects} and \cref{sec:droplets}, respectively, and the original bubble wall velocity $\xi_w$. The colors represent different values of the phase transition strength, $\alpha_N = \{ 0.05, \, 0.15, \, 0.5 \}$. As we can see, the suppression of the velocity is larger according to the droplet analysis than for the one in \cref{sec: temp and kin effects} and the difference becomes more prominent for larger $\alpha_N$. 
    Notice also that not for every initial bubble solution there exists a droplet that can consistently form in the late stage of the phase transition, as explained in the text.}
    \label{fig: ratio_velocities_hydro_droplet}
\end{figure}

In \cref{fig: ratio_velocities_hydro_droplet} we show a comparison between the correction to the original wall velocity, $\xi_w$, given by the heating effects discussed in \cref{sec: temp and kin effects}, and by considering droplets as the late--time stage of the bubble evolution. When a droplet solution exists, the suppression of the velocity is always more severe for the droplet case. For strong transitions, $\alpha_N =0.5$, the strongest suppression is found for the slowest deflagration (here $\xi_w \simeq 0.4$), whereas for weak and intermediate transitions the suppression mildly depends on $\xi_w$.
In this sense, let us notice that the numerical simulation in \cite{Correia:2025qif}, for which $\xi_w = 0.44$ and $\alpha_N = 0.5$, is among the most extreme cases of slow-down both in terms of $\xi_w$ and $\alpha_N$.

\begin{figure}[t!]
    \centering
    \includegraphics[width=0.65\linewidth]{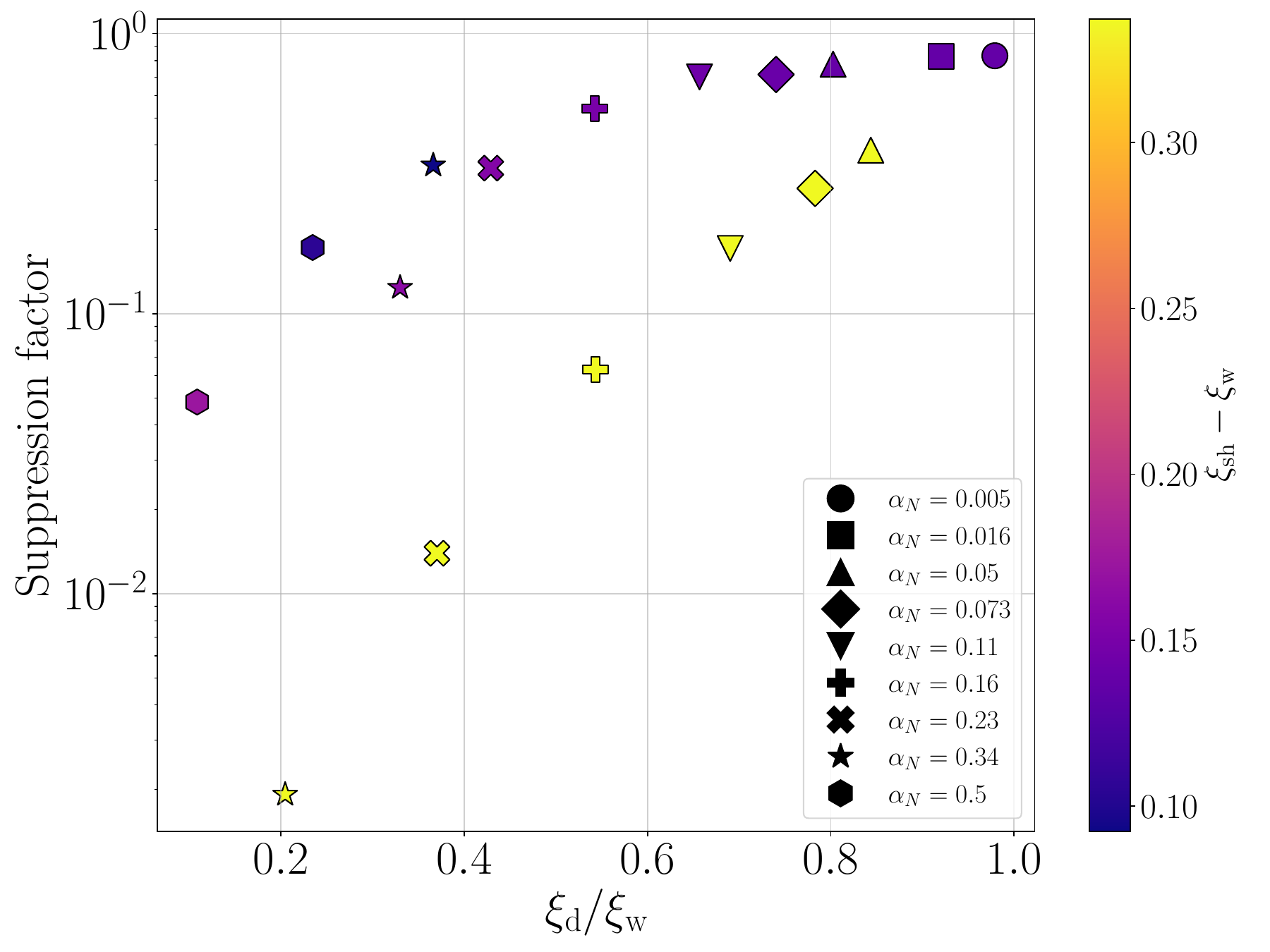}
    \caption{
    Suppression of GW emission with respect to the theoretical expected values from\,\cite{Caprini:2019egz} as shown in Table 1 of \cite{Cutting:2019zws} against the ratio of droplet velocity to bubble wall velocity, $\xi_d/\xi_w$, for different values of $\alpha_N$. The different shapes denote the values of $\alpha_N$, and the color-coding indicates the shock width, $\xi_{\rm sh} - \xi_w$. As we can see, a larger suppression in the GWs relates to slower droplets and larger shock widths. All points correspond to deflagration solutions.
   } 
    \label{fig: suppression_velocity_vs_shock}
\end{figure}

So far, we have only considered slow-down effects on the bubble walls. These alone, however, cannot determine the corresponding suppression of the GW emission, which would at least require an estimate of the fraction of volume in droplets.
In order to better understand the factors influencing the GW emission,
in \cref{fig: suppression_velocity_vs_shock} we plot the GW suppression factor reported in \cite{Cutting:2019zws} for phase transitions which start as deflagrations with different velocities.
This suppression factor is determined relatively to the prediction in Refs.\,\cite{Hindmarsh:2017gnf, Caprini:2019egz}, which relies on the assumption that the hydrodynamic solution of a single bubble is predictive for the GW spectrum of the full ensemble of bubbles.
We plot the suppression factor against $\xi_d/\xi_w$, and show the width of the shock of the expanding bubble solution in color coding for each phase transition.
We see that the GW suppression does indeed correlate with the slow-down due to droplet formation given here by $\xi_d/\xi_w$, but we observe that the shock width plays an important role as well: the suppression is stronger for wider shocks. This can be understood by noticing that larger shocks can potentially affect a bigger fraction of the entire volume, leading to the formation of bigger droplets which implies a stronger reduction of the fluid kinetic energy after percolation. This interpretation is consistent with the fact that none of these suppression effects should be relevant for detonations, which support no shocks ahead of the wall and can be smoothly obtained from hybrids precisely when the shock width vanishes. 

In conclusion, the existence of very slow droplet solutions indicates the possibility of a strongly suppressed GW emission, but the overall effect is parametrically controlled by the size of the shock due to its importance for the fraction of the volume that can end up in droplets. Also, the formation of droplets relies on 
some non-linear dynamics of the fluid which is hard to assess quantitatively.

\section{Conclusion}
In this work we explore several effects that slow down bubbles and can suppress the gravitational wave spectrum as compared to estimates based on single-bubble solutions. 
We focus on hydrodynamical modes featuring a shock, i.e. deflagrations and hybrids. 
For these modes, the plasma in front of the bubble wall can heat up significantly, and the strongest suppression of the gravitational wave emission was in fact found for deflagrations in \cite{Cutting:2019zws}.

In \cref{sec: temp and kin effects} we explore slow-down of the bubble walls due to heating and motion of the surrounding plasma, due to the presence of other bubbles.
Without doing a proper many-bubble simulation, we can of course not know the real temperature and fluid motion in front of the bubbles when they approach each other. 
Here, we take as an approximation the temperature and fluid velocity in front of one bubble with $\xi_w$, $T_+$ and $v_+$, and use these as the boundary conditions to recompute the wall velocity $\tilde \xi_w$ of another bubble.
We expect the real slowdown effect to be less strong than the result of this computation, since the plasma can not everywhere heat up all the way to $T_+$.
Nevertheless, we obtain the following conclusions, that will likely also apply to a full many-bubble simulation.

First of all, we find that the slow-down effect barely depends on the shape of the friction term, suggesting that the constant friction term used in the simulations gives a reasonable description for the wall-plasma interactions, despite its deviation from the form expected from the Boltzmann equations.
Second, for fixed $\alpha_N$, we find that the slow-down effect is most significant for the fastest wall velocities, which could be expected because the plasma temperature and velocity in front of the bubble are largest in those cases.
This implies that the slow-down from heating and kinematic effects observed here does \emph{not} explain the suppression effect observed in \cite{Cutting:2019zws}, which was strongest for smaller wall velocities.
Third, we also evaluated the slow-down effect due to heating by solving for $\xi_w$ and $\tilde \xi_w$ with a Boltzmann description for the heavy plasma particles, using {\tt WallGo}.
We find that the results are qualitatively similar. By varying the strength of the collision terms, we also concluded that the dependence of the heating effect on the shape of the friction is small, but not completely negligible, in a scenario with small $\delta a/a$.
Finally, we find that the slow-down effect is stronger for a larger difference in degrees of freedom between the broken and symmetric phase. 
In fact, the value of $\delta a/a$ in \cite{Cutting:2019zws, Correia:2025qif} is smaller than what is expected for SM-like phase transitions, and we thus expect a stronger effect in more realistic models.
Our finding suggests that the gravitational wave spectrum can not just be parameterized in terms of $\alpha_N$ and $\xi_w$ (and the speeds of sounds \cite{Giese:2020rtr, Giese:2020znk}), but that it also depends on the value of $\delta a/a$.

For a fixed particle content, the local friction was found to be always smaller than the B\"odeker-Moore limit in \cite{Ekstedt:2025awx}.
In \cref{sec:upperbound}, we use this relation to find lower bounds on the Yukawa couplings given a value of $\xi_w$, $\alpha_N$ and $\delta a/a$.
We find that, for strong phase transitions (e.g. $\alpha_N = 0.5$) with small $\delta a/a$, the required friction is so large, that it requires Yukawa couplings $y \gtrsim 10$, which raises serious concerns about the perturbativity of such a model.

Since the slow-down due to heating and kinematic effects does not give a complete explanation of the suppression observed in \cite{Cutting:2019zws, Correia:2025qif}, we turn our attention to droplet solutions in \cref{sec:droplets}.
The droplets are spherically symmetric solutions of heated symmetric phase, that form between the bubbles, observed in multi-bubble simulations in \cite{Cutting:2019zws, Correia:2025qif}.
We use energy-momentum conservation, and assume that the broken phase reaches a uniform temperature far away from the droplets when the droplets are formed.
This allows us to determine the velocity of the droplets $\xi_d$, and we find remarkably good agreement with the droplet velocity measured in 3D simulations.
By computing $\xi_{d}$ for different values of $\alpha_N$ and $\xi_w$, we indeed recover behavior consistent with the 3D simulations: the suppression of the velocity becomes most significant for small $\xi_w$ and large $\alpha_N$.
As demonstrated in \cref{fig: suppression_velocity_vs_shock}, the relevant parameters to predict the suppression of GWs are droplet velocity and the shock width. 
In contrast to the slow-down effect due to heating, the droplet velocity depends only weakly on the change in degrees of freedom. This suggests that the GW suppression factor from droplet formation does not depend significantly on $\delta a/a$, although such a conclusion also requires an estimate of the volume fraction of droplets, which is beyond the scope of this work.
Also, this limits the applicability of Higgsless simulations~\cite{Jinno:2022mie, Caprini:2024gyk} 
to detonations and other regions of the parameter space where 
these effects are weak. 

We also notice that, given a phase transition with a certain $\alpha_N$ and $\xi_w$, it is not always possible to find consistent droplet solutions in the final stage of the transition. It is then unclear what the suppression of the GW emission would be, even though this occurs at large values of $\xi_w$ where the suppression of the velocity due to droplet formation is typically less important and we may expect this trend to continue. 

It is also important to stress that the prediction for the resulting GW spectrum from cosmological first order phase transitions depends on various effects, and not only on the velocity of the expanding bubbles, and (as is shown in \cite{Ajmi:2022nmq}) the thermal suppression of nucleation could also change other parameters entering in the GW production mechanism, like the mean bubble spacing $R_*$.

In summary, we explore the dynamics of the primordial fluid right before percolation. We find that besides the conventional parameters of the phase transition (wall velocity, PT duration, PT strength, temperature) also the details of scalar damping through friction 
and entropy production will influence the system. 
It was observed in simulations that percolation can proceed via a formation of shrinking droplets that will absorb part of the kinetic energy of the fluid and suppresses the gravitational wave production. These simulations operated in a regime 
with strong friction and large entropy production. It will be interesting to study if the suppression persists for a larger class of models with sizable changes in the number of degrees of freedom.

\paragraph*{Acknowledgments}
We thank Andreas Ekstedt for fruitful discussions at the initial stages of this project. We thank David Weir and Mark Hindmarsh for helpful communication about the simulations of \cite{Cutting:2019zws, Correia:2025qif}.

We acknowledge support by the Deutsche Forschungsgemeinschaft (DFG, German Research Foundation) under Germany’s Excellence Strategy – EXC 2121 „Quantum Universe“ – 390833306. NB acknowledges support by the Deutsche Forschungsgemeinschaft (DFG, German Research Foundation) under the DFG Emmy Noether Grant No. PI 1933/1-1.

\newpage
\appendix
\section{Hydrodynamics}\label{app:hydro}
\subsection{Matching conditions} \label{subsec. matching conditions}

If we consider the plasma undergoing a phase transition as a perfect fluid and the phase transition front as a discontinuity surface moving with some velocity $v$ and in some direction, e.g along the $z$-axis, its energy--momentum tensor is given by \cref{eq:perfectFluid}.
If we look at distances much larger than the wall width $L_w$, we can integrate the conservation equation \( \partial_\mu T^{\mu\nu} = 0 \) across the infinitesimally thin wall to obtain continuity conditions for the energy and momentum fluxes. Denoting quantities in front of and behind the wall by \(+\) and \(-\), respectively, one obtains the standard matching conditions:
 \begin{align}
    \omega_+ \gamma_+^2 v_+ &= \omega_- \gamma_-^2 v_- , \label{eq:match1} \\
    \omega_+ \gamma_+^2 v_+^2 + p_+ &= \omega_- \gamma_-^2 v_-^2 + p_- . \label{eq:match2}
\end{align}
Usually the set of unknowns is constituted of $\{v_+, v_-, T_+, T_-\}$, while there are only two matching conditions relating them. 
Once the wall velocity $\xi_w$ and the nucleation temperature $T_N$ are fixed, the solution is uniquely defined. 

A third relation can be found involving the entropy flux \cite{Laine:1993ey},
\begin{equation}
    s_- \gamma_- v_- - s_+ \gamma_+ v_+ = \Delta S,
\end{equation}
which can be rewritten as
\begin{equation}\label{eq: third matching cond}
    \frac{T_+}{T_-} = \frac{\gamma_-}{\gamma_+} \left ( 1 + \frac{\Delta S}{ s_+ \gamma_+ v_+} \right ),
\end{equation}
using the thermodynamic relation between enthalpy and entropy density, i.e. $\omega = T s$.
This means that if we know the entropy production at the wall we can solve directly for the correct velocity $\xi_w$. Unfortunately, the entropy production is typically not known apriori, and needs to be computed, for example through a system of Boltzmann equations. 

In the special case of local thermal equilibrium, $\Delta S = 0$ and \cref{eq: third matching cond} becomes simply
\begin{equation}\label{eq. 3rd matching lte}
 T_+ \gamma_+ = T_- \gamma_-.
\end{equation}
The local thermal equilibrium approximation works well when neglecting dissipative effects is a good approximation but may lead to the impossibility of finding stable (non-runaway) solutions for some models.

\subsection{Relativistic fluid equations} \label{subsec. relat. fluid eqns}
Now let us discuss the solutions to the continuity equation $\partial_\mu T^{\mu\nu}=0$ far away from the bubble wall, i.e. $\partial^\nu \phi = 0$,
\be 
\partial_\mu T^{\mu \nu}_{\text{fl}} = u^\nu \partial_\mu ( u^\mu \, \omega ) + u^\mu \, \omega \, \partial_\mu u^\nu - \partial^\nu \pfl = 0 \, ,
\ee
where $\pfl$ denotes the pressure of fluid that is related to the free 
energy of the system via $\calF = V_0 - \pfl$.
This equation can be projected along the flow $u^\mu$ and along a vector $\bar{u}^\mu$ perpendicular to the flow such that $u_\mu \, \bar{u}^\mu = 0$, resulting in 
\be \label{eq. continuity, spherical}
    \partial_\mu ( u^\mu \, \omega) - u_\mu \, \partial^\mu \pfl = 0 \, , 
\ee
and 
\be \label{eq. Euler, sperical}
    \bar{u}^\nu \, u^\mu \, \omega \, \partial_\nu u_\nu - \bar{u}^\nu \, \partial_\nu \pfl = 0 \, .
\ee

If we now assume that the solution has spherical symmetry and is self-similar, we can find solutions to this system as a function of a parameter $\xi = r/t$. $r$ is the distance from the center of the bubble and $t$ is the time since nucleation, so $\xi$ represents the velocity of a given point in the wave profile and the particles at a point $\xi$ move at $v(\xi)$. The gradients can then be written as 
\be
    u_\mu \partial^\mu = - \frac{\gamma}{t} (\xi - v) \partial_\xi, \hspace{2 em} \bar{u}_\mu \partial^\mu = \frac{\gamma}{t} ( 1 - \xi v ) \partial_\xi\,,
\ee
and the energy-momentum conservation equations \eqref{eq. continuity, spherical}, \eqref{eq. Euler, sperical} become
\be
    (\xi - v) \frac{\partial_\xi \efl}{\omega} = 2 \frac{v}{\xi} + \left[ 1 - \gamma^2 v (\xi - v)\right] \partial_\xi v,
\ee
\be
    (1 - v \xi) \frac{\partial_\xi \pfl}{\omega} = \gamma^2 (\xi - v ) \partial_\xi v\,,
\ee
where $e_{\rm fl}$ is related to the pressure as
$e_{\rm fl} = T d p_{\rm fl}/dT - p_{\rm fl}$.

These equations can be further simplified by noticing that $\partial_\xi p_{\rm fl}$ and $\partial_\xi e_{\rm fl}$ can be related through the speed of sound in the plasma as 
\be 
    c_s^2 \equiv \frac{d\pfl/dT}{d\efl/dT}.
\ee
Now we obtain the central equation describing fluid velocity profiles
\be \label{eq. spherical velocity profile}
    2 \frac{v}{\xi} = \gamma^2 ( 1 - v\xi) \left [ \frac{\mu^2}{c_s^2} - 1 \right ] \, \partial_\xi v,
\ee
where $\mu$ relates velocities in the plasma frame to the ones in the wall frame through the Lorentz velocity addition formula,
\begin{equation}
    \mu(\xi, v) = \frac{\xi - v}{1 - \xi v}.
\end{equation}
Eq.~\eqref{eq. spherical velocity profile} can be solved with the proper boundary conditions to find the velocity and temperature profiles inside and outside the bubble wall.

\subsection{Detonations, deflagrations and hybrids} \label{subsec. hydro solutions}

Not all the solutions to the matching conditions are physical \cite{Steinhardt:1981ct, Laine:1993ey, Ignatius:1993qn, Kurki-Suonio:1995rrv, Kurki-Suonio:1995yaf, Espinosa:2010hh}; the correct ones need to satisfy some particular boundary conditions and generally fall into three physical classes:

\vspace{0.5em}
\textbf{Detonations.} 
The wall propagates at supersonic speed ($\xi_w > c_s^-$) into the plasma at rest in front of it. Since the fluid has no time to respond, it enters the broken phase with a velocity $v_+=\xi_w$ in the wall frame, then slowing down behind it such that $v_- < v_+$. This also implies that the temperature just in front of the wall is the same as the temperature far away from it, i.e $T_+ = T_N$. For this class of solutions to take place, it is not enough to have a supersonic $\xi_w$, we also need the wall to be faster than the so-called Jouguet velocity, which can be determined from \cref{eq. spherical velocity profile} 
using the condition $\mu(\xi_w, v_-) = c_{s,b}$.  
Mathematically this can be expressed as
\begin{equation} \label{eq. joug. vel.}
    \xi_w > \xi_J = \frac{1 + \sqrt{3 \, \alpha_N \left( 1 - c_{s,b}^2 + 3 \, c_{s,b}^2 \, \alpha_N \right)}}{1/c_{s,b} + 3 \, c_{s,b} \, \alpha_N},
\end{equation}
where $\alpha_N$ is the strength parameter defined through the pseudo-trace of the energy-momentum tensor as \cite{Giese:2020rtr, Giese:2020znk}
\begin{equation} \label{eq. alpha def.}
    \alpha_N \equiv \frac{D \bar{\theta}(T_N)}{3 \, \omega_N},
\end{equation}
with $\bar{\theta} = e - p/c_{s,b}^2$ and $D\bar{\theta} \equiv \bar{\theta}_s(T_N) - \bar{\theta}_b(T_N)$. The subscripts $s$ and $b$ refer to the symmetric and broken phase respectively.

\vspace{0.5em}
\textbf{Deflagrations.} 
This class of solutions, in contrast, occurs when the wall moves subsonically with respect to the plasma, i.e $\xi_w < c_{s,b}$. In these cases, the plasma is at rest behind the wall, such that $v_- = \xi_w$ in the wall frame, while the plasma in front of it is pushed away, implying $v_- > v_+$. The solutions to \cref{eq. spherical velocity profile} for the plasma velocity in front of the bubble can become double valued before going smoothly to zero, which implies the necessity of a shock-front. 
Since there is no change in vacuum expectation value of the field $\phi$ at this front, the corresponding $\alpha_{\rm sh} = 0$ and the condition for the shock can be written as $\mu({\xi_{\rm sh}}, v_{\rm sh}) \, \xi_{\rm sh} = c_{s,s}^2$. For these solutions, in general $v_+ \neq \xi_w$ and $T_+ \neq T_N$, so one needs to find the conditions at the wall such that the temperature in front of the shock equals $T_N$. This is usually done via a shooting procedure.

\vspace{0.5em}
\textbf{Hybrids.}
A third class of solutions, known as \emph{hybrids} or \emph{supersonic deflagrations}, interpolate
between detonations and deflagrations. These arise when the wall propagates supersonically 
with respect to the plasma ahead of it,
$\xi_w > c_{s,b}$,
but slow enough to perturb it and push it away. 
At the same time, a stable solution in this regime requires that a rarefaction wave forms behind the wall, identical to the one appearing in a detonation profile.
Hydrodynamic analyses and numerical simulations show that such configurations occur
because supersonic deflagrations are unstable: the fluid behind the wall cannot remain at rest, and a rarefaction wave necessarily develops \cite{Kurki-Suonio:1995rrv}. Entropy considerations require that this rarefaction wave be of Jouguet type, so that the fluid velocity immediately behind the wall saturates the sound speed,
\begin{equation}
    v_- = c_{s,b} .
\end{equation}
In a hybrid solution, therefore, the wall velocity $\xi_w$ satisfies
\begin{equation}
    \xi_w > v_- = c_{s,b} > v_+ ,
\end{equation}
and the matching conditions at the wall do not identify $\xi_w$ with either $v_+$ or $v_-$. Here again, the method to solve the matching conditions is similar to the one for deflagrations, with the only difference that $v_- \neq \xi_w$.

\vspace{0.5 em}

As the wall velocity increases, the deflagration (shock–heated) portion of the solution
becomes progressively thinner. In the limit that the shock becomes arbitrarily weak, the
hybrid solution approaches a Jouguet detonation, in which the heated region in front of the bubble
disappears entirely.

\subsection{Droplets in local thermal equilibrium}
\label{sec:dropletLTE}

For certain values of $\alpha_N$ and $\xi_w$, one may find a consistent solution to the Higgs and hydrodynamical equations without invoking out--of--equilibrium contributions, which in the local-friction approximation is equivalent to setting $\eta=0$. In such a scenario, the entropy current is conserved and fixes the wall velocity via a third independent matching condition, \cref{eq. 3rd matching lte}, in addition to the conservation of the energy--momentum tensor. Within the bag model, one has:
\begin{equation}
\label{eq:LTEmatch}
  v_+ v_- = \frac{1-(1-3 \alpha_+)r}{3-3(1+\alpha_+)r}, \quad \frac{v_+}{v_-} = \frac{3+(1-3 \alpha_+)r}{1+3(1+\alpha_+)r}, \quad \Psi \equiv \frac{a_-}{a_+} = \frac{1}{r} \left( \frac{\gamma_-}{\gamma_+}\right)^4,  
\end{equation}
where $\alpha_+$ and $r$ are the usual dimensionless variables given by
\begin{equation}
    \alpha_+ \equiv \frac{V_0}{a_+ T_+^4}, \quad r \equiv \frac{\omega_+}{\omega_-} = \frac{a_+ T_+^4}{a_- T_-^4}.
\end{equation} 
$\Psi$ describes the change in the effective relativistic degrees of freedom from the $+$ to the $-$ phase 
and is related to $\delta a/a$ as $\Psi  = 1- \delta a/ a$.

By fixing the value of $\Psi$, the strength of the phase transition via $\alpha_+$, and the hydrodynamical mode (\emph{e.g.}, whether it is an expanding bubble or a droplet), one can determine the resulting wall velocity. 
For standard deflagrations, for instance, one has $\xi_w = v_-$, while for a droplet one identifies its velocity with $\xi_d = -v_+$, as the fluid is at rest inside the droplet. Once a solution to the system in \eqref{eq:LTEmatch} is determined, it can in principle describe both droplets and expanding bubbles within the same particle physics model.

The question we wish to address here is whether droplets may form in the final stage of the phase transition also within the LTE approximation.
In other words, we would like to determine whether or not friction is a necessary ingredient for the formation of droplets in this context.
This should be considered as a special case of the analysis in \cref{sec:etanonzerodroplets} with $\eta=0$.

Rather than presenting a general study, we shall focus on a particular class of solutions that can be considered a proof of principle for the formation of droplets without friction. To this end, let us focus on the LTE solution for the original deflagration, and consider our candidate droplet to be described by the \emph{same} value of $\alpha_+$ and $v_{\pm}$. As mentioned above, this droplet solution will have $|\xi_d| = v_+$, while the original deflagration is such that $\xi_w = v_-$. Notice that, since the deflagration needs to satisfy 
\begin{equation}
v_+ < v_- < c_s,
\end{equation}
we are guaranteed that our candidate droplet solution will automatically fall in the consistent region for the hydrodynamics indicated by the white region in \cref{fig:droplet_self_similar}.

In addition to hydrodynamical consistency, however, droplets forming towards the end of the phase transition also need to satisfy energy conservation as given in \cref{eq:encons}.
For the droplet under consideration, such condition can be met by an appropriate choice of $\Psi$ and $\alpha_+$ (or equivalently, $\alpha_N$) for the original deflagration.
A concrete example is provided by the following benchmark point:
\begin{equation}
    \Psi = 1 - \frac{\delta a}{a} = 0.84, \quad \alpha_+ = 0.05 \quad \Rightarrow \quad \xi_w = v_- = 0.42, \quad -\xi_d = v_+ = 0.32,
\end{equation}
where the original LTE deflagration has $\alpha_N \simeq 0.06$. Let us stress again that this example has been constructed by taking $\alpha_+$ to be the same for \emph{both} the droplet and the original deflagration, and we expect more LTE solutions to appear when relaxing this assumption.

We thus conclude that the formation of droplets is not necessarily linked to the presence of out--of--equilibrium dynamics for the bubble walls, but can also be realized, in scenarios with negligible friction that can be described within the LTE approximation, as long as hydrodynamical consistency and energy conservation is concerned.

\newpage
\bibliography{LG_refs}

@article{Espinosa:2010hh,
    author = "Espinosa, Jose R. and Konstandin, Thomas and No, Jose M. and Servant, Geraldine",
    title = "{Energy Budget of Cosmological First-order Phase Transitions}",
    eprint = "1004.4187",
    archivePrefix = "arXiv",
    primaryClass = "hep-ph",
    reportNumber = "CERN-PH-TH-2010-027",
    doi = "10.1088/1475-7516/2010/06/028",
    journal = "JCAP",
    volume = "06",
    pages = "028",
    year = "2010"
}

@article{Konstandin:2014zta,
    author = "Konstandin, Thomas and Nardini, Germano and Rues, Ingo",
    title = "{From Boltzmann equations to steady wall velocities}",
    eprint = "1407.3132",
    archivePrefix = "arXiv",
    primaryClass = "hep-ph",
    reportNumber = "DESY-14-127, NSF-KITP-14-089",
    doi = "10.1088/1475-7516/2014/09/028",
    journal = "JCAP",
    volume = "09",
    pages = "028",
    year = "2014"
}

@article{Dorsch:2024jjl,
    author = "Dorsch, Gl{\'a}uber C. and Konstandin, Thomas and Perboni, Enrico and Pinto, Daniel A.",
    title = "{Non-singular solutions to the Boltzmann equation with a fluid Ansatz}",
    eprint = "2412.09266",
    archivePrefix = "arXiv",
    primaryClass = "hep-ph",
    reportNumber = "DESY-24-193",
    doi = "10.1088/1475-7516/2025/04/033",
    journal = "JCAP",
    volume = "04",
    pages = "033",
    year = "2025"
}

@article{Cutting:2022zgd,
    author = "Cutting, Daniel and Vilhonen, Essi and Weir, David J.",
    title = "{Droplet collapse during strongly supercooled transitions}",
    eprint = "2204.03396",
    archivePrefix = "arXiv",
    primaryClass = "astro-ph.CO",
    reportNumber = "HIP-2022-5/TH",
    doi = "10.1103/PhysRevD.106.103524",
    journal = "Phys. Rev. D",
    volume = "106",
    number = "10",
    pages = "103524",
    year = "2022"
}

@article{Cutting:2019zws,
    author = "Cutting, Daniel and Hindmarsh, Mark and Weir, David J.",
    title = "{Vorticity, kinetic energy, and suppressed gravitational wave production in strong first order phase transitions}",
    eprint = "1906.00480",
    archivePrefix = "arXiv",
    primaryClass = "hep-ph",
    reportNumber = "HIP-2019-15/TH",
    doi = "10.1103/PhysRevLett.125.021302",
    journal = "Phys. Rev. Lett.",
    volume = "125",
    number = "2",
    pages = "021302",
    year = "2020"
}

@article{Ai:2021kak,
    author = "Ai, Wen-Yuan and Garbrecht, Bjorn and Tamarit, Carlos",
    title = "{Bubble wall velocities in local equilibrium}",
    eprint = "2109.13710",
    archivePrefix = "arXiv",
    primaryClass = "hep-ph",
    reportNumber = "CP3-21-53, TUM-HEP-1365-21",
    doi = "10.1088/1475-7516/2022/03/015",
    journal = "JCAP",
    volume = "03",
    number = "03",
    pages = "015",
    year = "2022"
}

@article{Ai:2023see,
    author = "Ai, Wen-Yuan and Laurent, Benoit and van de Vis, Jorinde",
    title = "{Model-independent bubble wall velocities in local thermal equilibrium}",
    eprint = "2303.10171",
    archivePrefix = "arXiv",
    primaryClass = "astro-ph.CO",
    reportNumber = "KCL-PH-TH/2023-19",
    doi = "10.1088/1475-7516/2023/07/002",
    journal = "JCAP",
    volume = "07",
    pages = "002",
    year = "2023"
}

@article{Grojean:2004xa,
    author = "Grojean, Servant, Wells",
    title= "First-order electroweak phase
			transition in the standard model with a low cutoff",
    doi = "10.1103/PhysRevD.71.036001"

}

@article{Correia:2025qif,
    author = "Correia, Jos{\'e} and Hindmarsh, Mark and Rummukainen, Kari and Weir, David J.",
    title = "{Gravitational waves from strong first-order phase transitions}",
    eprint = "2505.17824",
    archivePrefix = "arXiv",
    primaryClass = "astro-ph.CO",
    doi = "10.1103/8wmq-f635",
    journal = "Phys. Rev. D",
    volume = "112",
    number = "12",
    pages = "123546",
    year = "2025"
}

@article{Ajmi:2022nmq,
    author = "Ajmi, Mudhahir Al and Hindmarsh, Mark",
    title = "{Thermal suppression of bubble nucleation at first-order phase transitions in the early Universe}",
    eprint = "2205.04097",
    archivePrefix = "arXiv",
    primaryClass = "gr-qc",
    reportNumber = "HIP-2022-3/TH",
    doi = "10.1103/PhysRevD.106.023505",
    journal = "Phys. Rev. D",
    volume = "106",
    number = "2",
    pages = "023505",
    year = "2022"
}

@article{Barni:2024lkj,
    author = "Barni, Giulio and Blasi, Simone and Vanvlasselaer, Miguel",
    title = "{The hydrodynamics of inverse phase transitions}",
    eprint = "2406.01596",
    archivePrefix = "arXiv",
    primaryClass = "hep-ph",
    doi = "10.1088/1475-7516/2024/10/042",
    journal = "JCAP",
    volume = "10",
    pages = "042",
    year = "2024"
}

@article{Ekstedt:2025awx,
    author = "Ekstedt, Andreas and Konstandin, Thomas and van de Vis, Jorinde",
    title = "{Scalar damping in cosmological phase transitions}",
    eprint = "2512.16663",
    archivePrefix = "arXiv",
    primaryClass = "hep-ph",
    reportNumber = "DESY-25-195, CERN-TH-2025-262",
    month = "12",
    year = "2025"
}

@misc{amaroseoane2017laserinterferometerspaceantenna,
      title={Laser Interferometer Space Antenna}, 
      author={Pau Amaro-Seoane and Heather Audley and Stanislav Babak and John Baker and Enrico Barausse and Peter Bender and Emanuele Berti and Pierre Binetruy and Michael Born and Daniele Bortoluzzi and Jordan Camp and Chiara Caprini and Vitor Cardoso and Monica Colpi and John Conklin and Neil Cornish and Curt Cutler and Karsten Danzmann and Rita Dolesi and Luigi Ferraioli and Valerio Ferroni and Ewan Fitzsimons and Jonathan Gair and Lluis Gesa Bote and Domenico Giardini and Ferran Gibert and Catia Grimani and Hubert Halloin and Gerhard Heinzel and Thomas Hertog and Martin Hewitson and Kelly Holley-Bockelmann and Daniel Hollington and Mauro Hueller and Henri Inchauspe and Philippe Jetzer and Nikos Karnesis and Christian Killow and Antoine Klein and Bill Klipstein and Natalia Korsakova and Shane L Larson and Jeffrey Livas and Ivan Lloro and Nary Man and Davor Mance and Joseph Martino and Ignacio Mateos and Kirk McKenzie and Sean T McWilliams and Cole Miller and Guido Mueller and Germano Nardini and Gijs Nelemans and Miquel Nofrarias and Antoine Petiteau and Paolo Pivato and Eric Plagnol and Ed Porter and Jens Reiche and David Robertson and Norna Robertson and Elena Rossi and Giuliana Russano and Bernard Schutz and Alberto Sesana and David Shoemaker and Jacob Slutsky and Carlos F. Sopuerta and Tim Sumner and Nicola Tamanini and Ira Thorpe and Michael Troebs and Michele Vallisneri and Alberto Vecchio and Daniele Vetrugno and Stefano Vitale and Marta Volonteri and Gudrun Wanner and Harry Ward and Peter Wass and William Weber and John Ziemer and Peter Zweifel},
      year={2017},
      eprint={1702.00786},
      archivePrefix={arXiv},
      primaryClass={astro-ph.IM},
      url={https://arxiv.org/abs/1702.00786}, 
}

@article{Laurent:2022jrs,
    author = "Laurent, Benoit and Cline, James M.",
    title = "{First principles determination of bubble wall velocity}",
    eprint = "2204.13120",
    archivePrefix = "arXiv",
    primaryClass = "hep-ph",
    doi = "10.1103/PhysRevD.106.023501",
    journal = "Phys. Rev. D",
    volume = "106",
    number = "2",
    pages = "023501",
    year = "2022"
}

@article{Ekstedt:2024fyq,
    author = "Ekstedt, Andreas and Gould, Oliver and Hirvonen, Joonas and Laurent, Benoit and Niemi, Lauri and Schicho, Philipp and van de Vis, Jorinde",
    title = "{How fast does the WallGo? A package for computing wall velocities in first-order phase transitions}",
    eprint = "2411.04970",
    archivePrefix = "arXiv",
    primaryClass = "hep-ph",
    reportNumber = "CERN-TH-2024-174, DESY-24-162, HIP-2024-21/TH",
    doi = "10.1007/JHEP04(2025)101",
    journal = "JHEP",
    volume = "04",
    pages = "101",
    year = "2025"
}

@article{Ai:2024btx,
    author = "Ai, Wen-Yuan and Laurent, Benoit and van de Vis, Jorinde",
    title = "{Bounds on the bubble wall velocity}",
    eprint = "2411.13641",
    archivePrefix = "arXiv",
    primaryClass = "hep-ph",
    reportNumber = "CERN-TH-2024-198, KCL-PH-TH/2024-57",
    doi = "10.1007/JHEP02(2025)119",
    journal = "JHEP",
    volume = "02",
    pages = "119",
    year = "2025"
}

@article{Hindmarsh:2013xza,
    author = "Hindmarsh, Mark and Huber, Stephan J. and Rummukainen, Kari and Weir, David J.",
    title = "{Gravitational waves from the sound of a first order phase transition}",
    eprint = "1304.2433",
    archivePrefix = "arXiv",
    primaryClass = "hep-ph",
    reportNumber = "HIP-2013-07-TH",
    doi = "10.1103/PhysRevLett.112.041301",
    journal = "Phys. Rev. Lett.",
    volume = "112",
    pages = "041301",
    year = "2014"
}

@article{Hindmarsh:2015qta,
    author = "Hindmarsh, Mark and Huber, Stephan J. and Rummukainen, Kari and Weir, David J.",
    title = "{Numerical simulations of acoustically generated gravitational waves at a first order phase transition}",
    eprint = "1504.03291",
    archivePrefix = "arXiv",
    primaryClass = "astro-ph.CO",
    reportNumber = "HIP-2015-13-TH",
    doi = "10.1103/PhysRevD.92.123009",
    journal = "Phys. Rev. D",
    volume = "92",
    number = "12",
    pages = "123009",
    year = "2015"
}

@article{Hindmarsh:2017gnf,
    author = "Hindmarsh, Mark and Huber, Stephan J. and Rummukainen, Kari and Weir, David J.",
    title = "{Shape of the acoustic gravitational wave power spectrum from a first order phase transition}",
    eprint = "1704.05871",
    archivePrefix = "arXiv",
    primaryClass = "astro-ph.CO",
    reportNumber = "HIP-2017-02-TH, HIP-2017-02/TH",
    doi = "10.1103/PhysRevD.96.103520",
    journal = "Phys. Rev. D",
    volume = "96",
    number = "10",
    pages = "103520",
    year = "2017",
    note = "[Erratum: Phys.Rev.D 101, 089902 (2020)]"
}

@article{Huber:2013kj,
    author = "Huber, Stephan J. and Sopena, Miguel",
    title = "{An efficient approach to electroweak bubble velocities}",
    eprint = "1302.1044",
    archivePrefix = "arXiv",
    primaryClass = "hep-ph",
    month = "2",
    year = "2013"
}

@article{Croon:2020cgk,
    author = "Croon, Djuna and Gould, Oliver and Schicho, Philipp and Tenkanen, Tuomas V. I. and White, Graham",
    title = "{Theoretical uncertainties for cosmological first-order phase transitions}",
    eprint = "2009.10080",
    archivePrefix = "arXiv",
    primaryClass = "hep-ph",
    reportNumber = "HIP-2020-26/TH",
    doi = "10.1007/JHEP04(2021)055",
    journal = "JHEP",
    volume = "04",
    pages = "055",
    year = "2021"
}

@article{Caprini:2019egz,
    author = "Caprini, Chiara and others",
    title = "{Detecting gravitational waves from cosmological phase transitions with LISA: an update}",
    eprint = "1910.13125",
    archivePrefix = "arXiv",
    primaryClass = "astro-ph.CO",
    reportNumber = "DESY-19-159, IPPP/19/27, HIP-2019-14/TH, MITP/19-066, IFT-UAM/CSIC-19-139",
    doi = "10.1088/1475-7516/2020/03/024",
    journal = "JCAP",
    volume = "03",
    pages = "024",
    year = "2020"
}

@article{Chala:2025aiz,
    author = "Chala, Mikael and Guedes, Guilherme",
    title = "{The high-temperature limit of the SM(EFT)}",
    eprint = "2503.20016",
    archivePrefix = "arXiv",
    primaryClass = "hep-ph",
    doi = "10.1007/JHEP07(2025)085",
    journal = "JHEP",
    volume = "07",
    pages = "085",
    year = "2025"
}

@article{Caprini:2024hue,
    author = "Caprini, Chiara and Jinno, Ryusuke and Lewicki, Marek and Madge, Eric and Merchand, Marco and Nardini, Germano and Pieroni, Mauro and Roper Pol, Alberto and Vaskonen, Ville",
    collaboration = "LISA Cosmology Working Group",
    title = "{Gravitational waves from first-order phase transitions in LISA: reconstruction pipeline and physics interpretation}",
    eprint = "2403.03723",
    archivePrefix = "arXiv",
    primaryClass = "astro-ph.CO",
    reportNumber = "LISA-COSWG-24-01, CERN-TH-2024-029",
    doi = "10.1088/1475-7516/2024/10/020",
    journal = "JCAP",
    volume = "10",
    pages = "020",
    year = "2024"
}

@article{Jinno:2022mie,
    author = "Jinno, Ryusuke and Konstandin, Thomas and Rubira, Henrique and Stomberg, Isak",
    title = "{Higgsless simulations of cosmological phase transitions and gravitational waves}",
    eprint = "2209.04369",
    archivePrefix = "arXiv",
    primaryClass = "astro-ph.CO",
    reportNumber = "DESY 22-148, IFT-UAM/CSIC-22-100, TUM-HEP-1416/22",
    doi = "10.1088/1475-7516/2023/02/011",
    journal = "JCAP",
    volume = "02",
    pages = "011",
    year = "2023"
}

@article{Caprini:2024gyk,
    author = "Caprini, Chiara and Jinno, Ryusuke and Konstandin, Thomas and Roper Pol, Alberto and Rubira, Henrique and Stomberg, Isak",
    title = "{Gravitational waves from first-order phase transitions: from weak to strong}",
    eprint = "2409.03651",
    archivePrefix = "arXiv",
    primaryClass = "gr-qc",
    reportNumber = "KOBE-COSMO-24-03, TUM-HEP-1522/24, DESY-24-131, CERN-TH-2025-072",
    doi = "10.1007/JHEP07(2025)217",
    journal = "JHEP",
    volume = "07",
    pages = "217",
    year = "2025"
}

@article{Giese:2020rtr,
    author = "Giese, Felix and Konstandin, Thomas and van de Vis, Jorinde",
    title = "{Model-independent energy budget of cosmological first-order phase transitions{\textemdash}A sound argument to go beyond the bag model}",
    eprint = "2004.06995",
    archivePrefix = "arXiv",
    primaryClass = "astro-ph.CO",
    reportNumber = "DESY-20-064",
    doi = "10.1088/1475-7516/2020/07/057",
    journal = "JCAP",
    volume = "07",
    number = "07",
    pages = "057",
    year = "2020"
}

@article{Giese:2020znk,
    author = "Giese, Felix and Konstandin, Thomas and Schmitz, Kai and van de Vis, Jorinde",
    title = "{Model-independent energy budget for LISA}",
    eprint = "2010.09744",
    archivePrefix = "arXiv",
    primaryClass = "astro-ph.CO",
    reportNumber = "DESY-20-173, DESY 20-173, CERN-TH-2020-170",
    doi = "10.1088/1475-7516/2021/01/072",
    journal = "JCAP",
    volume = "01",
    pages = "072",
    year = "2021"
}

@article{Moore:1995si,
    author = "Moore, Guy D. and Prokopec, Tomislav",
    title = "{How fast can the wall move? A Study of the electroweak phase transition dynamics}",
    eprint = "hep-ph/9506475",
    archivePrefix = "arXiv",
    reportNumber = "PUPT-1544, PUP-TH-1544, LANCS-TH-9517",
    doi = "10.1103/PhysRevD.52.7182",
    journal = "Phys. Rev. D",
    volume = "52",
    pages = "7182--7204",
    year = "1995"
}

@article{Bodeker:2009qy,
    author = "Bodeker, Dietrich and Moore, Guy D.",
    title = "{Can electroweak bubble walls run away?}",
    eprint = "0903.4099",
    archivePrefix = "arXiv",
    primaryClass = "hep-ph",
    doi = "10.1088/1475-7516/2009/05/009",
    journal = "JCAP",
    volume = "05",
    pages = "009",
    year = "2009"
}

@article{deVries:2017ncy,
    author = "de Vries, Jordy and Postma, Marieke and van de Vis, Jorinde and White, Graham",
    title = "{Electroweak Baryogenesis and the Standard Model Effective Field Theory}",
    eprint = "1710.04061",
    archivePrefix = "arXiv",
    primaryClass = "hep-ph",
    reportNumber = "Nikhef-2017-044",
    doi = "10.1007/JHEP01(2018)089",
    journal = "JHEP",
    volume = "01",
    pages = "089",
    year = "2018"
}

@article{Damgaard:2015con,
    author = "Damgaard, P. H. and Haarr, A. and O'Connell, D. and Tranberg, A.",
    title = "{Effective Field Theory and Electroweak Baryogenesis in the Singlet-Extended Standard Model}",
    eprint = "1512.01963",
    archivePrefix = "arXiv",
    primaryClass = "hep-ph",
    doi = "10.1007/JHEP02(2016)107",
    journal = "JHEP",
    volume = "02",
    pages = "107",
    year = "2016"
}

@article{Postma:2020toi,
    author = "Postma, Marieke and White, Graham",
    title = "{Cosmological phase transitions: is effective field theory just a toy?}",
    eprint = "2012.03953",
    archivePrefix = "arXiv",
    primaryClass = "hep-ph",
    doi = "10.1007/JHEP03(2021)280",
    journal = "JHEP",
    volume = "03",
    pages = "280",
    year = "2021"
}

@article{LIGOScientific:2025slb,
    author = "Abac, A. G. and others",
    collaboration = "LIGO Scientific, VIRGO, KAGRA",
    title = "{GWTC-4.0: Updating the Gravitational-Wave Transient Catalog with Observations from the First Part of the Fourth LIGO-Virgo-KAGRA Observing Run}",
    eprint = "2508.18082",
    archivePrefix = "arXiv",
    primaryClass = "gr-qc",
    reportNumber = "LIGO-P2400386",
    month = "8",
    year = "2025"
}

@article{KAGRA:2021vkt,
    author = "Abbott, R. and others",
    collaboration = "KAGRA, VIRGO, LIGO Scientific",
    title = "{GWTC-3: Compact Binary Coalescences Observed by LIGO and Virgo during the Second Part of the Third Observing Run}",
    eprint = "2111.03606",
    archivePrefix = "arXiv",
    primaryClass = "gr-qc",
    reportNumber = "LIGO-P2000318",
    doi = "10.1103/PhysRevX.13.041039",
    journal = "Phys. Rev. X",
    volume = "13",
    number = "4",
    pages = "041039",
    year = "2023"
}

@article{LIGOScientific:2018mvr,
    author = "Abbott, B. P. and others",
    collaboration = "LIGO Scientific, Virgo",
    title = "{GWTC-1: A Gravitational-Wave Transient Catalog of Compact Binary Mergers Observed by LIGO and Virgo during the First and Second Observing Runs}",
    eprint = "1811.12907",
    archivePrefix = "arXiv",
    primaryClass = "astro-ph.HE",
    reportNumber = "LIGO-P1800307",
    doi = "10.1103/PhysRevX.9.031040",
    journal = "Phys. Rev. X",
    volume = "9",
    number = "3",
    pages = "031040",
    year = "2019"
}

@article{NANOGrav:2023gor,
    author = "Agazie, Gabriella and others",
    collaboration = "NANOGrav",
    title = "{The NANOGrav 15 yr Data Set: Evidence for a Gravitational-wave Background}",
    eprint = "2306.16213",
    archivePrefix = "arXiv",
    primaryClass = "astro-ph.HE",
    doi = "10.3847/2041-8213/acdac6",
    journal = "Astrophys. J. Lett.",
    volume = "951",
    number = "1",
    pages = "L8",
    year = "2023"
}

@article{EPTA:2023fyk,
    author = "Antoniadis, J. and others",
    collaboration = "EPTA, InPTA:",
    title = "{The second data release from the European Pulsar Timing Array - III. Search for gravitational wave signals}",
    eprint = "2306.16214",
    archivePrefix = "arXiv",
    primaryClass = "astro-ph.HE",
    doi = "10.1051/0004-6361/202346844",
    journal = "Astron. Astrophys.",
    volume = "678",
    pages = "A50",
    year = "2023"
}

@article{Reardon:2023gzh,
    author = "Reardon, Daniel J. and others",
    title = "{Search for an Isotropic Gravitational-wave Background with the Parkes Pulsar Timing Array}",
    eprint = "2306.16215",
    archivePrefix = "arXiv",
    primaryClass = "astro-ph.HE",
    doi = "10.3847/2041-8213/acdd02",
    journal = "Astrophys. J. Lett.",
    volume = "951",
    number = "1",
    pages = "L6",
    year = "2023"
}

@article{Xu:2023wog,
    author = "Xu, Heng and others",
    title = "{Searching for the Nano-Hertz Stochastic Gravitational Wave Background with the Chinese Pulsar Timing Array Data Release I}",
    eprint = "2306.16216",
    archivePrefix = "arXiv",
    primaryClass = "astro-ph.HE",
    doi = "10.1088/1674-4527/acdfa5",
    journal = "Res. Astron. Astrophys.",
    volume = "23",
    number = "7",
    pages = "075024",
    year = "2023"
}

@article{Witten:1984rs,
    author = "Witten, Edward",
    title = "{Cosmic Separation of Phases}",
    reportNumber = "PRINT-84-0400 (IAS,PRINCETON)",
    doi = "10.1103/PhysRevD.30.272",
    journal = "Phys. Rev. D",
    volume = "30",
    pages = "272--285",
    year = "1984"
}

@article{Kosowsky:1991ua,
    author = "Kosowsky, Arthur and Turner, Michael S. and Watkins, Richard",
    title = "{Gravitational Radiation from Colliding Vacuum Bubbles}",
    reportNumber = "FERMILAB-PUB-91-323-A",
    doi = "10.1103/PhysRevD.45.4514",
    journal = "Phys. Rev. D",
    volume = "45",
    pages = "4514--4535",
    year = "1992"
}

@article{vandeVis:2025plm,
    author = "van de Vis, Jorinde and Schicho, Philipp and Niemi, Lauri and Laurent, Benoit and Hirvonen, Joonas and Gould, Oliver",
    title = "{WallGo investigates: Theoretical uncertainties in the bubble wall velocity}",
    eprint = "2510.27691",
    archivePrefix = "arXiv",
    primaryClass = "hep-ph",
    reportNumber = "CERN-TH-2025-221",
    month = "10",
    year = "2025"
}

@article{Laine:1993ey,
    author = "Laine, M.",
    title = "{Bubble growth as a detonation}",
    eprint = "hep-ph/9309242",
    archivePrefix = "arXiv",
    reportNumber = "HU-TFT-93-44",
    doi = "10.1103/PhysRevD.49.3847",
    journal = "Phys. Rev. D",
    volume = "49",
    pages = "3847--3853",
    year = "1994"
}

@article{Kurki-Suonio:1995rrv,
    author = "Kurki-Suonio, H. and Laine, M.",
    title = "{Supersonic deflagrations in cosmological phase transitions}",
    eprint = "hep-ph/9501216",
    archivePrefix = "arXiv",
    reportNumber = "HU-TFT-95-3",
    doi = "10.1103/PhysRevD.51.5431",
    journal = "Phys. Rev. D",
    volume = "51",
    pages = "5431--5437",
    year = "1995"
}

@article{Kamionkowski:1993fg,
    author = "Kamionkowski, Marc and Kosowsky, Arthur and Turner, Michael S.",
    title = "{Gravitational radiation from first order phase transitions}",
    eprint = "astro-ph/9310044",
    archivePrefix = "arXiv",
    reportNumber = "IASSNS-HEP-93-44, FERMILAB-PUB-93-235-A",
    doi = "10.1103/PhysRevD.49.2837",
    journal = "Phys. Rev. D",
    volume = "49",
    pages = "2837--2851",
    year = "1994"
}

@article{Ignatius:1993qn,
    author = "Ignatius, J. and Kajantie, K. and Kurki-Suonio, H. and Laine, M.",
    title = "{The growth of bubbles in cosmological phase transitions}",
    eprint = "astro-ph/9309059",
    archivePrefix = "arXiv",
    reportNumber = "HU-TFT-93-43",
    doi = "10.1103/PhysRevD.49.3854",
    journal = "Phys. Rev. D",
    volume = "49",
    pages = "3854--3868",
    year = "1994"
}

@article{Kuzmin:1985mm,
    author = "Kuzmin, V. A. and Rubakov, V. A. and Shaposhnikov, M. E.",
    title = "{On the Anomalous Electroweak Baryon Number Nonconservation in the Early Universe}",
    reportNumber = "IC/85/8",
    doi = "10.1016/0370-2693(85)91028-7",
    journal = "Phys. Lett. B",
    volume = "155",
    pages = "36",
    year = "1985"
}

@article{Kurki-Suonio:1995yaf,
    author = "Kurki-Suonio, H. and Laine, M.",
    title = "{On bubble growth and droplet decay in cosmological phase transitions}",
    eprint = "hep-ph/9512202",
    archivePrefix = "arXiv",
    reportNumber = "HU-TFT-95-71, HD-THEP-95-52",
    doi = "10.1103/PhysRevD.54.7163",
    journal = "Phys. Rev. D",
    volume = "54",
    pages = "7163--7171",
    year = "1996"
}

@article{Rezzolla:1995kv,
    author = "Rezzolla, L. and Miller, J. C. and Pantano, O.",
    title = "{Evaporation of quark drops during the cosmological quark - hadron transition}",
    eprint = "astro-ph/9502064",
    archivePrefix = "arXiv",
    reportNumber = "SISSA-15-95-A-EP",
    doi = "10.1103/PhysRevD.52.3202",
    journal = "Phys. Rev. D",
    volume = "52",
    pages = "3202--3213",
    year = "1995"
}

@article{Steinhardt:1981ct,
    author = "Steinhardt, Paul Joseph",
    title = "{Relativistic Detonation Waves and Bubble Growth in False Vacuum Decay}",
    reportNumber = "UPR-0181T",
    doi = "10.1103/PhysRevD.25.2074",
    journal = "Phys. Rev. D",
    volume = "25",
    pages = "2074",
    year = "1982"
}

@article{Rezzolla:1995br,
    author = "Rezzolla, Luciano and Miller, John C.",
    title = "{Evaporation of cosmological quark drops and relativistic radiative transfer}",
    eprint = "astro-ph/9510039",
    archivePrefix = "arXiv",
    reportNumber = "SISSA-116-95-A",
    doi = "10.1103/PhysRevD.53.5411",
    journal = "Phys. Rev. D",
    volume = "53",
    pages = "5411--5425",
    year = "1996"
}

@article{Vachaspati:1991nm,
    author = "Vachaspati, T.",
    title = "{Magnetic fields from cosmological phase transitions}",
    doi = "10.1016/0370-2693(91)90051-Q",
    journal = "Phys. Lett. B",
    volume = "265",
    pages = "258--261",
    year = "1991"
}

@article{Kajantie:1996mn,
    author = "Kajantie, K. and Laine, M. and Rummukainen, K. and Shaposhnikov, Mikhail E.",
    title = "{Is there a~ hot electroweak phase transition at $m_H \gtrsim m_W$?}",
    eprint = "hep-ph/9605288",
    archivePrefix = "arXiv",
    reportNumber = "CERN-TH-96-126, HD-THEP-96-15, IUHET-333",
    doi = "10.1103/PhysRevLett.77.2887",
    journal = "Phys. Rev. Lett.",
    volume = "77",
    pages = "2887--2890",
    year = "1996"
}

@article{Gurtler:1997hr,
    author = "Gurtler, M. and Ilgenfritz, Ernst-Michael and Schiller, A.",
    title = "{Where the electroweak phase transition ends}",
    eprint = "hep-lat/9704013",
    archivePrefix = "arXiv",
    reportNumber = "UL-NTZ-10-97, HUB-EP-97-24, DESY-97-086",
    doi = "10.1103/PhysRevD.56.3888",
    journal = "Phys. Rev. D",
    volume = "56",
    pages = "3888--3895",
    year = "1997"
}

@article{Blasi:2022woz,
    author = "Blasi, Simone and Mariotti, Alberto",
    title = "{Domain Walls Seeding the Electroweak Phase Transition}",
    eprint = "2203.16450",
    archivePrefix = "arXiv",
    primaryClass = "hep-ph",
    doi = "10.1103/PhysRevLett.129.261303",
    journal = "Phys. Rev. Lett.",
    volume = "129",
    number = "26",
    pages = "261303",
    year = "2022"
}

@article{Agrawal:2023cgp,
    author = "Agrawal, Prateek and Blasi, Simone and Mariotti, Alberto and Nee, Michael",
    title = "{Electroweak phase transition with a double well done doubly well}",
    eprint = "2312.06749",
    archivePrefix = "arXiv",
    primaryClass = "hep-ph",
    reportNumber = "DESY-23-208",
    doi = "10.1007/JHEP06(2024)089",
    journal = "JHEP",
    volume = "06",
    pages = "089",
    year = "2024"
}

@article{Blasi:2023rqi,
    author = "Blasi, Simone and Jinno, Ryusuke and Konstandin, Thomas and Rubira, Henrique and Stomberg, Isak",
    title = "{Gravitational waves from defect-driven phase transitions: domain walls}",
    eprint = "2302.06952",
    archivePrefix = "arXiv",
    primaryClass = "astro-ph.CO",
    doi = "10.1088/1475-7516/2023/10/051",
    journal = "JCAP",
    volume = "10",
    pages = "051",
    year = "2023"
}

@article{Bai:2025qch,
    author = "Bai, Yang and Xu, Yifu and Yang, Yiming",
    title = "{Heterogeneous Cosmological Phase Transitions: Seeded by Domain Walls and Junctions}",
    eprint = "2512.10917",
    archivePrefix = "arXiv",
    primaryClass = "hep-ph",
    month = "12",
    year = "2025"
}

@article{Li:2023yzq,
    author = "Li, Yang and Jia, Yongtao and Bian, Ligong",
    title = "{Numerical simulation of domain wall and first-order phase transition in an expanding universe}",
    eprint = "2304.05220",
    archivePrefix = "arXiv",
    primaryClass = "hep-ph",
    doi = "10.1088/1475-7516/2025/02/038",
    journal = "JCAP",
    volume = "02",
    pages = "038",
    year = "2025"
}
\bibliographystyle{JHEP}

\end{document}